\documentclass[12pt,thmsa]{article}
\usepackage{amssymb}
\usepackage{epsfig}

\usepackage{sw20jart}

\input{tcilatex}
\begin{document}

\title{Continuous Self-Similar Evaporation of a Rotating Cosmic String Loop}
\author{Malcolm Anderson \\
Department of Mathematics\\
Universiti Brunei Darussalam\\
Jalan Tungku Link, Gadong BE 1410\\
Negara Brunei Darussalam}
\date{}
\maketitle

\begin{abstract}
\noindent A solution of the linearized Einstein and Nambu-Goto equations is
constructed which describes the evaporation of a certain type of rotating
cosmic string -- the Allen-Casper-Ottewill loop -- under the action of its
own self-gravity. The solution evaporates self-similarly, and radiates away
all its mass-energy and momentum in a finite time. Furthermore, the
corresponding weak-field metric can be matched to a remnant Minkowski
spacetime at all points on the future light cone of the final evaporation
point of the loop.\bigskip

Short Title: Evaporation of a String Loop\bigskip

\noindent PACS\ numbers: 04.25.Nx, 98.80.Cq
\end{abstract}

\section{Introduction}

Cosmic strings are thin filaments of topologically-trapped Higgs field
energy whose dynamics, in the zero-thickness limit, are controlled by the
Nambu-Goto action. In this limit cosmic strings are effectively line
singularities which move under the action of their own elastic tension, and
radiate gravitational energy and momentum in the process. They may have
played a role in the formation of large-scale structure in the early
Universe (see \cite{Vile-Shell} for a review), but their gravitational
properties are still only poorly understood, and they consequently remain
problematic objects for general relativity. In particular, a rigorous
distributional description of line singularities has not yet been developed
within the framework of general relativity, nor has it been demonstrated
that zero-thickness cosmic strings can always be interpreted as the unique
limits of well-behaved finite-thickness solutions \cite{Geroch}.

At present, all known exact self-gravitating string solutions in a Minkowski
background describe either infinite straight strings \cite{Gott, Hiscock,
Xanth, Garri} or infinite strings interacting with plane-fronted
gravitational waves \cite{Garf3, Frolov}, and the only general result
available on the dynamics of string \emph{loops} in strong-field gravity is
Hawking's proof that a collapsing circular loop would form an event horizon
after radiating away at most 29\% of its original energy \cite{Hawking}. (In
addition, the full 3-metric outside a circular loop at a moment of time
symmetry has been constructed in \cite{FIU}.)

Given the complexity of the general strong-field problem, therefore, the
most promising arena for the study of loop self-gravity would appear to be
the weak-field limit. Fortunately, it can be shown that the self-gravity of
a GUT string would almost everywhere be small enough to justify a weak-field
treatment \cite{Quash}, although this treatment typically breaks down in the
vicinity of certain common pathological features known as kinks and cusps.

In \cite{Anderson3} (henceforth referred to as Paper I) it was shown that a
certain type of rigidly-rotating cosmic string loop -- the
Allen-Casper-Ottewill solution \cite{ACO1} -- evolves by self-similar
shrinkage in the weak-field limit. This is the first explicit back-reaction
calculation to have been published for a cosmic string loop, as all previous
calculations have relied on numerical approximation \cite{Quash}. However,
the calculation relies on what might be called an adiabatic approximation,
as the total flux of gravitational energy and momentum from the loop is
integrated over a single oscillation period of the loop's unperturbed
trajectory, and is shown to induce no changes other than a small decrease in
the loop's energy and length, plus a comparable rotational phase shift.

In this paper I extend the analysis of Paper I\ by relaxing the adiabatic
assumption and constructing a self-similar evaporating solution of the
linearized field equations which is identical to the Allen-Casper-Ottewill
solution on spacelike sections and ultimately radiates away all its
mass-energy. Although the trajectory of the string loop, and the
perturbations it induces in the background metric, are only minimally
different from the corresponding quantities in the adiabatic case, the
evaporating solution represents an incremental advance in the understanding
of loop self-gravity. In particular, the metric perturbations can be
calculated at all points outside the string's worldsheet, and the perturbed
(radiating) spacetime can be matched to a flat spacetime which occupies the
causal future of the final point of evaporation of the loop. The naive
expectation that the loop evaporates completely to leave behind a vacuum
remnant spacetime therefore appears to be consistent at this level of
approximation.

\section{Preliminaries}

The mathematical formalism used to describe zero-thickness cosmic strings
has been discussed in some detail in Paper I, and in this section I will
offer only a brief summary.

Throughout the paper, $x^{a}\equiv [x^{0},x^{1},x^{2},x^{3}]=[t,x,y,z]^{a}$
are local coordinates on the four-dimensional background spacetime $(\mathbf{%
M},g_{ab})$, and the metric tensor $g_{ab}$ has signature $(+,-,-,-)$, so
that timelike vectors have positive norm. The world sheet $\mathbf{T}$ of
the string is the two-dimensional surface it traces out as it moves, and is
described parametrically by a set of equations of the form $%
x^{a}=X^{a}(\zeta ^{A})$, where the parameters $\zeta ^{A}\equiv (\zeta
^{0},\zeta ^{1})$ are gauge coordinates.

The intrinsic two-metric induced on $\mathbf{T}$ by a given choice of gauge
coordinates is:
\begin{equation}
\gamma _{AB}=g_{ab}\,X^{a},_{A}\,X^{b},_{B}
\end{equation}
where $X^{a},_{A}$ is shorthand for $\partial X^{a}/\partial \zeta ^{A}$.
The intrinsic metric $\gamma _{AB}$ is assumed to be non-degenerate with
signature $(+,-)$ almost everywhere on $\mathbf{T}$. If $\gamma $ denotes $%
|\det (\gamma _{AB})|$ the Nambu-Goto action \cite{Nambu, Goto} has the form
\begin{equation}
I=-\mu \int \gamma ^{1/2}d^{2}\zeta  \label{action}
\end{equation}
where $\mu $ is the (constant) mass per unit length of the string.

The stress-energy tensor $T^{ab}$ induced by a given trajectory is found by
varying the Nambu-Goto action with respect to the background metric $g_{ab}$%
, according to the standard prescription $\delta I=-\tfrac{1}{2}\int
T^{ab}\delta g_{ab}g^{1/2}d^{4}x$, where $g\equiv |\det (g_{ab})|$. The
equation for $\delta I$ inverts to give
\begin{equation}
T^{ab}(x^{c})=-2g^{-1/2}\int (\delta \mathcal{L}/\delta g_{ab})\,\delta
^{(4)}(x^{c}-X^{c})d^{2}\zeta  \label{se1}
\end{equation}
where $\mathcal{L}\equiv -\mu \gamma ^{1/2}$ is the Lagrangian density in (%
\ref{action}), and
\begin{equation}
\delta \mathcal{L}/\delta g_{ab}=-\tfrac{1}{2}\mu \gamma ^{1/2}\gamma
^{AB}X^{a},_{A}X^{b},_{B}  \label{se2}
\end{equation}
with $\gamma ^{AB}$ the inverse of $\gamma _{AB}$.

In Paper I the gauge choice $\gamma _{AB}=\kappa \eta _{AB}$ was imposed
from the outset, where $\eta _{AB}=\,$diag$\,(1,-1)$ is the Minkowski
2-tensor. However, it turns out that this choice is not as convenient when
the string is evaporating (and the trajectory is no longer time-periodic),
and for this reason the more general versions of the string stress-energy
and equation of motion will be retained in what follows.

The extrinsic curvature tensor of the world sheet $\mathbf{T}$ has the
general form
\begin{equation}
K^{abc}=p_{m}^{a}p^{bn}\nabla _{n}p^{mc}
\end{equation}
where $\nabla _{n}$ is the derivative operator associated with $g_{ab}$, and
$p^{ab}\equiv \gamma ^{AB}X^{a},_{A}X^{b},_{B}$ is the projection tensor
corresponding to $\gamma _{AB}$. In terms of the derivatives of the position
function $X^{a}$ the curvature tensor $K^{abc}$ and its trace $K^{c}\equiv
g_{ab}K^{abc}$ can be written as
\begin{equation}
K^{abc}=q_{d}^{c}(\gamma ^{AC}\gamma
^{BD}X^{a},_{A}X^{b},_{B}X^{d},_{CD}+p^{am}p^{bn}\Gamma _{mn}^{d})
\end{equation}
and
\begin{equation}
K^{c}=q_{d}^{c}(\gamma ^{CD}X^{d},_{CD}+p^{mn}\Gamma _{mn}^{d})
\end{equation}
with $q^{ab}\equiv g^{ab}-p^{ab}$ the orthogonal complement of $p^{ab}$, and
$\Gamma _{bc}^{a}$ the Christoffel symbol associated with $g_{ab}$. The
equation of motion of the string is recovered by setting the functional
derivative $\delta \mathcal{L}/\delta X^{a}$ equal to zero, and reads
simply:
\begin{equation}
K^{c}=0  \label{eqmo1}
\end{equation}

In the absence of any sources of stress-energy other than the string itself,
the system of equations for the metric and the string's trajectory is closed
by imposing the Einstein equation $G^{ab}=-8\pi T^{ab}$, where $%
G^{ab}=R^{ab}-\frac{1}{2}g^{ab}R$ is the Einstein tensor.\footnote{%
The convention adopted here for the Riemann tensor is that
\[
R_{abcd}=\tfrac{1}{2}
(g_{ac},_{bd}+g_{bd},_{ac}-g_{ad},_{bc}-g_{bc},_{ad})+g_{ef}\,(\Gamma
_{ac}^{e}\Gamma _{bd}^{f}-\Gamma _{ad}^{e}\Gamma _{bc}^{f}).
\]
Also, geometrized units have been chosen, so that the gravitational constant
$G$ and the speed of light $c$ are set to $1$.} Equations (\ref{se1}), (\ref
{se2}) and (\ref{eqmo1}), together with the Einstein equation, constitute
the strong-field back-reaction problem. As was mentioned above, the only
known solutions to the strong-field back-reaction problem describe either
infinite straight strings (with or without an envelope of cylindrical
gravitational waves \cite{Gott, Hiscock, Xanth, Garri}) or infinite strings
interacting with plane-fronted gravitational waves (the so-called
travelling-wave solutions \cite{Garf3, Frolov}). There are no known exact
loop solutions in a vacuum background, and an analysis of loop evolution is
currently feasible only at the level of the weak-field approximation.

In the weak-field formalism the field and trajectory variables are truncated
at linear order in the mass per unit length $\mu $ of the string, which is
assumed to be small compared to $1$. This is certainly thought to be true of
GUT\ strings, for which $\mu \sim 10^{-6}$. The underlying assumption is
that the solution pair $(g_{ab},X^{a})$ to the strong-field back-reaction
problem can be expanded as perturbative series of the form
\begin{equation}
g_{ab}=\sum_{k=0}^{\infty }\mu ^{k}g_{ab}^{(k)}\text{\qquad and\qquad }%
X^{a}=\sum_{k=0}^{\infty }\mu ^{k}X_{(k)}^{a}
\end{equation}
where the functions $g_{ab}^{(k)}$ and $X_{(k)}^{a}$ are independent of $\mu
$. The weak-field back-reaction problem consists in setting $g_{ab}^{(0)}$
equal to $\eta _{ab}$, the Minkowski metric tensor, and solving the Einstein
equation and equation of motion (\ref{eqmo1}) at linear order in $\mu $ to
obtain the metric perturbation $g_{ab}^{(1)}$ and the trajectory functions $%
X_{(0)}^{a}$ and $X_{(1)}^{a}$.

In what follows, the Minkowski 4-tensor will be taken to have its rigid form
$\eta _{ab}=\,$diag$\,(1,-1,-1,-1)$. In solving the weak-field problem, a
gauge must first be chosen for the metric perturbation $g_{ab}^{(1)}$, and
also for the functions $X_{(0)}^{a}$ and $X_{(1)}^{a}$. If $h_{ab}\equiv \mu
g_{ab}^{(1)}$ the standard harmonic gauge conditions read $h_{a}^{b},_{b}=%
\frac{1}{2}h,_{a}$ (where $h\equiv h_{b}^{b}$ and indices are everywhere
lowered and raised using $\eta _{ab}$ and its inverse $\eta ^{ab}$). The
Einstein equation $G^{ab}=-8\pi T^{ab}$ then reduces to $\square
h_{ab}=-16\pi S_{ab}$, where $S_{ab}=T_{ab}-\frac{1}{2}\eta _{ab}T_{c}^{c}$
and $\square \equiv \partial _{t}^{2}-\nabla ^{2}$ is the flat-space
d'Alembertian.

With $x^{a}\equiv [t,\mathbf{x}]^{a}$ the retarded solution for $h_{ab}$ is:
\begin{equation}
h_{ab}(t,\mathbf{x})=-4\int \frac{S_{ab}(t^{\prime },\mathbf{x}^{\prime })}{|%
\mathbf{x}-\mathbf{x}^{\prime }|}\,d^{3}x^{\prime }  \label{metpert}
\end{equation}
where $t^{\prime }=t-|\mathbf{x}-\mathbf{x}^{\prime }|$ is the retarded time
at the source point $\mathbf{x}^{\prime }$. Since $g^{1/2}=1$ to leading
order in $h_{ab}$, equations (\ref{se1}) and (\ref{se2}) together give
\begin{equation}
S_{ab}(x^{\prime \,d})=\mu \int \gamma ^{1/2}\gamma
^{AB}(X_{(0)a},_{A}X_{(0)b},_{B}-\tfrac{1}{2}\eta
_{ab}X_{(0)}^{c},_{A}X_{(0)c},_{B})\,\delta ^{(4)}(x^{\prime
\,d}-X_{(0)}^{d})d^{2}\zeta  \label{source1}
\end{equation}
with $\gamma ^{1/2}$ and $\gamma ^{AB}$ evaluated using $\eta _{ab}$ in
place of $g_{ab}$ and $X_{(0)}^{a}$ in place of $X^{a}$.

Furthermore, if $g_{ab}^{(0)}=\eta _{ab}$ then $\Gamma _{mn}^{d}=0$ to
leading order, and at linear order in $h_{ab}$ the equation of motion $%
K^{c}=0$ reads:
\begin{equation}
q_{d}^{c}(\gamma ^{CD}X^{d},_{CD}-\gamma ^{AC}\gamma
^{BD}h_{ab}X^{a},_{A}X^{b},_{B}X^{d},_{CD})=-q^{cd}\gamma
^{AB}X^{a},_{A}X^{b},_{B}(h_{bd},_{a}-\tfrac{1}{2}h_{ab},_{d})
\label{Bat-Cart}
\end{equation}
where $q_{d}^{c}$ and $\gamma ^{AC}$ are evaluated using $\eta _{ab}$ rather
than $g_{ab}$ (but $X^{a}$ is retained in full). As was explained in Paper
I, this equation is just the flat-space version of the Battye-Carter
equation \cite{Cart1, Cart2}.

If $X^{a}$ is now replaced by $X_{(0)}^{a}+\delta X^{a}$ where $\delta
X^{a}\equiv \mu X_{(1)}^{a}$ then, with $q_{d}^{c}$ and $\gamma ^{AB}$
evaluated using $X_{(0)}^{a}$ in place of $X^{a}$, equation (\ref{Bat-Cart})
splits into two parts:
\begin{equation}
q_{d}^{c}\gamma ^{CD}X_{(0)}^{d},_{CD}=0  \label{eqmo-zero}
\end{equation}
and
\begin{eqnarray}
&&q_{d}^{c}\gamma ^{CD}\delta X^{d},_{CD}-2\eta _{db}\gamma
^{AB}q_{a}^{(b}X_{(0)}^{c)},_{A}\delta X^{a},_{B}\gamma
^{CD}X_{(0)}^{d},_{CD}  \nonumber \\
&&-q_{d}^{c}\gamma ^{AC}\gamma ^{BD}(2\eta _{ab}X_{(0)}^{a},_{A}\delta
X^{b},_{B}+h_{ab}X_{(0)}^{a},_{A}X_{(0)}^{b},_{B})X_{(0)}^{d},_{CD}
\nonumber \\
&=&-q^{cd}\gamma ^{AB}X_{(0)}^{a},_{A}X_{(0)}^{b},_{B}(h_{bd},_{a}-\tfrac{1}{%
2}h_{ab},_{d})  \label{eqmo-first}
\end{eqnarray}
at zeroth order and first order in $\mu $ respectively. (Here, round
brackets on indices denote symmetrization, so that $%
q_{a}^{(b}X_{(0)}^{c)},_{A}\equiv \frac{1}{2}[%
q_{a}^{b}X_{(0)}^{c},_{A}+q_{a}^{c}X_{(0)}^{b},_{A}]$.)

The weak-field back-reaction problem, in its most general form, therefore
reduces to equations (\ref{metpert}), (\ref{eqmo-zero}) and (\ref{eqmo-first}%
) for $h_{ab}$, $X_{(0)}^{a}$ and $\delta X^{a}$. In order to specialize
these equations to the problem in hand, it is necessary to first explain the
choice that will be made for $X_{(0)}^{a}$, and the corresponding choice of
gauge coordinates $\zeta ^{0}$ and $\zeta ^{1}$. This will be done in
Sections 3 and 4 below.

\section{The Allen-Casper-Ottewill Loop}

The equation of motion at leading order (\ref{eqmo-zero}) can be solved
exactly by making a suitable choice of gauge coordinates. A common choice is
the so-called standard gauge $\zeta ^{A}=[\tau ,\sigma ]^{A}$, in which $%
\gamma _{AB}$ is everywhere proportional to the Minkowski tensor $\eta
_{AB}= $diag$\,(1,-1)$. Since $\gamma _{AB}=\eta
_{ab}X_{(0)}^{a},_{A}X_{(0)}^{b},_{B}$ the geometric significance of this
gauge choice is that the tangent vectors $X_{(0)}^{a},_{\tau }$ (which is
timelike) and $X_{(0)}^{a},_{\sigma }$ (which is spacelike) are orthogonal
and of equal magnitude. As was demonstrated in Paper I, the equation of
motion (\ref{eqmo-zero}) reduces in the standard gauge to the wave equation
\begin{equation}
X_{(0)}^{a},_{\tau \tau }-X_{(0)}^{a},_{\sigma \sigma }=0
\end{equation}

If the subsidiary gauge alignment condition $X_{(0)}^{0}(\tau ,\sigma )=\tau
$ is imposed then all loop solutions of (\ref{eqmo-zero}) can be written in
the form $X_{(0)}^{a}=[\tau ,\mathbf{X}(\tau ,\sigma )]^{a}$, where

\begin{equation}
\mathbf{X}(\tau ,\sigma )=\tfrac{1}{2}[\mathbf{a}(\tau +\sigma )+\mathbf{b}%
(\tau -\sigma )]  \label{worldsheet}
\end{equation}
for some choice of vector functions $\mathbf{a}$ and $\mathbf{b}$. The gauge
constraints $\gamma _{\tau \tau }+\gamma _{\sigma \sigma }=0$ and $\gamma
_{\sigma \tau }=0$ together imply that $|\mathbf{a}^{\prime }|^{2}=|\mathbf{b%
}^{\prime }|^{2}=1$, where a prime denotes differentiation with respect to
the relevant argument. If the loop is in its centre-of-momentum frame then $%
\mathbf{a}$ and $\mathbf{b}$ are each periodic functions of their arguments
with some parametric period $L$, and since $\mathbf{X}(\tau +L/2,\sigma
+L/2)=\mathbf{X}(\tau ,\sigma )$ for any values of $\tau $ and $\sigma $
when $\mathbf{a}$ and $\mathbf{b}$ are periodic, the fundamental oscillation
period $t_{p}$ of the loop is $L/2$.

In the centre-of-momentum frame, the total four-momentum of the string loop
on any surface of constant $t=\tau $ is:
\begin{equation}
p^{a}=\mu \int_{0}^{L}X_{(0)}^{a},_{\tau }\,d\sigma =\mu \int_{0}^{L}[1,%
\tfrac{1}{2}\mathbf{a}^{\prime }(\tau +\sigma )+\tfrac{1}{2}\mathbf{b}%
^{\prime }(\tau -\sigma )]\,d\sigma =\mu L[1,\mathbf{0}]
\end{equation}
and in particular the energy of the loop is $E=\mu L$. The corresponding
angular momentum of the loop is:
\begin{equation}
\mathbf{J}=\tfrac{1}{4}\mu \int_{0}^{L}(\mathbf{a}\times \mathbf{a}^{\prime
}+\mathbf{b}\times \mathbf{b}^{\prime })\,d\sigma .
\end{equation}

In the weak-field limit, the gravitational power $P$ radiated by an
oscillating compact source can be calculated from the standard expression
for the power per unit solid angle in the wave zone (at distances large
compared to the characteristic size $L$ of the source):

\begin{equation}
\frac{dP}{d\Omega }=\frac{\omega ^{2}}{\pi }\sum_{m=1}^{\infty }m^{2}[\bar{T}%
^{ab}\bar{T}_{ab}^{*}\mathbf{-}\tfrac{1}{2}|\bar{T}_{a}^{a}\mathbf{|}^{2}]
\end{equation}
where $\omega =2\pi /t_{p}$ is the circular frequency of the source, and
\begin{equation}
\bar{T}^{ab}(m,\mathbf{n})=t_{p}^{-1}\int_{\Bbb{R}^{3}}\int_{0}^{t_{p}}%
\,T^{ab}(t,\mathbf{x})\,e^{im\omega (t-\mathbf{n\,}\cdot \,\mathbf{x}%
)}\,dt\,d^{3}x
\end{equation}
is the Fourier transform of its integrated stress-energy, with $\mathbf{n}$
the unit vector in the direction of the field point.

For a string loop in a flat background in the standard gauge, the
stress-energy tensor (\ref{se1}) becomes
\begin{equation}
T^{ab}(x^{c})=\mu \int (X_{(0)}^{a},_{\tau }\,X_{(0)}^{b},_{\tau
}-X_{(0)}^{a},_{\sigma }\,X_{(0)}^{b},_{\sigma })\,\delta
^{(4)}(x^{c}-X_{(0)}^{c})\,d\tau \,d\sigma
\end{equation}
and the power per unit solid angle reduces to
\begin{equation}
\frac{dP}{d\Omega }=8\pi \mu ^{2}\sum_{m=1}^{\infty }m^{2}[(A\cdot
A^{*})(B\cdot B^{*})+|A\cdot B^{*}|^{2}-|A\cdot B|^{2}].
\end{equation}
where
\begin{equation}
A^{c}(m,\mathbf{n})=\tfrac{1}{L}\int_{0}^{L}e^{2\pi im[\sigma _{+}-\,\mathbf{%
n\,}\cdot \,\mathbf{a}(\sigma _{+})]/L}\,[1,\mathbf{a}^{\prime }(\sigma
_{+})]^{c}\,d\sigma _{+}
\end{equation}
and
\begin{equation}
B^{c}(m,\mathbf{n})=\tfrac{1}{L}\int_{0}^{L}e^{2\pi im[\sigma _{-}-\,\mathbf{%
n\,}\cdot \,\mathbf{b}(\sigma _{-})]/L}\,[1,\mathbf{b}^{\prime }(\sigma
_{-})]^{c}\,d\sigma _{-}.
\end{equation}
with the gauge coordinates $\sigma _{+}$ and $\sigma _{-}$ defined by $%
\sigma _{\pm }\equiv \tau \pm \sigma $. (Here and elsewhere the dot product
is taken with respect to the flat-space metric $\eta _{ab}$.)

The radiative efficiency $\gamma ^{0}$ of a string loop is defined to be $%
\gamma ^{0}\equiv \mu ^{-2}P$, where $P=\int \frac{dP}{d\Omega }d\Omega $ is
the total power of the loop. The radiative efficiency has been calculated
analytically for a large class of simple loop trajectories by Allen, Casper
and Ottewill \cite{ACO1}, and numerically for the loop by-products of string
network simulations by Allen and Shellard \cite{All-Shell} and Allen and
Casper \cite{Casp-All}. The simulation studies found that the value of $%
\gamma ^{0}$ is typically of order $65$-$70$, and seems to be bounded below
by about $40$. This observation is consistent with the results of the
analytic study, which found that $\gamma ^{0}$ has a minimum value of about $%
39.0025$, and occurs if the mode functions have the forms
\begin{equation}
\mathbf{a}(\sigma _{+})=(|\sigma _{+}|-\tfrac{1}{4}L)\,\mathbf{\hat{z}}\quad
\quad \text{for\quad }-\tfrac{1}{2}L\leq \sigma _{+}\leq \tfrac{1}{2}L
\label{a-mode}
\end{equation}
and
\begin{equation}
\mathbf{b}(\sigma _{-})=\tfrac{L}{2\pi }[\cos (2\pi \sigma _{-}/L)\,\mathbf{%
\hat{x}}+\sin (2\pi \sigma _{-}/L)\,\mathbf{\hat{y}}]\text{,}  \label{b-mode}
\end{equation}
with $\mathbf{a}$ represented by its even periodic extension when $\sigma
_{+}$ lies outside $[-\frac{1}{2}L,\frac{1}{2}L]$.\footnote{%
Strictly speaking, the $\mathbf{a}$ mode function was represented in both
\cite{ACO1} and Paper I in the form
\[
\mathbf{a}(\sigma _{+})=\left\{
\begin{array}{ll}
\sigma _{+}\,\,\mathbf{\hat{z}} & \text{if\qquad }0\leq \sigma _{+}<\frac{1}{%
2}L \\
(L-\sigma _{+})\,\mathbf{\hat{z}} & \text{if\qquad }\frac{1}{2}L\leq \sigma
_{+}<L
\end{array}
\right.
\]
in which case the centroid of the loop lies at $(x,y,z)=(0,0,\frac{1}{8}L)$.
For the purposes of the analysis presented below it is more convenient to
translate the loop down the $z$-axis by an amount $\frac{1}{8}L$ so that the
centroid lies at the origin. The resulting mode function is then given by (%
\ref{a-mode}).}

The corresponding loop, which I will henceforth refer to as the
Allen-Casper-Ottewill or ACO loop, is rigidly rotating about the $z$-axis
with an angular speed $\omega =4\pi /L$. The evolution of the loop is
illustrated in Figure 1, which shows the $y$-$z$ projection of the loop at
times $\tau -\varepsilon =0$, $L/16$, $L/8$ and $3L/16$ (top row) and $\tau
-\varepsilon =L/4$, $5L/16$, $3L/8$ and $7L/16$ (bottom row), where the time
offset $\varepsilon $ is $0.02L$. The string has been artificially thickened
for the sake of visibility, and the $z$-axis is also shown. The projections
of the loop onto the $x$-$y$ plane are circles of radius $L/(4\pi )$. The
points at the extreme top and bottom of the loop, where the helical segments
meet and the modal tangent vector $\mathbf{a}^{\prime }$ is discontinuous,
are technically known as kinks. They trace out circles in the planes $z=L/8$
and $z=-L/8$. All other points on the loop trace out identical circles,
although with varying phase lags. The net angular momentum of the loop is $%
\mathbf{J}=\tfrac{1}{8\pi }\mu L^{2}\mathbf{\hat{z}}$.

I should stress that it is not currently known whether the ACO solution has
the lowest radiative efficiency of all possible string loops. Allen, Casper
and Ottewill \cite{ACO1} showed only that the ACO solution minimizes $\gamma
^{0}$ over a special class of solutions with the mode function $\mathbf{a}$
given by (\ref{a-mode}) and the mode function $\mathbf{b}$ confined to the $%
x $-$y$ plane but otherwise arbitrary. Nonetheless, in Paper I the
first-order equation of motion (\ref{eqmo-first}) was solved for a
leading-order trajectory $X_{(0)}^{a}$ of ACO\ form, and it was shown that
the net perturbation $\delta X^{a}$ induced in the string trajectory over a
single oscillation period $\Delta \tau =t_{p}$ corresponds to no more than a
change $\Delta L=-\frac{1}{2}\gamma ^{0}\mu L\approx -19.501\mu L$ in the
length and parametric period $L$ of the loop, and a rotational phase shift
with magnitude about $38.92\mu $ acting to advance the overall pattern of
the loop. In other words, as it radiates the ACO loop retains its shape and
therefore its original radiative efficiency. If the ACO\ loop does indeed
possess the lowest possible radiative efficiency, it follows that it would
be the longest-lived loop in any ensemble of loops with the same energy.

\section{The Trajectory of the Evaporating ACO\ Loop}

Whether or not the ACO\ loop minimizes $\gamma ^{0}$, the fact that $\gamma
^{0}$ remains constant as the loop evaporates is the key to a more extended
description of the loop's evolution. Over the course of a double period $%
\Delta t=2t_{p}=L$ the total energy $E=\mu L$ of the loop is reduced by an
amount $\Delta E=-PL=-\mu ^{2}\gamma ^{0}L$ and so $\Delta E/\Delta t=-\mu
^{2}\gamma ^{0}$, or equivalently $\Delta L/\Delta t=-\mu \gamma ^{0}$.
Since the loop is rotating uniformly, this relation is also true
instantaneously, so that $dL/dt=-\mu \gamma ^{0}$ and therefore $%
L(t)=L_{0}-\mu \gamma ^{0}(t-t_{0})$, where $L_{0}$ is the value of $L$ at
some initial time $t_{0}$. In particular, if the time coordinate is chosen
so that $t_{0}=0$ then
\begin{equation}
L(t)=L_{0}(1-t/t_{L})  \label{scalefact}
\end{equation}
for $t<t_{L}$, where $t_{L}\equiv L_{0}/(\gamma ^{0}\mu )$ is the total
future lifetime of the loop.

If the adiabatic calculation performed in Paper I accurately represents the
evolution of the ACO loop, it is expected that as it evaporates the loop
retains the same spatial cross-section, apart from a uniform reduction in
scale proportional to $L(t)$, and a rigid rotation about the $z$-axis. What
follows now is a heuristic argument leading to a plausible form for the
trajectory of the evaporating loop. It should be cautioned that the argument
below relies implicitly on an assumption that the background spacetime is
flat, and consideration of the first-order equation of motion (\ref
{eqmo-first}) quickly shows that this assumption is overly optimistic.
Nonetheless, it is also clear on physical grounds what form the relativistic
corrections to the trajectory should have, and the first-order equation of
motion will be solved to give these corrections explicitly in Section 6.

The most general form of the position function $\mathbf{X}$ of the
evaporating loop (with $\tau $ henceforth replaced by the time coordinate $t$%
, and $\xi $ a general parameter) is
\begin{equation}
\mathbf{X}(t,\xi )=L(t)\,\mathbf{\bar{X}}(t,\xi )
\end{equation}
where $\mathbf{\bar{X}}$ is rigidly rotating, so that if $t<t_{L}$ and $%
t^{*}<t_{L}$ are two different times then
\begin{equation}
\mathbf{\bar{X}}(t,\xi )=\left[
\begin{array}{ccc}
\cos \phi & -\sin \phi & 0 \\
\sin \phi & \cos \phi & 0 \\
0 & 0 & 1
\end{array}
\right] \,\mathbf{\bar{X}}(t^{*},\xi ^{*})
\end{equation}
with $\phi (t^{*},t)$ a phase shift and $\xi ^{*}(t,t^{*},\xi )$ a
reparametrization of $\xi $.

Here, if $t^{*}$ is fixed, $\mathbf{\bar{X}}(t^{*},\xi ^{*})$ is a reference
function that at each value of $t$ must trace out the same loop in $\Bbb{R}%
^{3}$ as $\xi $ varies over its range. In particular, if $t\rightarrow
t+\delta t$ then $\xi ^{*}(t+\delta t,t^{*},\xi +\delta \xi )=\xi
^{*}(t,t^{*},\xi )$ for some function $\delta \xi (\xi )$. That is,

\begin{equation}
\partial _{t}\xi ^{*}\,\delta t+\partial _{\xi }\xi ^{*}\,\delta \xi =0
\end{equation}
to linear order in $\delta t$ and $\delta \xi $, and since this is a
homogeneous linear first-order PDE with characteristic $\xi =\alpha (t)$ for
some function $\alpha $, it follows that $\xi ^{*}$ can be any function of $%
\xi -\alpha (t)$. Hence, if the fixed reference time $t^{*}$ is suppressed,
\begin{equation}
\mathbf{X}(t,\xi )=L(t)\left[
\begin{array}{ccc}
\cos \phi (t) & -\sin \phi (t) & 0 \\
\sin \phi (t) & \cos \phi (t) & 0 \\
0 & 0 & 1
\end{array}
\right] \,\mathbf{\bar{X}}_{0}(\xi -\alpha (t))
\end{equation}
for some choice of functions $\phi $, $\mathbf{\bar{X}}_{0}$ and $\alpha $.

Furthermore, because the parametric period of a cosmic string loop scales
with its length $L$, all timescales on the loop do likewise, and in
particular the velocity field $\mathbf{X}_{t}$ of the loop must at all times
be rotating rigidly. Now,
\begin{equation}
\mathbf{X}_{t}=\left[
\begin{array}{ccc}
\cos \phi & -\sin \phi & 0 \\
\sin \phi & \cos \phi & 0 \\
0 & 0 & 1
\end{array}
\right] [L^{\prime }\mathbf{\bar{X}}_{0}+L(\phi ^{\prime }M\mathbf{\bar{X}}%
_{0}-\alpha ^{\prime }\mathbf{\bar{X}}_{0}^{\prime })]
\end{equation}
where
\begin{equation}
M=\left[
\begin{array}{ccc}
\cos \phi & \sin \phi & 0 \\
-\sin \phi & \cos \phi & 0 \\
0 & 0 & 1
\end{array}
\right] \left[
\begin{array}{ccc}
-\sin \phi & -\cos \phi & 0 \\
\cos \phi & -\sin \phi & 0 \\
0 & 0 & 0
\end{array}
\right] =\left[
\begin{array}{ccc}
0 & -1 & 0 \\
1 & 0 & 0 \\
0 & 0 & 0
\end{array}
\right]
\end{equation}
and the primes denote derivatives with respect to the relevant arguments.

If the vector functions $\mathbf{\bar{X}}_{0}$, $M\mathbf{\bar{X}}_{0}$ and $%
\mathbf{\bar{X}}_{0}^{\prime }$ are linearly independent, the requirement
that $\mathbf{X}_{t}$ be rigidly rotating constrains $L^{\prime }$, $L\phi
^{\prime }$ and $L\alpha ^{\prime }$ to all be constant. The first
condition, that $L^{\prime }$ be constant, was derived independently at the
beginning of this section, and is equivalent to requiring the radiative
efficiency $\gamma ^{0}$ to be constant. Since $L(t)=L_{0}(1-t/t_{L})$ for $%
t<t_{L}$, the remaining constraints can be integrated immediately to give
\begin{equation}
\phi (t)=-p\ln (1-t/t_{L})+\phi _{0}\text{\qquad and\qquad }\alpha (t)=-q\ln
(1-t/t_{L})+\alpha _{0}
\end{equation}
where $p$, $q$, $\phi _{0}$ and $\alpha _{0}$ are constants.

By suitably rotating the $x$-$y$ coordinate axes and rezeroing the gauge
coordinate $\xi $ it is always possible to set $\phi _{0}$ and $\alpha _{0}$
to zero. Furthermore, if $\bar{\sigma}\equiv p\xi /q$ then $\mathbf{\bar{X}}%
_{0}$ is a function of $\bar{\sigma}+p\ln (1-t/t_{L})$ and so
\begin{equation}
\mathbf{X}(t,\bar{\sigma})=L_{0}(1-t/t_{L})\left[
\begin{array}{ccc}
\cos \phi (t) & -\sin \phi (t) & 0 \\
\sin \phi (t) & \cos \phi (t) & 0 \\
0 & 0 & 1
\end{array}
\right] \,\mathbf{\bar{X}}_{0}(\bar{\sigma}-\phi (t))  \label{evappos}
\end{equation}
with $\phi (t)=-p\ln (1-t/t_{L})$.

If this trajectory is to describe the evaporation of an ACO\ loop, the
position function (\ref{evappos}) and its time derivative should, to leading
order in $\mu $, match the position function $\mathbf{X}$ and velocity $%
\mathbf{X}_{t}$ of the unperturbed ACO loop at time $t=0$. The ACO\ loop is
described by the mode functions (\ref{a-mode}) and (\ref{b-mode}), so its
position function has the explicit form
\begin{equation}
\mathbf{X}(t,\sigma )=\tfrac{1}{4\pi }L\{\cos [2\pi (t-\sigma )/L]\,\mathbf{%
\hat{x}}+\sin [2\pi (t-\sigma )/L]\,\mathbf{\hat{y}}\}+\tfrac{1}{2}%
(|t+\sigma |-\tfrac{1}{4}L)\,\mathbf{\hat{z}}  \label{ACOpos}
\end{equation}
for $-\tfrac{1}{2}L\leq t+\sigma \leq \tfrac{1}{2}L$.

Here, the constant scale factors $L_{0}$ and $L$ are taken to be identical.
Since $\phi (0)=0$, the position functions (\ref{evappos}) and (\ref{ACOpos}%
) will agree at $t=0$ if
\begin{equation}
\mathbf{\bar{X}}_{0}(\bar{\sigma})=\tfrac{1}{4\pi }L[\cos (2\pi \sigma /L)\,%
\mathbf{\hat{x}}-\sin (2\pi \sigma /L)\,\mathbf{\hat{y}}]+\tfrac{1}{2}%
(|\sigma |-\tfrac{1}{4}L)\,\mathbf{\hat{z}}\text{.}  \label{X0}
\end{equation}
for $-\tfrac{1}{2}L\leq \sigma \leq \tfrac{1}{2}L$. Furthermore, $L^{\prime
}(t)=-L_{0}/t_{L}$ is of order $\mu $ and can be ignored when calculating $%
\mathbf{X}_{t}$, while $\phi ^{\prime }(t)=p/t_{L}$ at $t=0$. Equating the
time derivatives of (\ref{evappos}) and (\ref{ACOpos}) at $t=0$ then leads
to the consistency conditions $\bar{\sigma}=-4\pi \sigma /L$ and $p=4\pi
t_{L}/L_{0}\equiv 4\pi /(\gamma ^{0}\mu )$. After substituting these
conditions and equation (\ref{X0}) into (\ref{evappos}), the position
function of the evaporating loop becomes:
\begin{equation}
\mathbf{X}(t,\bar{\sigma})=\tfrac{1}{4\pi }L_{0}(1-t/t_{L})\{\cos [\tfrac{1}{%
2}(\phi (t)+\bar{\sigma})]\,\mathbf{\hat{x}}+\sin [\tfrac{1}{2}(\phi (t)+%
\bar{\sigma})]\,\mathbf{\hat{y}}+\tfrac{1}{2}[|\phi (t)-\bar{\sigma}|-\pi ]\,%
\mathbf{\hat{z}}\}  \label{evap0}
\end{equation}
for $-2\pi \leq \phi (t)-\bar{\sigma}\leq 2\pi $, where $\phi (t)=-(4\pi
t_{L}/L_{0})\ln (1-t/t_{L})$.

Equation (\ref{evap0}) suggests that a convenient choice of gauge
coordinates $(\zeta ^{0},\zeta ^{1})=(u,v)$ for the trajectory of the
evaporating ACO\ loop is
\begin{equation}
u=\tfrac{1}{2}(\phi (t)-\bar{\sigma})\text{\qquad and\qquad }v=\tfrac{1}{2}%
(\phi (t)+\bar{\sigma}).
\end{equation}
Then $1-t/t_{L}=e^{-\kappa \mu (u+v)}$, where $\kappa \mu \equiv (4\pi
t_{L}/L_{0})^{-1}$ or equivalently $\kappa =\gamma ^{0}/(4\pi )$. This in
turn means that the time coordinate $t$ on the loop can be written as $t=$ $%
t_{L}[1-e^{-\kappa \mu (u+v)}]$, and the world sheet $\mathbf{T}$ of the
evaporating loop can be represented by the 4-function
\begin{equation}
X^{a}(u,v)=[t_{L},\mathbf{0}]^{a}+\tfrac{1}{4\pi }L_{0}\,e^{-\kappa \mu
(u+v)}[-\tfrac{1}{\kappa \mu },\mathbf{r}(u,v)]^{a}  \label{evap1}
\end{equation}
where
\begin{equation}
\mathbf{r}(u,v)=\cos (v)\,\mathbf{\hat{x}}+\sin (v)\,\mathbf{\hat{y}}+(|u|-%
\tfrac{1}{2}\pi )\,\mathbf{\hat{z}}  \label{r-vector}
\end{equation}
for $-\pi \leq u\leq \pi $, and $t_{L}=\tfrac{1}{4\pi \kappa \mu }L_{0}$.

Strictly speaking, the vector function $\mathbf{r}$ appearing in the
4-function (\ref{evap1}) is the $2\pi $-periodic extension (in $u$) of the
function defined in (\ref{r-vector}). The gauge coordinate pairs $(u,v)$ and
$(u+2\pi k,v-2\pi k)$ therefore correspond to the same spacetime point for
any integer $k$. However, for the sake of simplicity the range of the
parameter $u$ will henceforth be restricted to $(-\pi ,\pi ]$. The final
evaporation point $x^{a}=[t_{L},\mathbf{0}]^{a}$ of the loop then
corresponds to the limit $v\rightarrow \infty $. Note that in this limit the
scale factor $L(t)=L_{0}e^{-\kappa \mu (u+v)}$ goes to zero, while the
angular speed $dv/dt$ of the loop diverges and $\mathbf{r}(u,v)$ is
undefined (although the loop's 4-velocity $X_{t}^{a}$, being scale-free,
remains subluminal everywhere).

If $|\kappa \mu (u+v)|\ll 1$ the loop's position function (\ref{evap1}) can
be expanded in powers of $\mu $ to give
\begin{equation}
X_{(0)}^{a}(u,v)=\tfrac{1}{4\pi }L_{0}[u+v,\mathbf{r}(u,v)]^{a}
\label{pos-zero}
\end{equation}
and
\begin{equation}
\delta X^{a}(u,v)=-\tfrac{1}{4\pi }L_{0}\kappa \mu [\tfrac{1}{2}%
(u+v)^{2},(u+v)\,\mathbf{r}(u,v)]^{a}\text{.}  \label{pos-first}
\end{equation}
It is these functions that need to satisfy the zeroth- and first-order
equations of motion (\ref{eqmo-zero}) and (\ref{eqmo-first}). Note that the
leading-order position function (\ref{pos-zero}) is effectively generated by
an expansion of $X^{a}$ about the curve $u+v=0$ on the loop's world sheet,
or equivalently about the $t=0$ spacelike cross-section.

The position function $X^{a}$ can be expanded about other cross-sections of
constant $t<t_{L}$ by writing $v=v_{*}+\bar{v}$ where $v_{*}$ is any fixed
real number. Then equation (\ref{pos-zero}) becomes
\begin{equation}
X^{a}(u,\bar{v})=[t_{L},\mathbf{0}]^{a}+\tfrac{1}{4\pi }L_{*}\,e^{-\kappa
\mu (u+\bar{v})}[-\tfrac{1}{\kappa \mu },\mathbf{r}(u,v_{*}+\bar{v})]^{a}
\end{equation}
with $L_{*}\,\equiv L_{0}\,e^{-\kappa \mu v_{*}}$, and if $|\kappa \mu (u+%
\bar{v})|\ll 1$ the zeroth-order position function is
\begin{equation}
X_{(0)}^{a}(u,\bar{v})=[t_{*},\mathbf{0}]^{a}+\tfrac{1}{4\pi }L_{*}\,[u+\bar{%
v},\mathbf{r}(u,v_{*}+\bar{v})]^{a}
\end{equation}
where $t_{*}\equiv \tfrac{1}{4\pi \kappa \mu }L_{0}(1-e^{-\kappa \mu v_{*}})$%
. The analogue of equation (\ref{pos-first}) for $\delta X^{a}$ is found by
replacing $L_{0}$ with $L_{*}$, $u+v$ with $u+\bar{v}$, and $\mathbf{r}(u,v)$
with $\mathbf{r}(u,v_{*}+\bar{v})$. Clearly, any two expansions of this type
will be related by a time translation, a constant dilation, and a rotation
in the $x$-$y$ plane, and if any of the expansions satisfies the equations
of motion (\ref{eqmo-zero}) and (\ref{eqmo-first}) then they all will.

The equations of motion will be considered more closely in Section 6. It
will come as no surprise that the zeroth-order position function (\ref
{pos-zero}) does satisfy the zeroth-order equation of motion (\ref{eqmo-zero}%
), which reads simply $X_{(0)}^{a},_{uv}=0$. Before considering the
first-order equation of motion (\ref{eqmo-first}), however, it is necessary
to calculate the metric perturbations $h_{ab}$, which depend only on $%
X_{(0)}^{a}$. This will be done in Section 5.

After a long calculation described in detail in Section 6, it turns out that
the first-order position function $\delta X^{a}$ given by (\ref{pos-first})
does \emph{not} satisfy the first-order equation of motion (\ref{eqmo-first}%
). The reason for this is not difficult to appreciate. As was mentioned at
the beginning of this section, the calculation leading to the evaporating
loop trajectory (\ref{evap1}) was predicated on the assumption that the
background metric is flat. It is evident from (\ref{eqmo-first}), however,
that the first-order equation of motion involves the first-order metric
perturbation $h_{ab}$. Once the background metric $g_{ab}$ ceases to be
flat, the spacetime coordinates $[t,x,y,z]$ no longer measure proper times
or proper lengths. In particular, the self-similar shrinkage driven by the
scale factor $L_{0}e^{-\kappa \mu (u+v)}$ may proceed at different rates in
different directions at various points on the world sheet $\mathbf{T}$.

This possibility is easily accounted for by adding scale corrections, of
order $\mu $, to the components of the evaporating loop's position function (%
\ref{evap1}). The form of these scale corrections is of course constrained
by the equation of motion (\ref{eqmo-first}), but even so there remains
considerable freedom in the choice of the correction functions,
corresponding not only to possible transformations (at order $\mu $) of the
gauge coordinates $(u,v)$ and the spacetime coordinates $[t,x,y,z]$, but
also to the propagation of free vibrations (with amplitude $\mu $) around
the loop.

In what follows, the freedom in the choice of gauge coordinates will be
partially fixed by taking the position vector to have the form
\begin{equation}
X^{a}(u,v)=[t_{L},\mathbf{0}]^{a}+\tfrac{1}{4\pi }L_{0}\,e^{-\kappa \mu
(u+v)}[-\tfrac{1}{\kappa \mu }e^{\kappa \mu ^{2}F(u,v)},\mathbf{\bar{r}}%
(u,v)]^{a}\text{,}  \label{evap2}
\end{equation}
where now
\begin{equation}
\mathbf{\bar{r}}(u,v)=e^{\mu G(u,v)}[\cos (v)\,\mathbf{\hat{x}}+\sin (v)\,%
\mathbf{\hat{y}]}+[(|u|-\tfrac{1}{2}\pi )+\mu A(u,v)]\,\mathbf{\hat{z}}
\end{equation}
for $-\pi \leq u\leq \pi $. Here, the correction functions $F$, $G$ and $A$
are assumed to be continuous and $2\pi $-periodic in both $u$ and $v$, so
that $(u,v)$ and $(u+2\pi ,v-2\pi )$ again correspond to the same spacetime
point, and all three functions remain small compared to $\mu ^{-1}$ as $%
u+v\rightarrow \infty $. The assumption that the trajectory is dilated by
the same factor $e^{\mu G}$ in the $x$ and $y$ directions effectively
removes any freedom to redefine the gauge coordinate $v$.\footnote{%
To see this, suppose that the term $G(u,v)\{\cos (v)\,\mathbf{\hat{x}}+\sin
(v)\,\mathbf{\hat{y}}\}$ in equation (\ref{pos-first2}) has the more general
form $G(u,v)\cos (v)\,\mathbf{\hat{x}}+H(u,v)\sin (v)\,\mathbf{\hat{y}}$.
Then if $v$ is replaced by a new coordinate $v^{\prime }$ defined by $%
v=v^{\prime }+\mu V(u,v^{\prime })$ for some function $V$, the correction
term $G\cos (v)\,\mathbf{\hat{x}}+H\sin (v)\,\mathbf{\hat{y}}$ becomes $%
(G\cos v^{\prime }-V\sin v^{\prime })\,\mathbf{\hat{x}}+(H\sin v^{\prime
}+V\cos v^{\prime })\,\mathbf{\hat{y}}$ to leading order in $\mu $. The
unique gauge choice $V=(G-H)\sin v^{\prime }\cos v^{\prime }$ then reduces
this term to the isotropic form $G^{\prime }\{\cos (v^{\prime })\,\mathbf{%
\hat{x}}+\sin (v^{\prime })\,\mathbf{\hat{y}}\}$, with $G^{\prime }=G\cos
^{2}v^{\prime }+H\sin ^{2}v^{\prime }$.}

The presence of free vibrations which break the self-similarity of the
world-sheet can be removed by requiring the intrinsic 2-metric $\gamma
_{AB}=g_{ab}X^{a},_{A}X^{b},_{B}$ to have the strictly self-similar form
\begin{equation}
\gamma _{AB}(u,v)=(\tfrac{1}{4\pi }L_{0})^{2}e^{-2\kappa \mu (u+v)}\bar{%
\gamma}_{AB}(u)  \label{self-sim}
\end{equation}
to linear order in $\mu $, with $\bar{\gamma}_{AB}$ a 2-metric to be
determined. This constraint will be considered in more detail in Section 6.2.

Given (\ref{evap2}), the leading-order position function $X_{(0)}^{a}$ near $%
t=0$ retains the form (\ref{pos-zero}), while the first-order position
function becomes
\begin{eqnarray}
\delta X^{a}(u,v) &=&-\tfrac{1}{4\pi }L_{0}\kappa \mu [\tfrac{1}{2}%
(u+v)^{2},(u+v)\,\mathbf{r}(u,v)]^{a}  \nonumber \\
&&+\tfrac{1}{4\pi }L_{0}\mu [-F(u,v),G(u,v)\{\cos (v)\,\mathbf{\hat{x}}+\sin
(v)\,\mathbf{\hat{y}}\}\mathbf{+}A(u,v)\,\mathbf{\hat{z}}]^{a}\text{.}
\nonumber \\
&&  \label{pos-first2}
\end{eqnarray}
The actual form of the correction functions $F$, $G$ and $A$ will be
determined once the first-order equation of motion (\ref{eqmo-first}) has
been solved, and additional gauge conditions have been imposed.

Two remarks about the final form (\ref{evap2}) of the loop trajectory are in
order here. The first is that although (\ref{evap2})\ was developed to
describe an evaporating ACO\ loop over the time interval $0\leq t<t_{L}$ (or
equivalently $0\leq u+v<\infty $), it can without difficulty be assumed to
extend over the entire interval $-\infty <t<t_{L}$ (or $-\infty <u+v<\infty $%
). The trajectory then describes a loop that has been evaporating
``eternally'', with its energy and length scaling (to leading order in $\mu $%
) as $1-t/t_{L}$. This is obviously just a mathematical idealization, but it
does side-step the need to construct initial data for the metric
perturbations $h_{ab}$ at $t=0$.

The second point is that the value of the constant $\kappa $ appearing in
the corrected trajectory (\ref{evap2}) is initially undetermined. Although
it is to be expected in view of the results of Paper I that $\kappa \approx
39.0025/(4\pi )\approx 3.1037$, no particular value of $\kappa $ is assumed
in the analysis that follows. It is only after the first-order equation of
motion is solved that it becomes evident that the periodicity constraint on $%
G$ forces $\kappa $ to take on its expected value.

A schematic representation of the evaporation of the loop is shown in Figure
2. The thickened line corresponds to the outer envelope of the loop, whose
radius shrinks to zero at the final evaporation point $x^{a}=[t_{L},\mathbf{0%
}]^{a}$. Also shown are the future light cones $F_{0}$ and $F_{L}$ of the
origin and the evaporation point respectively. Note that, because of the
first-order corrections to the background metric and the loop trajectory $%
X^{a}$, the envelope of the loop and the light cones depicted in Figure 2
should deviate from straight lines by terms of order $\mu $.

Now that the choice of the evaporating loop trajectory (\ref{evap2}) has
been explained in some depth, the remainder of the paper is a
straightforward application of the solution methods discussed in Section 2.
The metric perturbations $h_{ab}$ generated by $X_{(0)}^{a}$ are calculated
in Section 5, the equation of motion for the perturbation functions $\delta
X^{a}$ is constructed and solved in Section 6, and the metric perturbations
on the future light cone $F_{L}$ of the final evaporation point are shown to
match onto an empty, flat spacetime in Section 7.

\section{The Metric Perturbations}

\subsection{Rotating self-similarity of the metric perturbations}

The metric perturbations $h_{ab}$ at any field point $x^{a}\equiv [t,\mathbf{%
x}]^{a}$ away from the world sheet of the evaporating loop can be calculated
from equation (\ref{metpert}). Because the analysis of Section 7 depends on
an understanding of the behavior of $h_{ab}$ arbitrarily close to the light
cone $F_{L}$, it will be necessary to consider source points on the
evaporating loop close to the final evaporation point (where $u+v\rightarrow
\infty $) and so it cannot be assumed that $|\kappa \mu (u+v)|\ll 1$. The
scale factor $L_{0}\,e^{-\kappa \mu (u+v)}$ will therefore be retained in
full in this section, although the correction functions $F$, $G$ and $A$ --
which are bounded and appear only at order $\mu $ in (\ref{evap2}) -- will
be ignored. In other words, the trajectory will in this section be assumed
to have its flat-space form (\ref{evap1}).

The first step in the calculation is to write down the leading-order
expressions for the intrinsic 2-metric $\gamma _{AB}$ and the source
function $S_{ab}$. Since
\begin{equation}
X^{a},_{u}=\tfrac{1}{4\pi }L_{0}\,e^{-\kappa \mu (u+v)}[1,\text{sgn}(u)\,%
\mathbf{\hat{z}}]^{a}  \label{tan-vectors}
\end{equation}
and
\begin{equation}
X^{a},_{v}=\tfrac{1}{4\pi }L_{0}\,e^{-\kappa \mu (u+v)}[1,-\sin (v)\,\mathbf{%
\hat{x}}+\cos (v)\,\mathbf{\hat{y}}]^{a}  \label{tan-vectors2}
\end{equation}
to leading order in $\mu $, it follows that the intrinsic 2-metric $\gamma
_{AB}=X,_{A}\,\cdot X,_{B}$ on the world sheet has components
\begin{equation}
\gamma _{uu}=\gamma _{vv}=0\qquad \text{and\qquad }\gamma _{uv}=(\tfrac{1}{%
4\pi }L_{0})^{2}e^{-2\kappa \mu (u+v)}
\end{equation}
to the same order. Hence, $\gamma ^{1/2}=(\tfrac{1}{4\pi }%
L_{0})^{2}e^{-2\kappa \mu (u+v)}$, while the inverse 2-metric $\gamma ^{AB}$
has components
\begin{equation}
\gamma ^{uu}=\gamma ^{vv}=0\qquad \text{and\qquad }\gamma ^{uv}=(\tfrac{1}{%
4\pi }L_{0})^{-2}e^{2\kappa \mu (u+v)}\text{.}  \label{gamma-inv}
\end{equation}

The source function defined in equation (\ref{source1}) can be written as
\begin{equation}
S_{ab}(x^{\prime \,d})=\mu \int \Psi _{ab}(u,v)\,\delta ^{(4)}(x^{\prime
\,d}-X^{d})du\,dv  \label{source2}
\end{equation}
where
\begin{equation}
\Psi _{ab}\equiv \gamma ^{1/2}\gamma ^{AB}(X_{a},_{A}\,X_{b},_{B}-\tfrac{1}{2%
}\eta _{ab}\,X,_{A}\,\cdot X,_{B})\text{,}
\end{equation}
evaluated again at leading order in $\mu $. Note here that $\gamma
^{AB}X_{a},_{A}\,X_{b},_{B}$ is just the world-sheet projection tensor $%
p_{ab}$, which in matrix form is
\begin{equation}
p_{ab}=\left[
\begin{array}{cccc}
2 & \sin v & -\cos v & -s \\
\sin v & 0 & 0 & -s\sin v \\
-\cos v & 0 & 0 & s\cos v \\
-s & -s\sin v & s\cos v & 0
\end{array}
\right] _{ab}
\end{equation}
where $s\equiv \,$sgn$(u)$. The function $\Psi _{ab}$ is therefore equal to $%
-\gamma ^{1/2}q_{ab}$, where $q_{ab}\equiv \eta _{ab}-p_{ab}$ was introduced
in Section 2 as the orthogonal complement of $p_{ab}$. Hence, $\Psi _{ab}=(%
\tfrac{1}{4\pi }L_{0})^{2}e^{-2\kappa \mu (u+v)}\bar{\Psi}_{ab}$ with
\begin{equation}
\bar{\Psi}_{ab}(s,v)\equiv \left[
\begin{array}{cccc}
1 & \sin v & -\cos v & -s \\
\sin v & 1 & 0 & -s\sin v \\
-\cos v & 0 & 1 & s\cos v \\
-s & -s\sin v & s\cos v & 1
\end{array}
\right] _{ab}\text{.}  \label{big-psi}
\end{equation}

Given (\ref{source2}), the integral expression (\ref{metpert}) for $h_{ab}$
reads
\begin{equation}
h_{ab}(t,\mathbf{x})=-4\mu \int \frac{d^{3}x^{\prime }}{|\mathbf{x}-\mathbf{x%
}^{\prime }|}\,\int \Psi _{ab}(u,v)\,\delta ^{(4)}(x^{\prime
\,d}-X^{d})\,du\,dv  \label{metpert2}
\end{equation}
where the source point is $x^{\prime \,d}=[t^{\prime },\mathbf{x}^{\prime
}]^{d}$, with $t^{\prime }\equiv t-|\mathbf{x}-\mathbf{x}^{\prime }|$ the
retarded time\footnote{%
Note that, at a physical level, the equation $t^{\prime }=t-|\mathbf{x}-%
\mathbf{x}^{\prime }|$ imposes the requirement that the source point $%
[t^{\prime },\mathbf{x}^{\prime }]$ lie on the backwards light cone of the
field point $[t,\mathbf{x}]$ to leading order in $\mu .$}. The
distributional factor $\delta ^{(4)}(x^{\prime \,d}-X^{d})$ in (\ref
{metpert2}) can be integrated out by first transforming from $y^{a}\equiv [u,%
\mathbf{x}^{\prime }]^{a}$ to $z^{a}\equiv [t^{\prime },\mathbf{x}^{\prime
}]^{a}-X^{a}(u,v)$, with $v$, $t$ and $\mathbf{x}$ held constant. The
Jacobian of this transformation is:
\begin{equation}
\left| \partial \,y^{a}/\partial \,z^{b}\right| =\left| \partial
\,z^{b}/\partial \,y^{a}\right| ^{-1}=|X^{0},_{u}-\mathbf{n}\cdot \mathbf{X}%
,_{u}|^{-1},
\end{equation}
where $\mathbf{n}=(\mathbf{x}-\mathbf{x}^{\prime })/|\mathbf{x}-\mathbf{x}%
^{\prime }|$ is the unit vector from the source point $\mathbf{x}^{\prime }$
to the field point $\mathbf{x}$.

On integrating over $z^{a}$, $t^{\prime }$ is everywhere replaced by $%
X^{0}\equiv $ $t_{L}[1-e^{-\kappa \mu (u+v)}]$ and $\mathbf{x}^{\prime }$ by
$\mathbf{X}\equiv \tfrac{1}{4\pi }L_{0}\,e^{-\kappa \mu (u+v)}\mathbf{r}%
(u,v) $, while $u$ becomes an implicit function of $v$, $t$ and $\mathbf{x}$
through the equation $X^{0}(u,v)=t-|\mathbf{x}-\mathbf{X}(u,v)|$. The
integral for $h_{ab}$ therefore reads
\begin{equation}
h_{ab}(t,\mathbf{x})=-4\mu \int ||\mathbf{x}-\mathbf{X}|\,X^{0},_{u}-(%
\mathbf{x}-\mathbf{X})\cdot \mathbf{X},_{u}|^{-1}\Psi _{ab}(u,v)\,dv\text{.}
\label{metpert3}
\end{equation}
where, as before, $X^{0},_{u}=\tfrac{1}{4\pi }L_{0}\,e^{-\kappa \mu (u+v)}$
and $\mathbf{X},_{u}=\tfrac{1}{4\pi }L_{0}\,e^{-\kappa \mu (u+v)}s\,\mathbf{%
\hat{z}}$ to leading order in $\mu $.

Evaluation of the integral (\ref{metpert3}) is simplified by the fact that $%
h_{ab}$ satisfies a very elegant self-similarity relation. If $[t,\mathbf{x}%
] $ is any field point with $t-t_{L}<|\mathbf{x}|$ (where $t-t_{L}=|\mathbf{x%
}| $ is just the equation, at leading order, of the future light cone $F_{L}$
of the final evaporation point), first define
\begin{equation}
\psi (t,\mathbf{x})\equiv -(\kappa \mu )^{-1}\ln [(|\mathbf{x}%
|-t+t_{L})/t_{L}]  \label{psi}
\end{equation}
then consider a second field point $[\bar{t},\mathbf{\bar{x}}]$ with
\begin{equation}
\bar{t}\equiv e^{\kappa \mu \psi }|\mathbf{x}|\text{\qquad and\qquad }%
\mathbf{\bar{x}}\equiv e^{\kappa \mu \psi }R(-\psi )\mathbf{x}  \label{image}
\end{equation}
where
\begin{equation}
R(\theta )\equiv \left[
\begin{array}{ccc}
\cos \theta & -\sin \theta & 0 \\
\sin \theta & \cos \theta & 0 \\
0 & 0 & 1
\end{array}
\right]  \label{rot-mat}
\end{equation}
is the standard rotation matrix about the $z$-axis. Note that $[\bar{t},%
\mathbf{\bar{x}}]$ lies on the future light cone $F_{0}$ of the origin, as $%
\bar{t}=|\mathbf{\bar{x}}|$.

Now, the equation $t-X^{0}(u,v)=|\mathbf{x}-\mathbf{X}(u,v)|$ which
specifies $u$ as a function of $v$ in the integral (\ref{metpert3}) -- or
equivalently on the backwards light cone of the original field point $[t,%
\mathbf{x}]$ -- reads explicitly
\begin{equation}
t-t_{L}[1-e^{-\kappa \mu (u+v)}]=|\mathbf{x}-\tfrac{1}{4\pi }%
L_{0}\,e^{-\kappa \mu (u+v)}\,\mathbf{r}(u,v)|\text{.}  \label{backlight}
\end{equation}
Multiplying this equation by $e^{\kappa \mu \psi (t,\mathbf{x})}$ gives:
\begin{equation}
(t-t_{L})e^{\kappa \mu \psi }+t_{L}e^{-\kappa \mu (u+v-\psi )}=|e^{\kappa
\mu \psi }\,\mathbf{x}-\tfrac{1}{4\pi }L_{0}\,e^{-\kappa \mu (u+v-\psi )}\,%
\mathbf{r}(u,v)|  \label{backlight2}
\end{equation}
where $(t-t_{L})e^{\kappa \mu \psi }=e^{\kappa \mu \psi }|\mathbf{x}|-t_{L}$
and
\begin{eqnarray}
\mathbf{r}(u,v) &\equiv &\cos (v)\,\mathbf{\hat{x}}+\sin (v)\,\mathbf{\hat{y}%
}+(|u|-\tfrac{1}{2}\pi )\,\mathbf{\hat{z}}  \nonumber \\
&=&R(\psi )[\cos (v-\psi )\,\mathbf{\hat{x}}+\sin (v-\psi )\,\mathbf{\hat{y}}%
+(|u|-\tfrac{1}{2}\pi )\,\mathbf{\hat{z}}]  \nonumber \\
&\equiv &R(\psi )\,\mathbf{r}(u,v-\psi )\text{.}
\end{eqnarray}
Since $|\mathbf{v}|=|R(-\psi )\mathbf{v}|$ for any 3-vector $\mathbf{v}$,
and $R(-\psi )R(\psi )=I$, equation (\ref{backlight2}) can be rewritten as
\begin{equation}
\bar{t}-t_{L}[1-e^{-\kappa \mu (u+v-\psi )}]=|\mathbf{\bar{x}}-\tfrac{1}{%
4\pi }L_{0}\,e^{-\kappa \mu (u+v-\psi )}\,\mathbf{r}(u,v-\psi )|\text{,}
\end{equation}
or equivalently $\bar{t}-X^{0}(u,v-\psi )=|\mathbf{\bar{x}}-\mathbf{X}%
(u,v-\psi )|$.

Hence, if $u=U(v)$ is the equation of the curve of intersection of the world
sheet $\mathbf{T}$ with the backwards light cone of a given field point $[t,%
\mathbf{x}]$, then the equation for $u$ on the backwards light cone of the
corresponding image point $[\bar{t},\mathbf{\bar{x}}]$ on $F_{0}$ is $%
u=U(v-\psi (t,\mathbf{x}))$. Similarly, the integrand in equation (\ref
{metpert3}) for $h_{ab}$, which is equivalent to
\begin{equation}
\tfrac{1}{4\pi }L_{0}\,e^{-\kappa \mu (u+v)}||\mathbf{x}-\mathbf{X}%
(u,v)|\,-s\,[\mathbf{x}-\mathbf{X}(u,v)]\cdot \mathbf{\hat{z}\,}|^{-1}\bar{%
\Psi}_{ab}(s,v)\text{,}  \label{int1}
\end{equation}
can be re-expressed as\footnote{%
Note here that
\[
e^{\kappa \mu \psi }[\mathbf{x}-\mathbf{X}(u,v)]\cdot \mathbf{\hat{z}\,}%
=[e^{\kappa \mu \psi }R(-\psi )\mathbf{x}-\mathbf{X}(u,v-\psi )]\cdot
\mathbf{\hat{z}}
\]
for the simple reason that the $z$-component of $\mathbf{x}$ is unaffected
by the rotation matrix $R$, while the $z$-component of $\mathbf{X}$ is $%
\tfrac{1}{4\pi }L_{0}\,e^{-\kappa \mu (u+v)}(|u|-\tfrac{1}{2}\pi )$.}
\begin{eqnarray}
&&\tfrac{1}{4\pi }L_{0}\,e^{-\kappa \mu (u+v-\psi )}||e^{\kappa \mu \psi
}R(-\psi )\mathbf{x}-\mathbf{X}(u,v-\psi )|\,-s\,[e^{\kappa \mu \psi
}R(-\psi )\mathbf{x}-\mathbf{X}(u,v-\psi )]\cdot \mathbf{\hat{z}\,}|^{-1}
\nonumber \\
&&\times \bar{\Psi}_{ab}(s,v)  \nonumber \\
&=&\tfrac{1}{4\pi }L_{0}\,e^{-\kappa \mu (u+v-\psi )}||\mathbf{\bar{x}}-%
\mathbf{X}(u,v-\psi )|\,-s\,[\mathbf{\bar{x}}-\mathbf{X}(u,v-\psi )]\cdot
\mathbf{\hat{z}\,}|^{-1}  \nonumber \\
&&\times R_{a}^{\;c}(\psi )\bar{\Psi}_{cd}(s,v-\psi )R_{b}^{\;d}(\psi )
\label{int2}
\end{eqnarray}
where
\begin{equation}
R_{a}^{\;c}(\psi )\equiv (1\oplus R(\psi ))_{a}^{\;c}=\left[
\begin{array}{llll}
1 & 0 & 0 & 0 \\
0 & \cos \psi & -\sin \psi & 0 \\
0 & \sin \psi & \cos \psi & 0 \\
0 & 0 & 0 & 1
\end{array}
\right] \text{.}
\end{equation}

Comparison of equations (\ref{int1}) and (\ref{int2}), together with the
fact that if $u=U(v)$ on the backwards light cone of $[t,\mathbf{x}]$ then $%
u=U(v-\psi )$ on the backwards light cone of $[\bar{t},\mathbf{\bar{x}}]$,
indicates that
\begin{equation}
h_{ab}(t,\mathbf{x})=R_{a}^{\;c}(\psi )\,h_{cd}(\bar{t},\mathbf{\bar{x}}%
)\,R_{b}^{\;d}(\psi )\text{,}  \label{rotation}
\end{equation}
where $\psi (t,\mathbf{x})$ is given by equation (\ref{psi}). The geometric
meaning of this relation is illustrated in Figure 3. To calculate $h_{ab}$
at a field point $[t,\mathbf{x}]$ in the past of $F_{L}$, simply dilate the
point by a factor $e^{\kappa \mu \psi }$ through $[t_{L},\mathbf{0}]$ and
rotate it by an angle $-\psi $ in the $x$-$y$ plane to map it to the image
point $[\bar{t},\mathbf{\bar{x}}]$ on $F_{0}$. Then calculate $h_{ab}$ at $[%
\bar{t},\mathbf{\bar{x}}]$ and rotate each of the components of the
resulting tensor by an angle $\psi $ in the $x$-$y$ plane to give $h_{ab}$
at $[t,\mathbf{x}]$.

From a computational viewpoint, equation (\ref{rotation}) allows us to
calculate $h_{ab}$ at any point away from the world sheet of the loop from a
knowledge of $h_{ab}$ on the null surface $F_{0}$ alone. Furthermore, it
points to the existence of a symmetry in the background metric induced by
the self-similarity of the loop's trajectory. This symmetry can be better
appreciated by writing the components (\ref{image}) of the image point
explicitly in terms of $t$ and $\mathbf{x}$:
\begin{equation}
\bar{t}\equiv t_{L}(1+|\mathbf{w}|)^{-1}|\mathbf{w}|\text{\qquad and\qquad }%
\mathbf{\bar{x}}\equiv t_{L}(1+|\mathbf{w}|)^{-1}R(-\psi )\mathbf{w}
\label{image2}
\end{equation}
where
\begin{equation}
\mathbf{w}\equiv \mathbf{x}/(t_{L}-t)\text{.}  \label{vector-w}
\end{equation}
If it were not for the presence of the rotation operators in (\ref{rotation}%
) and (\ref{image2}), which depend on $\psi $ rather than $\mathbf{w}$, the
metric $\eta _{ab}+h_{ab}$ would be a function of $\mathbf{w}$ alone and so
would be strictly self-similar. As it is, the rotation operators act to
enforce a slightly more complicated symmetry in the metric which could
perhaps be termed ``rotating self-similarity''.

\subsection{Explicit form of the metric perturbations on $F_{0}$}

Suppose now that $[|\mathbf{\bar{x}}|,\mathbf{\bar{x}}]$ is a general point
on the future light cone $F_{0}$ of the origin. The equation (\ref{backlight}%
) for $u$ as a function of $v$ on the backwards light cone of $[|\mathbf{%
\bar{x}}|,\mathbf{\bar{x}}]$ reads:
\begin{equation}
|\mathbf{\bar{x}}|-t_{L}[1-e^{-\kappa \mu (u+v)}]=|\mathbf{\bar{x}}-\tfrac{1%
}{4\pi }L_{0}\,e^{-\kappa \mu (u+v)}\,\mathbf{r}(u,v)|\text{.}
\label{backlight3}
\end{equation}
Application of the triangle inequality to the norm on the right-hand side of
this equation then gives:
\begin{equation}
t_{L}|1-e^{-\kappa \mu (u+v)}|\leq \tfrac{1}{4\pi }L_{0}\,e^{-\kappa \mu
(u+v)}[1+(|u|-\tfrac{1}{2}\pi )^{2}]^{1/2}\text{,}
\end{equation}
as $|\mathbf{r}(u,v)|=[1+(|u|-\tfrac{1}{2}\pi )^{2}]^{1/2}$. Since $t_{L}=%
\tfrac{1}{4\pi \kappa \mu }L_{0}$ and the maximum value of $[1+(|u|-\tfrac{1%
}{2}\pi )^{2}]^{1/2}$ for $u\in (-\pi ,\pi ]$ is $(1+\frac{1}{4}\pi
^{2})^{1/2}\approx 1.\,\allowbreak 8621$, this inequality can be
re-expressed as
\begin{equation}
|e^{\kappa \mu (u+v)}-1|\leq (1+\tfrac{1}{4}\pi ^{2})^{1/2}\kappa \mu \text{.%
}  \label{ineq1}
\end{equation}

It is therefore evident that $|\kappa \mu (u+v)|$ will remain small at all
points on the backwards light cone of $[|\mathbf{\bar{x}}|,\mathbf{\bar{x}}]$
provided that $\kappa \mu $ itself is sufficiently small. For example, $%
|\kappa \mu (u+v)|$ will be everywhere no greater than $10^{-2}$ if $\ln
[1-(1+\tfrac{1}{4}\pi ^{2})^{1/2}\kappa \mu ]\geq -10^{-2}$, and this in
turn will be true if $\kappa \mu \lesssim 5.34\times 10^{-3}$. Given that we
expect that $\kappa \approx 3.1037$ and $\mu $ will be of order $10^{-6}$ or
smaller for a GUT string, the assumption that $|\kappa \mu (u+v)|$ remains
small is not an unreasonable one.

Expanding $e^{-\kappa \mu (u+v)}$ in powers of $\mu $ in (\ref{backlight3})
and retaining only the leading-order terms gives the following equation for $%
u$ as a function of $v$:
\begin{equation}
T-(u+v)=[(X-\cos v)^{2}+(Y-\sin v)^{2}+\{Z-(|u|-\tfrac{1}{2}\pi
)\}^{2}]^{1/2}\text{,}  \label{backlight4}
\end{equation}
where
\begin{equation}
(X,Y,Z)\equiv 4\pi L_{0}^{-1}(\bar{x},\bar{y},\bar{z})\text{\qquad and\qquad
}T\equiv (X^{2}+Y^{2}+Z^{2})^{1/2}\text{.}  \label{image3}
\end{equation}
Equation (\ref{backlight4}) can easily be solved to give an explicit formula
for $u$, but it turns out that in calculating $h_{ab}$ it is more useful to
regard $u$ as the independent variable.

Similarly, for small values of $|\kappa \mu (u+v)|$ the integrand (\ref{int1}%
) in the formula (\ref{metpert3}) for $h_{ab}$ at $[|\mathbf{\bar{x}}|,%
\mathbf{\bar{x}}]$ becomes:
\begin{equation}
\{[(X-\cos v)^{2}+(Y-\sin v)^{2}+\{Z-(|u|-\tfrac{1}{2}\pi
)\}^{2}]^{1/2}-s\,[Z-(|u|-\tfrac{1}{2}\pi )]\,\}^{-1}\bar{\Psi}_{ab}(s,v)%
\text{.}
\end{equation}
Note that the denominator in this expression is manifestly positive or zero.
In view of equation (\ref{backlight4}) and the fact that $s|u|=u$, the
denominator can be expressed in the simpler form
\begin{equation}
\lbrack (X-\cos v)^{2}+(Y-\sin v)^{2}+\{Z-(|u|-\tfrac{1}{2}\pi
)\}^{2}]^{1/2}-s\,[Z-(|u|-\tfrac{1}{2}\pi )]\,=T-s(Z+\tfrac{1}{2}\pi )\text{
}-v\text{,}  \label{int3}
\end{equation}
and (\ref{metpert3}) reduces to
\begin{equation}
h_{ab}(|\mathbf{\bar{x}}|,\mathbf{\bar{x}})=-4\mu \int [T-s(Z+\tfrac{1}{2}%
\pi )-v]^{-1}\bar{\Psi}_{ab}(s,v)\,dv\text{.}  \label{metpert4}
\end{equation}

All that remains now is to determine the limits of integration. The entire
curve of intersection $\Gamma $ of the world sheet with the backwards light
cone of $[|\mathbf{\bar{x}}|,\mathbf{\bar{x}}]$ will be traced out as $u$
varies from $-\pi $ to $\pi $. The integral (\ref{metpert4}) therefore
breaks into two parts: one with $s=-1$ and $u\in [-\pi ,0]$, and one with $%
s=+1$ and $u\in [0,\pi ]$. The corresponding values of $v$, which I will
denote as $V_{k}\equiv v|_{u=k\pi }$ for $k=-1,0$ and $1$, are roots of the
light-cone equation (\ref{backlight4}):
\begin{equation}
T-(k\pi +V_{k})=[(X-\cos V_{k})^{2}+(Y-\sin V_{k})^{2}+\{Z-(|k|-\tfrac{1}{2}%
)\pi \}^{2}]^{1/2}\text{.}  \label{roots}
\end{equation}
Although it is not possible to solve for $V_{k}$ explicitly except in
certain special cases, it is clear by inspection that $V_{-1}=V_{1}+2\pi $.
Furthermore, implicit differentiation of (\ref{backlight4}) with $[T,X,Y,Z]$
constant indicates that
\begin{equation}
\frac{dv}{du}=-\frac{T-s(Z+\tfrac{1}{2}\pi )-v}{T-(u+v)+X\sin v-Y\cos v}%
\text{.}  \label{dv/du}
\end{equation}

Some of the mathematical consequences of equations (\ref{roots}) and (\ref
{dv/du}) are examined in Appendix A. In particular, it is shown that the
values of $V_{0}$ and $V_{1}$ are bounded on $F_{0}$, and that $v$ is a
monotonic decreasing function of $u$. The last property follows partly from
the fact that the numerator in (\ref{dv/du}) is positive or zero from (\ref
{int3}), while the denominator is positive or zero as a consequence of (\ref
{backlight4}) and the identity
\begin{equation}
(X-\cos v)^{2}+(Y-\sin v)^{2}=(X\sin v-Y\cos v)^{2}+(X\cos v+Y\sin v-1)^{2}%
\text{.}
\end{equation}

However, cases where either the numerator or denominator is zero for some
value of $u$ need to be considered separately, and it is shown in Appendices
A.2 and A.3 that the corresponding spacetime points have interesting
physical interpretations. Perhaps the most important result is that the
spacetime derivatives of $h_{ab}$ are undefined at all points with the
property that the denominator in (\ref{dv/du}) is zero at one of the extreme
values $u=0$ or $\pm \pi $.

Given that $V_{-1}=V_{1}+2\pi \geq V_{0}\geq V_{1}$, the equation for the
metric perturbations becomes:
\begin{eqnarray}
h_{ab}(|\mathbf{\bar{x}}|,\mathbf{\bar{x}}) &=&-4\mu
\int_{V_{0}}^{V_{1}+2\pi }[T+(Z+\tfrac{1}{2}\pi )-v]^{-1}\bar{\Psi}%
_{ab}(-1,v)\,dv  \nonumber \\
&&-4\mu \int_{V_{1}}^{V_{0}}[T-(Z+\tfrac{1}{2}\pi )-v]^{-1}\bar{\Psi}%
_{ab}(+1,v)\,dv\text{.}  \label{h-ints}
\end{eqnarray}
After referring back to the definition (\ref{big-psi}) of $\bar{\Psi}_{ab}$,
it is readily seen that
\begin{equation}
h_{ab}(|\mathbf{\bar{x}}|,\mathbf{\bar{x}})=4\mu \left[
\begin{array}{cccc}
I_{+}+I_{-} & S_{+}+S_{-} & -C_{+}-C_{-} & I_{+}-I_{-} \\
S_{+}+S_{-} & I_{+}+I_{-} & 0 & S_{+}-S_{-} \\
-C_{+}-C_{-} & 0 & I_{+}+I_{-} & -C_{+}+C_{-} \\
I_{+}-I_{-} & S_{+}-S_{-} & -C_{+}+C_{-} & I_{+}+I_{-}
\end{array}
\right] _{ab}  \label{metperts}
\end{equation}
where
\begin{equation}
I_{+}=\ln [(\chi _{+}-V_{1}-2\pi )/(\chi _{+}-V_{0})]\text{,\qquad }%
I_{-}=\ln [(\chi _{-}-V_{0})/(\chi _{-}-V_{1})]\text{,}  \label{I-ints}
\end{equation}
\begin{equation}
\left[
\begin{array}{c}
S_{+} \\
C_{+}
\end{array}
\right] =\left[
\begin{array}{cc}
-\cos \chi _{+} & \sin \chi _{+} \\
\sin \chi _{+} & \cos \chi _{+}
\end{array}
\right] \left[
\begin{array}{c}
\func{Si}(\chi _{+}-V_{1}-2\pi )-\func{Si}(\chi _{+}-V_{0}) \\
\func{Ci}(\chi _{+}-V_{1}-2\pi )-\func{Ci}(\chi _{+}-V_{0})
\end{array}
\right]  \label{S,C-ints1}
\end{equation}
and
\begin{equation}
\left[
\begin{array}{c}
S_{-} \\
C_{-}
\end{array}
\right] =\left[
\begin{array}{cc}
-\cos \chi _{-} & \sin \chi _{-} \\
\sin \chi _{-} & \cos \chi _{-}
\end{array}
\right] \left[
\begin{array}{c}
\func{Si}(\chi _{-}-V_{0})-\func{Si}(\chi _{-}-V_{1}) \\
\func{Ci}(\chi _{-}-V_{0})-\func{Ci}(\chi _{-}-V_{1})
\end{array}
\right] \text{.}  \label{S,C-ints2}
\end{equation}
with $\chi _{\pm }\equiv T\pm (Z+\tfrac{\pi }{2})$, and $\func{Si}%
(x)=\int_{0}^{x}w^{-1}\sin w\,dw$ and $\func{Ci}(x)=-\int_{x}^{\infty
}w^{-1}\cos w\,dw$ the usual sine and cosine integrals.

One case in which it is possible to solve explicitly for $V_{0}$ and $V_{1}$%
, and so to write explicit formulas for the metric perturbations $h_{ab}$,
occurs when the image point $[|\mathbf{\bar{x}}|,\mathbf{\bar{x}}]$ lies on
the $\bar{z}$-axis. Then $X=Y=0$ and $T=|Z|$, and equation (\ref{roots})
reduces to
\begin{equation}
V_{k}=|Z|-k\pi -[1+\{Z-(|k|-\tfrac{1}{2})\pi \}^{2}]^{1/2}\text{.}
\end{equation}
In particular,
\begin{equation}
V_{0}=|Z|-[1+(Z+\tfrac{\pi }{2})^{2}]^{1/2}\qquad \text{and\qquad }%
V_{1}=|Z|-\pi -[1+(Z-\tfrac{\pi }{2})^{2}]^{1/2}  \label{roots2}
\end{equation}
while $\chi _{\pm }=|Z|\pm (Z+\tfrac{\pi }{2})$.

Hence,
\begin{eqnarray}
&&(\chi _{+}-V_{1}-2\pi )/(\chi _{+}-V_{0})  \nonumber \\
&=&(\chi _{-}-V_{0})/(\chi _{-}-V_{1})  \nonumber \\
&=&\{Z-\tfrac{\pi }{2}+[1+(Z-\tfrac{\pi }{2})^{2}]^{1/2}\}\{-Z-\tfrac{\pi }{2%
}+[1+(Z+\tfrac{\pi }{2})^{2}]^{1/2}\}
\end{eqnarray}
and $I_{-}=I_{+}$. This means that $h_{tz}=0$, and that
\begin{eqnarray}
&&h_{tt}=h_{xx}=h_{yy}=h_{zz}  \nonumber \\
&=&8\mu \ln (\{Z-\tfrac{\pi }{2}+[1+(Z-\tfrac{\pi }{2})^{2}]^{1/2}\}\{-Z-%
\tfrac{\pi }{2}+[1+(Z+\tfrac{\pi }{2})^{2}]^{1/2}\})  \nonumber \\
&&
\end{eqnarray}
increases monotonically from approximately $-19.734\mu $ (at $Z=0$) to $0$
(as $|Z|\rightarrow \infty $). In fact, for large values of $|Z|$,
\begin{equation}
h_{tt}=h_{xx}=h_{yy}=h_{zz}=-8\pi \mu /|Z|+O(|Z|^{-2})\text{.}
\end{equation}
The variation of $h_{tt}/(4\mu )$ with $Z$ is plotted in Figure 4a.

Explicit formulas for the remaining non-zero metric perturbations, $%
h_{tx}=4\mu (S_{+}+S_{-})$, $h_{ty}=-4\mu (C_{+}+C_{-})$, $h_{zx}=4\mu
(S_{+}-S_{-})$ and $h_{zy}=-4\mu (C_{+}-C_{-})$ can be found by substituting
the expressions (\ref{roots2}) into (\ref{S,C-ints1}) and (\ref{S,C-ints2}).
However, there is little to be gained from writing these formulas out in
full. It is easily seen that if $Z$ is replaced by $-Z$ then $%
S_{+}\leftrightarrow -S_{-}$ and $C_{+}\leftrightarrow -C_{-}$, and so $%
h_{tx}\rightarrow -h_{tx}$, $h_{ty}\rightarrow -h_{ty}$, $h_{zx}\rightarrow
h_{zx}$ and $h_{zy}\rightarrow h_{zy}$. This set of symmetries is to be
expected on physical grounds, as the ACO loop is invariant under a complete
spatial inversion $(x\rightarrow -x$, $y\rightarrow -y$ and $z\rightarrow -z$%
), and therefore in this special case the space-space components of $h_{ab}$
should be even functions of $Z$, and the time-space components odd functions
of $Z$.

The variations of $h_{tx}/(4\mu )$, $h_{ty}/(4\mu )$, $h_{zx}/(4\mu )$ and $%
h_{zy}/(4\mu )$ are plotted in Figures 4b to 4e respectively. At $Z=0$, the
perturbations $h_{tx}$ and $h_{ty}$ are of course both zero, while $%
h_{zx}\approx -5.\,308\mu $ and $h_{zy}\approx 12.\,395\mu $. All the
perturbations tend to zero as $|Z|\rightarrow \infty $, with
\begin{equation}
h_{tx}=4\pi \mu /Z+O(|Z|^{-3})\text{,}
\end{equation}
\begin{equation}
h_{ty}=-4\pi \mu /(Z|Z|)+O(|Z|^{-4})\text{,}
\end{equation}
\begin{equation}
h_{zx}=-4\pi \mu /|Z|+O(|Z|^{-3})
\end{equation}
and
\begin{equation}
h_{zy}=\tfrac{13}{6}\pi \mu /Z^{4}+O(|Z|^{-6})\text{.}
\end{equation}

\section{Solving the String Equations of Motion}

\subsection{Near-field form of the metric perturbations}

For the purposes of setting out the first-order equation of motion (\ref
{eqmo-first}), it is necessary to generate expressions for the metric
perturbations $h_{ab}$ in the neighborhood of the world sheet of the
evaporating loop. As was mentioned in Section 4, when doing this it is
sufficient to restrict attention to points near the spacelike surface $t=0$
only. So consider a field point $x^{a}=[t,\mathbf{x}]^{a}$ of the form $%
X^{a}(u,v)+\delta x^{a}$, where $X^{a}(u,v)$ is a fixed point on the
flat-space trajectory (\ref{evap1}) with $|\kappa \mu (u+v)|\ll 1$, and the
components of the displacement $\delta x^{a}$ are all of the order of some
small parameter.

A first point to note is that if $X^{a}(\bar{u},\bar{v})$ is any source
point on the loop's trajectory on the backwards light cone of $X^{a}(u,v)$
then $|\kappa \mu (\bar{u}+\bar{v})|$ will also be small compared to $1$ for
reasonable values of $\kappa \mu $. This follows from equation (\ref
{backlight}), which in the current situation reads:
\begin{equation}
t_{L}[1-e^{-\kappa \mu (u+v)}]-t_{L}[1-e^{-\kappa \mu (\bar{u}+\bar{v})}]=|%
\tfrac{1}{4\pi }L_{0}\,e^{-\kappa \mu (u+v)}\,\mathbf{r}(u,v)-\tfrac{1}{4\pi
}L_{0}\,e^{-\kappa \mu (\bar{u}+\bar{v})}\,\mathbf{r}(\bar{u},\bar{v})|\text{%
.}
\end{equation}
An application of the triangle inequality gives
\begin{equation}
t_{L}|e^{-\kappa \mu (\bar{u}+\bar{v})}-e^{-\kappa \mu (u+v)}|\leq \tfrac{1}{%
4\pi }L_{0}\,e^{-\kappa \mu (u+v)}\,|\mathbf{r}(u,v)|+\tfrac{1}{4\pi }%
L_{0}\,e^{-\kappa \mu (\bar{u}+\bar{v})}\,|\mathbf{r}(\bar{u},\bar{v})|\text{%
,}
\end{equation}
or equivalently (in parallel with the derivation of (\ref{ineq1})),
\begin{equation}
|1-e^{\kappa \mu (\bar{u}+\bar{v})-\kappa \mu (u+v)}|\leq (1+e^{\kappa \mu (%
\bar{u}+\bar{v})-\kappa \mu (u+v)})(1+\tfrac{1}{4}\pi ^{2})^{1/2}\kappa \mu
\text{.}
\end{equation}
Hence,
\begin{equation}
|\kappa \mu (\bar{u}+\bar{v})-\kappa \mu (u+v)|\leq \ln \left[ \frac{1+(1+%
\tfrac{1}{4}\pi ^{2})^{1/2}\kappa \mu }{1-(1+\tfrac{1}{4}\pi
^{2})^{1/2}\kappa \mu }\right] \text{,}
\end{equation}
and so, for example, $|\kappa \mu (\bar{u}+\bar{v})|$ will be no greater
than $10^{-2}$ everywhere if $\kappa \mu \lesssim 2.69\times 10^{-3}$.

Since $|\kappa \mu (\bar{u}+\bar{v})|$ is guaranteed to be small, the
equations (\ref{roots}) for $V_{k}$ and (\ref{metperts})-(\ref{S,C-ints2})
for $h_{ab}$ continue to apply in this case, with
\begin{equation}
T=u+v+4\pi L_{0}^{-1}\delta t\qquad \text{and\qquad }(X,Y,Z)^{T}=\mathbf{r}%
(u,v)+4\pi L_{0}^{-1}\delta \mathbf{x}\text{.}
\end{equation}
where as before $\mathbf{r}(u,v)=\cos (v)\,\mathbf{\hat{x}}+\sin (v)\,%
\mathbf{\hat{y}}+(|u|-\tfrac{1}{2}\pi )\,\mathbf{\hat{z}}$. In particular,
if $\delta x^{a}=0$ the equation for $V_{k}$ becomes:
\begin{equation}
u+v-(k\pi +V_{k})=[2-2\cos (v-V_{k})+(|u|-|k|\pi )^{2}]^{1/2}\text{,}
\end{equation}
and, given that $u$ is restricted to the range $(-\pi ,\pi ]$, the trivial
solution $V_{k}=v$ holds in two cases: (i) if $u>0$ and $k=0$; and (ii) if $%
u<0$ and $k=-1$.

In the analysis that follows it will be assumed that $u>0$. Some remarks
about the behavior of $h_{ab}$ near points $X^{a}(u,v)$ with $u<0$ will be
made towards the end of this section. When $u>0$ the $z$-component of the
field point satisfies $Z+\tfrac{1}{2}\pi =u+4\pi L_{0}^{-1}\delta z$, and so
\begin{equation}
\chi _{\pm }\equiv v+(u\pm u)+4\pi L_{0}^{-1}(\delta t\pm \delta z)\text{.}
\end{equation}

Furthermore, if $\delta x^{a}=0$ the limits of integration in (\ref{h-ints})
are given by $V_{0}=v$ and $V_{1}=v-W(u)$, where $W$ is the unique root of
the equation
\begin{equation}
u-\pi +W=[2-2\cos W+(u-\pi )^{2}]^{1/2}\text{.}  \label{W-def}
\end{equation}
Note that because $V_{0}\geq V_{1}$ it follows that $W\geq 0$, while it is
clear from (\ref{W-def}) that $W=0$ only at the kink point $u=\pi $. For
future reference, it is useful to rearrange (\ref{W-def}) in the form
\begin{equation}
u-\pi =W^{-1}(1-\cos W)-\tfrac{1}{2}W\text{.}  \label{u-W}
\end{equation}
(Incidentally, one way to verify that (\ref{W-def}) has a unique solution $W$
is to note that the function on the right-hand side of (\ref{u-W}) is
strictly decreasing.)

Implicit differentiation of the defining equation (\ref{roots}) for $V_{k}$
gives
\[
\partial _{T}V_{k}=H_{k}^{-1}[T-(k\pi +V_{k})]
\]
and
\begin{equation}
(\partial _{X}V_{k},\partial _{Y}V_{k},\partial _{Z}V_{k})=H_{k}^{-1}(\cos
V_{k}-X,\sin V_{k}-Y,(|k|-\tfrac{1}{2})\pi -Z)\text{,}
\end{equation}
with
\begin{equation}
H_{k}\equiv T-(k\pi +V_{k})+X\sin V_{k}-Y\cos V_{k}\text{.}
\end{equation}
Note that if $\delta x^{a}=0$ then $H_{0}=u$ and $H_{1}=u-\pi +W-\sin W$.

Hence, to linear order in $\delta x^{a}$,
\begin{equation}
V_{0}=v+4\pi L_{0}^{-1}(\delta t-\delta z)\text{\qquad and\qquad }%
V_{1}=v-W+4\pi L_{0}^{-1}\delta V  \label{Vk}
\end{equation}
where
\begin{equation}
\delta V\equiv H_{1}^{-1}[(u-\pi +W)\delta t+\{\cos (v-W)-\cos v\}\delta
x+\{\sin (v-W)-\sin v\}\delta y+(\pi -u)\delta z]\text{.}  \label{del-V}
\end{equation}
Unfortunately, the linear expansion (\ref{Vk}) for $V_{0}$ is not sharp
enough to resolve the singular behavior of $h_{ab}$ near the world sheet, as
$\chi _{-}=v+4\pi L_{0}^{-1}(\delta t-\delta z)$ and so $\chi _{-}-V_{0}=0$
at this order.

At quadratic order in $\delta x^{a}$ it turns out that
\begin{equation}
V_{0}=v+4\pi L_{0}^{-1}(\delta t-\delta z)-\tfrac{1}{2}u^{-1}(4\pi
L_{0}^{-1})^{2}Q_{+}\text{,}
\end{equation}
where
\begin{eqnarray}
Q_{+} &\equiv &(\delta x)^{2}+(\delta y)^{2}+(\delta t-\delta z)(\delta
t-\delta z+2\delta x\sin v-2\delta y\cos v)  \nonumber \\
&\equiv &\bar{\Psi}_{ab}(+1,v)\,\delta x^{a}\delta x^{b}\text{.}
\label{Q-plus}
\end{eqnarray}
Since $\bar{\Psi}_{ab}$ is proportional to the orthogonal complement $q_{ab}$
of the projection tensor $p_{ab}$, the quadratic form $Q_{+}$ is zero if and
only if $\delta x^{a}$ lies in the tangent plane to the world sheet at $%
X^{a}(u,v)$. It should therefore be understood that in evaluating $h_{ab}$
and its derivatives the limit $\delta x^{a}\rightarrow 0$ is taken only from
directions outside the world sheet.

The algebraic computations whose results appear throughout the rest of this
section are long and very mechanical, and their details have been relegated
to Appendix B. Neglecting contributions that tend to zero in the limit as $%
\delta x^{a}\rightarrow 0$, the metric perturbations have the near-field
form:

\begin{eqnarray}
(4\mu )^{-1}h_{ab} &=&\{\ln [(4\pi L_{0}^{-1})^{2}Q_{+}]\,-\ln (2u)\}\bar{%
\Psi}_{ab}(+1,v)+(\gamma _{\text{E}}-\func{Ci}W)\,\Phi _{ab}(+1,v)  \nonumber
\\
&&+(\func{Si}W)\,\Omega _{ab}(+1,v)+K(u)\,\Phi _{ab}(-1,2u+v)-\Sigma
(u)\,\Omega _{ab}(-1,2u+v)  \nonumber \\
&&-(\ln W)\,\Pi _{ab}(+1)+[\ln \left( W+2u-2\pi \right) -\ln (2u)]\,\Pi
_{ab}(-1)\text{,}  \label{h-near}
\end{eqnarray}
where $\gamma _{\text{E}}\approx 0.5772$ is Euler's constant,
\begin{equation}
\Phi _{ab}(s,v)\equiv \left[
\begin{array}{cccc}
0 & \sin v & -\cos v & 0 \\
\sin v & 0 & 0 & -s\sin v \\
-\cos v & 0 & 0 & s\cos v \\
0 & -s\sin v & s\cos v & 0
\end{array}
\right] _{ab}\text{,}  \label{Phi}
\end{equation}
\begin{equation}
\Omega _{ab}(s,v)\equiv \partial _{v}\Phi _{ab}(s,v)  \label{Omega}
\end{equation}
and
\begin{equation}
\Pi _{ab}(s)\equiv \left[
\begin{array}{cccc}
1 & 0 & 0 & -s \\
0 & 1 & 0 & 0 \\
0 & 0 & 1 & 0 \\
-s & 0 & 0 & 1
\end{array}
\right] _{ab}\text{.}  \label{Pi}
\end{equation}
Note in particular that $\bar{\Psi}_{ab}(s,v)=\Phi _{ab}(s,v)+\Pi _{ab}(s)$.
Also, the functions $K$ and $\Sigma $ are shorthand for
\begin{equation}
K(u)\equiv \func{Ci}(W+2u-2\pi )-\func{Ci}\left( 2u\right) \text{\qquad
and\qquad }\Sigma (u)\equiv \func{Si}(W+2u-2\pi )-\func{Si}\left( 2u\right)
\text{.}
\end{equation}

Similarly, the leading-order contributions to the spacetime derivatives of $%
h_{ab}$ are:
\begin{eqnarray}
(4\mu )^{-1}(\tfrac{1}{4\pi }L_{0})\partial _{t}h_{ab} &=&(\tfrac{1}{4\pi }%
L_{0})\,(\partial _{t}\Lambda \,)\bar{\Psi}_{ab}(+1,v)+(\delta t-\delta
z)\,(\partial _{t}\Lambda \,)\,\Omega _{ab}(+1,v)  \nonumber \\
&&+[\Lambda +\gamma _{\text{E}}-\func{Ci}W-\ln (2u)]\,\Omega _{ab}(+1,v)-(%
\func{Si}W\,)\,\Phi _{ab}(+1,v)  \nonumber \\
&&+K(u)\,\Omega _{ab}(-1,2u+v)+\Sigma (u)\,\Phi _{ab}(-1,2u+v)  \nonumber \\
&&+\tfrac{\sin W}{u-\pi +W-\sin W}[\tfrac{1}{W}\,\bar{\Psi}_{ab}(+1,v-W)
\nonumber \\
&&-\tfrac{1}{W+2u-2\pi }\,\bar{\Psi}_{ab}(-1,v-W)]\text{,}  \label{ht-near}
\end{eqnarray}
\begin{eqnarray}
(4\mu )^{-1}(\tfrac{1}{4\pi }L_{0})\partial _{x}h_{ab} &=&(\tfrac{1}{4\pi }%
L_{0})\,(\partial _{x}\Lambda \,)\,\bar{\Psi}_{ab}(+1,v)+(\delta t-\delta
z)\,(\partial _{x}\Lambda \,)\,\Omega _{ab}(+1,v)  \nonumber \\
&&+\tfrac{\cos (v-W)-\cos v}{u-\pi +W-\sin W}[\tfrac{1}{W}\,\bar{\Psi}%
_{ab}(+1,v-W)  \nonumber \\
&&-\tfrac{1}{W+2u-2\pi }\,\bar{\Psi}_{ab}(-1,v-W)]\text{,}  \label{hx-near}
\end{eqnarray}
\begin{eqnarray}
(4\mu )^{-1}(\tfrac{1}{4\pi }L_{0})\partial _{y}h_{ab} &=&(\tfrac{1}{4\pi }%
L_{0})(\partial _{y}\Lambda \,)\,\bar{\Psi}_{ab}(+1,v)+(\delta t-\delta
z)\,(\partial _{y}\Lambda \,)\,\Omega _{ab}(+1,v)  \nonumber \\
&&+\tfrac{\sin (v-W)-\sin v}{u-\pi +W-\sin W}[\tfrac{1}{W}\,\bar{\Psi}%
_{ab}(+1,v-W)  \nonumber \\
&&-\tfrac{1}{W+2u-2\pi }\,\bar{\Psi}_{ab}(-1,v-W)]  \label{hy-near}
\end{eqnarray}
and
\begin{eqnarray}
(4\mu )^{-1}(\tfrac{1}{4\pi }L_{0})\partial _{z}h_{ab} &=&(\tfrac{1}{4\pi }%
L_{0})\,(\partial _{z}\Lambda \,)\,\bar{\Psi}_{ab}(+1,v)+(\delta t-\delta
z)\,(\partial _{z}\Lambda \,)\,\Omega _{ab}(+1,v)  \nonumber \\
&&-[\Lambda +\gamma _{\text{E}}-\func{Ci}W-\ln (2u)]\,\Omega _{ab}(+1,v)+(%
\func{Si}W)\,\Phi _{ab}(+1,v)  \nonumber \\
&&+K(u)\,\Omega _{ab}(-1,2u+v)+\Sigma (u)\,\Phi _{ab}(-1,2u+v)-u^{-1}\bar{%
\Psi}_{ab}(-1,v)  \nonumber \\
&&+\tfrac{1}{u-\pi +W-\sin W}[\tfrac{W-\sin W}{W}\,\bar{\Psi}_{ab}(+1,v-W)
\nonumber \\
&&+\tfrac{2u-2\pi +W-\sin W}{W+2u-2\pi }\,\bar{\Psi}_{ab}(-1,v-W)]\text{,}
\label{hz-near}
\end{eqnarray}
where
\begin{equation}
\Lambda \equiv \ln [(4\pi L_{0}^{-1})^{2}Q_{+}]\text{.}
\end{equation}

The expressions listed here for $h_{ab}$ and its derivatives neglect
additional terms of order $\delta x^{a}\ln [(4\pi L_{0}^{-1})^{2}Q_{+}]$ or
higher, which are zero in the limit $\delta x^{a}\rightarrow 0$, provided of
course that this limit is taken from directions outside the tangent plane to
the world sheet at $X^{a}(u,v)$. However, the expressions also contain terms
of order $\ln [(4\pi L_{0}^{-1})^{2}Q_{+}]$ and $\delta x^{a}\partial
_{b}\ln [(4\pi L_{0}^{-1})^{2}Q_{+}]$, which are undefined in the limit as $%
\delta x^{a}\rightarrow 0$. The presence of the singular term $4\mu \ln
[(4\pi L_{0}^{-1})^{2}Q_{+}]\bar{\Psi}_{ab}(+1,v)$ in $h_{ab}$ is a
reflection of the conical singularity that is expected the mark the location
of any zero-thickness cosmic string, as has been explained in more detail in
Paper I. Fortunately, none of the singular terms makes a contribution to the
first-order equation of motion, as will been seen shortly.

Another important point to note is that equations (\ref{h-near}) and (\ref
{ht-near})-(\ref{hz-near}) are valid only near points $X^{a}(u,v)$ with $%
u\in (0,\pi )$, which lie on one of the two smooth segments of the loop
shown in Figure 1. The metric perturbations near points on the second
segment, for which $u\in (-\pi ,0)$, can be calculated by making use of the
identities $I_{+}(u-\pi ,v)=I_{-}(u,v)$ and $I_{-}(u-\pi ,v)=I_{+}(u,v)$ for
$u\in (0,\pi )$, and the analogous identities satisfied by the pairs $%
(S_{+},S_{-})$ and $(C_{+},C_{-})$. In view of equation (\ref{metperts}) for
$h_{ab}$, it follows that all the components of $h_{ab}$ except $h_{tz}$, $%
h_{xz}$ and $h_{yz}$ are half-range periodic in $u$ near the world sheet
(that is, $h_{ab}(u-\pi ,v)=h_{ab}(u,v)$), while the components $h_{tz}$, $%
h_{xz}$ and $h_{yz}$ are half-range anti-periodic (that is, $h_{ab}(u-\pi
,v)=-h_{ab}(u,v)$). This behavior was previously mentioned of the tangential
projections $h_{ab}X_{(0)}^{a},_{A}X_{(0)}^{b},_{B}$ (which are all
half-range periodic in $u$) in Paper I. Although Paper I deals with a loop
that is strictly stationary, the statements it makes about the local
behavior of $h_{ab}$ are still applicable here, as the local effects of loop
evaporation first appear at order $\mu ^{2}$ in $g_{ab}$.

\subsection{The equations of motion}

It is now possible to write down and verify or solve the equations of motion
(\ref{eqmo-zero}) and (\ref{eqmo-first}). Since the only non-zero component
of the 2-metric $\gamma ^{AB}$ is $\gamma ^{uv}=(\tfrac{1}{4\pi }L_{0})^{-2}$
(to leading order in $\mu $), the zeroth-order equation of motion $%
q_{d}^{c}\gamma ^{CD}X_{(0)}^{d},_{CD}=0$ reads simply $%
q_{d}^{c}X_{(0)}^{d},_{uv}=0$. It is clear by inspection of (\ref{pos-zero})
that $X_{(0)}^{d},_{uv}=0$, and so (\ref{eqmo-zero}) is automatically
satisfied. The second term in the first-order equation of motion (\ref
{eqmo-first}), which is proportional to $\gamma ^{CD}X_{(0)}^{d},_{CD}$, is
consequently also zero.

The details of the calculation of the remaining terms in (\ref{eqmo-first})
can be found in Appendix C. Because all the terms in the first-order
equation of motion are projected onto $q_{d}^{c}$, and so lie in the
subspace orthogonal to the tangent plane at $X^{a}(u,v)$, the equation
contains only two independent components. In what follows, it proves to be
convenient to decompose each term in the directions of the vectors
\begin{equation}
M^{a}(s,v)\equiv [1,-\sin (v)\,\mathbf{\hat{x}}+\cos (v)\,\mathbf{\hat{y}}+s%
\hspace{0in}\,\mathbf{\hat{z}}]^{a}\text{\qquad and\qquad }N^{a}(v)\equiv
[0,\cos (v)\,\mathbf{\hat{x}}+\sin (v)\,\mathbf{\hat{y}}]^{a}\text{,}
\end{equation}
which span the orthogonal subspace.

The first and third terms in (\ref{eqmo-first}), which depend on the
perturbation $\delta X^{a}$ and therefore on the correction functions $F$, $%
G $ and $A$, are:
\begin{equation}
q_{d}^{c}\gamma ^{CD}\delta X^{d},_{CD}=2\mu (\tfrac{1}{4\pi }%
L_{0})^{-1}(F_{uv}+sA_{uv}+G_{u}-\kappa )\,M^{c}+2\mu (\tfrac{1}{4\pi }%
L_{0})^{-1}G_{uv}\,N^{c}  \label{em1}
\end{equation}
and
\begin{equation}
-2q_{d}^{c}\gamma ^{AC}\gamma ^{BD}\eta _{ab}X_{(0)}^{a},_{A}\delta
X^{b},_{B}X_{(0)}^{d},_{CD}=2\mu (\tfrac{1}{4\pi }L_{0})^{-1}[\kappa s(|u|-%
\tfrac{1}{2}\pi )-(F_{u}+sA_{u})]N^{c}\text{.}  \label{em2}
\end{equation}

The remaining terms depend on $h_{ab}$ and its derivatives, which were
evaluated earlier in the case $u\in (0,\pi )$ only. For positive values of $%
u $ (or equivalently, for $s=+1$) it turns out that
\begin{equation}
-q_{d}^{c}\gamma ^{AC}\gamma
^{BD}h_{ab}X_{(0)}^{a},_{A}X_{(0)}^{b},_{B}X_{(0)}^{d},_{CD}=16\mu (\tfrac{1%
}{4\pi }L_{0})^{-1}[\ln \left( W+2u-2\pi \right) -\ln (2u)]N^{c}\text{,}
\label{em3}
\end{equation}
\begin{eqnarray}
-q^{cd}\gamma ^{AB}X_{(0)}^{a},_{A}X_{(0)}^{b},_{B}h_{bd},_{a} &=&16\mu (%
\tfrac{1}{4\pi }L_{0})^{-1}\Sigma (u)[-(2\cos 2u)M^{c}+(\sin 2u)N^{c}]
\nonumber \\
&&+16\mu (\tfrac{1}{4\pi }L_{0})^{-1}K(u)[(2\sin 2u)M^{c}+(\cos 2u)N^{c}]
\nonumber \\
&&+4\mu (\tfrac{1}{4\pi }L_{0})^{-1}\tfrac{1}{u-\pi +W-\sin W}[(1-\cos
W)M^{c}-(\sin W)N^{c}]  \nonumber \\
&&+4\mu (\tfrac{1}{4\pi }L_{0})^{-1}\tfrac{W+2u-2\pi -2\sin W}{u-\pi +W-\sin
W}\tfrac{1}{W+2u-2\pi }  \nonumber \\
&&\times [3(1-\cos W)M^{c}-(\sin W)N^{c}]  \label{em4}
\end{eqnarray}
and
\begin{eqnarray}
\tfrac{1}{2}q^{cd}\gamma ^{AB}X_{(0)}^{a},_{A}X_{(0)}^{b},_{B}h_{ab},_{d}
&=&-16\mu (\tfrac{1}{4\pi }L_{0})^{-1}[K(u)\sin 2u-\Sigma (u)\cos 2u]M^{c}
\nonumber \\
&&-8\mu (\tfrac{1}{4\pi }L_{0})^{-1}\tfrac{W+2u-2\pi -\sin W}{u-\pi +W-\sin W%
}\tfrac{1-\cos W}{W+2u-2\pi }M^{c}  \nonumber \\
&&-8\mu (\tfrac{1}{4\pi }L_{0})^{-1}\tfrac{1-\cos W}{u-\pi +W-\sin W}\tfrac{%
1-\cos W}{W+2u-2\pi }N^{c}\text{.}  \label{em5}
\end{eqnarray}

Now, it was shown in Paper I that $h_{ab}X_{(0)}^{a},_{A}X_{(0)}^{b},_{B}$
is half-range periodic in $u$, and it is easily seen that taking the inner
product with $q_{d}^{c}\gamma ^{AC}\gamma ^{BD}X_{(0)}^{d},_{CD}$ preserves
this symmetry. Hence, the functional dependence of the term on the left of (%
\ref{em3}) for $u\in (-\pi ,0)$ is the half-range periodic extension of the
term on the right. Similarly, the sum of the terms in (\ref{em4}) and (\ref
{em5}) -- which is to say, the term on the right-hand side of the
first-order equation of motion (\ref{eqmo-first}) -- is just $2\alpha ^{a}$,
where $\alpha ^{a}$ is the acceleration vector introduced in Paper I. It was
demonstrated in Paper I that the $t$-, $x$- and $y$-components of $\alpha
^{a}$ are half-range periodic in $u$, while the $z$-component is half-range
anti-periodic, and this is also true of the vectors $M^{c}(s,v)$ and $%
N^{c}(v)$. Hence, the first-order equation of motion requires the sum of (%
\ref{em1}) and (\ref{em2}) to have the same symmetries, and since the term $%
\kappa s(|u|-\tfrac{1}{2}\pi )$ in (\ref{em2}) is half-range periodic this
means that $F+sA$ and $G$ must be half-range periodic as well.

These properties can also be predicted from the geometry of the problem. The
two segments of the evaporating loop should share the same geometry, so the
correction functions $F$, $G$ and $A$ are expected to have the same
symmetries as the corresponding components of the original ACO loop. That
is, $F$ and $G$ should be half-range periodic in $u$ as the $t$ and $x$-$y$
components of the ACO\ loop are, while $A$ should be half-range
anti-periodic as the $z$-component $|u|-\frac{1}{2}\pi $ is.

On setting $s=+1$ in the first two terms (\ref{em1}) and (\ref{em2}), the
equation of motion for $u\in (0,\pi )$ reduces to two partial differential
equations:
\begin{equation}
F_{uv}+A_{uv}+G_{u}-\kappa =8[K(u)\sin 2u-\Sigma (u)\cos 2u]+4\tfrac{1-\cos W%
}{u-\pi +W-\sin W}\tfrac{W+2u-2\pi -2\sin W}{W+2u-2\pi }  \label{eqmo-1a}
\end{equation}
and
\begin{eqnarray}
&&G_{uv}+\kappa (u-\tfrac{1}{2}\pi )-(F_{u}+A_{u})+8[\ln \left( W+2u-2\pi
\right) -\ln (2u)]  \nonumber \\
&=&8[K(u)\cos 2u+\Sigma (u)\sin 2u]-4\tfrac{1}{u-\pi +W-\sin W}(\sin W-2%
\tfrac{1-\cos W}{W+2u-2\pi }\cos W)  \nonumber \\
&&  \label{eqmo-2a}
\end{eqnarray}
which appear as the coefficients of $M^{c}$ and $N^{c}$ respectively.

At a schematic level, these equations read:
\begin{equation}
F_{uv}+A_{uv}+G_{u}=P_{1}(u)\qquad \text{and\qquad }%
G_{uv}-F_{u}-A_{u}=P_{2}(u)\text{,}  \label{schema}
\end{equation}
where the functions $P_{1}$ and $P_{2}$ represent all the remaining terms.
Taking the $v$ derivative of each equation in turn and adding or subtracting
the other leads to two forced simple harmonic equations:
\begin{equation}
G_{uvv}+G_{u}=P_{1}(u)\qquad \text{and\qquad }%
(F_{u}+A_{u})_{vv}+F_{u}+A_{u}=-P_{2}(u)\text{,}
\end{equation}
whose general solutions are:
\begin{equation}
G=c_{1}(v)+a(u)\cos v+b(u)\sin v+\int P_{1}(u)\,du  \label{G-1}
\end{equation}
and
\begin{equation}
F+A=c_{2}(v)+b(u)\cos v-a(u)\sin v-\int P_{2}(u)\,du\text{,}  \label{FA-1}
\end{equation}
where $c_{1}$ and $c_{2}$ are arbitrary continuous $2\pi $-periodic
functions of $v$, and $a$ and $b$ are the restrictions to $(0,\pi )$ of
arbitrary half-range periodic $C^{1}$ functions of $u$.

The physical significance of the arbitrary homogeneous contributions to (\ref
{G-1}) and (\ref{FA-1}) can be understood by referring to the intrinsic
2-metric $\gamma _{AB}$. The components of $\gamma _{AB}$ are expanded to
order $\mu $ in Appendix D.1, where it is shown that $\gamma _{AB}$ takes on
the self-similar form (\ref{self-sim}) at this order if and only if
\begin{equation}
F_{uv}+A_{uv}=0\text{,\qquad }F_{uv}+F_{vv}+A_{vv}=0\qquad \text{and\qquad }%
G_{v}+F_{vv}=0\text{.}  \label{constraints}
\end{equation}
The first of these constraints reduces the first equation in (\ref{schema})
to $G_{u}=P_{1}(u)$, and so from (\ref{G-1}) $a$ and $b$ are both zero. The
trigonometric functions of $v$ in (\ref{G-1}) and (\ref{FA-1}) therefore
represent free vibrations of the loop with amplitudes of order $\mu $, which
break the self-similarity of the evaporating loop and, if allowed to remain,
would contribute to the metric tensor $g_{ab}$ at order $\mu ^{2}$.

The second of the constraints (\ref{constraints}) now reduces to $%
F_{uv}=-c_{2}^{\prime \prime }(v)$, and since $F$ is by assumption
continuous and half-range periodic in $u$ it follows that $c_{2}^{\prime }=0$
and therefore that $c_{2}$ is a constant (and can be absorbed into $\int
P_{2}(u)\,du$). Consequently $F_{uv}=0$, and $F$ has the generic form $%
F_{1}(u)+F_{2}(v)$ for some choice of continuous and suitably periodic
functions $F_{1}$ and $F_{2}$. Furthermore, the third constraint now becomes
$G_{v}=-F_{2}^{\prime \prime }$, and so from (\ref{G-1}) $%
c_{1}(v)=-F_{2}^{\prime }$.

The possibility of further constraining $F$, $G$ and $A$ arises from a
consideration of transformations of the spacetime coordinates $%
x^{a}=[t,x,y,z]^{a}$ of the form $x^{a}\rightarrow \hat{x}^{a}$ where
\begin{equation}
\hat{x}^{a}=x^{a}+\mu y^{a}(x^{b})\text{.}
\end{equation}
The only transformations of this type that are admissible are those that
preserve the form of the metric tensor $g_{ab}=\eta _{ab}+\mu h_{ab}$ at
order $\mu $. Since
\begin{equation}
\hat{g}_{ab}=\eta _{ab}+\mu h_{ab}+\mu \eta _{ca}\partial _{b}y^{c}+\mu \eta
_{cb}\partial _{a}y^{c}
\end{equation}
at linear order in $\mu $, admissible transformations satisfy the condition $%
\eta _{c(a}\partial _{b)}y^{c}=0$ and so are generators of the Poincar\'{e}
transformations (that is, combinations of boosts, rotations and
translations).

Under such a transformation, the position vector $X^{a}$ of the loop is
replaced by
\begin{equation}
\hat{X}^{a}=X_{(0)}^{a}+\delta X^{a}+\mu Y^{a}(u,v)  \label{Poin-transf}
\end{equation}
where $X_{(0)}^{a}$ and $\delta X^{a}$ are given by (\ref{X0-new}) and (\ref
{delX-new}) respectively, and
\begin{equation}
Y^{a}(u,v)\equiv \tfrac{1}{4\pi }L_{0}[-\delta F(u,v),\delta G(u,v)\{\cos
(v)\,\mathbf{\hat{x}}+\sin (v)\,\mathbf{\hat{y}}\}+\delta A(u,v)\,\mathbf{%
\hat{z}}]^{a}
\end{equation}
for some choice of functions $\delta F$, $\delta G$ and $\delta A$.\footnote{%
Note that the $x$- and $y$-components of $Y^{a}$ have the forms shown so as
to preserve the isotropic gauge condition discussed in Section 4, following
equation (\ref{evap2}).}

It is only possible to impose the admissibility condition $\eta
_{c(a}\partial _{b)}y^{c}=0$ on $Y^{a}$ if the derivatives are taken tangent
to the worldsheet of the loop, and so $Y^{a}$ must satisfy the equations
\begin{equation}
X_{(0)},_{u}\cdot Y,_{u}=0\text{,\qquad }X_{(0)},_{v}\cdot Y,_{v}=0\text{%
\qquad and\qquad }X_{(0)},_{u}\cdot Y,_{v}+X_{(0)},_{v}\cdot Y,_{u}=0\text{.}
\end{equation}
In terms of $\delta F$, $\delta G$ and $\delta A$ these equations read, for $%
u\in (0,\pi )$,
\begin{equation}
\delta F_{u}+\delta A_{u}=0\qquad \delta G+\delta F_{v}=0\text{\qquad
and\qquad }\delta F_{v}+\delta A_{v}+\delta F_{u}=0\text{.}
\label{Poin-const}
\end{equation}
The choice $\delta F=-F_{2}(v)$, $\delta G=F_{2}^{\prime }(v)$ and $\delta
A=F_{2}(v)$ satisfies all three equations, and specifies a spacetime
transformation which reduces $F$ and $A$ to functions of $u$ only, and sets $%
c_{1}=0$ in (\ref{G-1}).

In a suitable set of spacetime coordinates, therefore, the correction
functions become
\begin{equation}
G=\int P_{1}(u)\,du\qquad \text{and\qquad }F+A=-\int P_{2}(u)\,du\text{,}
\end{equation}
with $F$ a continuous and half-range periodic function of $u$ alone. In
fact, $F$ is effectively an arbitrary function of $u$, as it is always
possible to define a transformation $u\rightarrow u^{\prime }=u+\mu U(u)$ of
the gauge coordinate $u$, where $U$ is continuous and half-range periodic
but otherwise arbitrary, under which $X^{a}\rightarrow X^{a}+\mu
UX_{(0)}^{a},_{u}$ and so $F\rightarrow F-U$ and $A\rightarrow A+U$ for $%
u\in (0,\pi )$, but $G$ remains unchanged.

\subsection{Solving for $F$, $G$ and $A$}

The equation (\ref{u-W}) specifying $u-\pi $ as a function of $W$ can be
used to eliminate $u$ from the terms in (\ref{eqmo-1a}) and (\ref{eqmo-2a})
which mix both $u$ and $W$. The resulting expressions for $P_{1}$ and $P_{2}$
read:
\begin{equation}
P_{1}(u)=\kappa +8[K(u)\sin 2u-\Sigma (u)\cos 2u]+4(1-\cos W-W\sin W)/J(W)
\end{equation}
and
\begin{eqnarray}
P_{2}(u) &=&-\kappa (u-\tfrac{1}{2}\pi )-8[\ln \{2W^{-1}(1-\cos W)\}-\ln
(2u)]  \nonumber \\
&&+8[K(u)\cos 2u+\Sigma (u)\sin 2u]-4(\sin W-W\cos W)/J(W)\text{,}  \nonumber
\\
&&
\end{eqnarray}
where
\begin{equation}
J(W)\equiv W^{-1}(1-\cos W)+\tfrac{1}{2}W-\sin W
\end{equation}
and now
\begin{equation}
K(u)=\func{Ci}\{2W^{-1}(1-\cos W)\}-\func{Ci}(2u)\text{\qquad and\qquad }%
\Sigma (u)=\func{Si}\{2W^{-1}(1-\cos W)\}-\func{Si}(2u)\text{.}
\end{equation}

Furthermore, the derivative of (\ref{u-W}) reads $du=-W^{-1}J(W)dW$, and so
\begin{eqnarray}
G &=&\kappa u+8\int [K(u)\sin 2u-\Sigma (u)\cos 2u]\,du-4\int W^{-1}(1-\cos
W-W\sin W)\,dW  \nonumber \\
&=&\kappa u-4[K(u)\cos 2u+\Sigma (u)\sin 2u]+4\ln (1-\cos W)-4\ln (uW)+C_{1}
\label{G-2}
\end{eqnarray}
and
\begin{eqnarray}
F+A &=&\tfrac{1}{2}\kappa u(u-\pi )+8\int [\ln \{2W^{-1}(1-\cos W)\}-\ln
(2u)]\,du  \nonumber \\
&&-8\int [K(u)\cos 2u+\Sigma (u)\sin 2u]\text{\thinspace }du-4\int
W^{-1}(\sin W-W\cos W)\,dW  \nonumber \\
&=&\tfrac{1}{2}\kappa u(u-\pi )-8W^{-1}[1+\ln W-\ln (1-\cos W)](1-\cos
W)-4W(2-\ln W)  \nonumber \\
&&-8(u\ln u-u)-4[K(u)\sin 2u-\Sigma (u)\cos 2u]-4\int_{2\pi }^{W}\ln (1-\cos
\omega )\,d\omega +C_{2}\text{.}  \nonumber \\
&&  \label{FA-2}
\end{eqnarray}
Here, the integral $\int \ln (1-\cos \omega )\,d\omega $ unfortunately
cannot be expressed in closed form as a combination of standard functions.
(The lower bound in the integral has been set to $\omega =2\pi $ because $%
W\rightarrow 2\pi ^{-}$ as $u\rightarrow 0^{+}$.) Equations (\ref{G-2}) and (%
\ref{FA-2}) can of course be verified directly by differentiation.. However,
in tracing out the original calculations -- that is, in integrating $P_{1}$
and $P_{2}$ -- it proves to be most efficient to integrate the terms $%
K(u)\sin 2u-\Sigma (u)\cos 2u$ in (\ref{G-2}) and $\ln (1-\cos W)$ and $%
\Sigma (u)\sin 2u+K(u)\cos 2u$ in (\ref{FA-2}) by parts with respect to $u$
first, then integrate the remainders with respect to $W$.

Turning to the initial conditions satisfied by $G$ and $F+A$, it should be
recalled that $G$ is continuous and half-range periodic in $u$ and so $%
G(0)=G(\pi )$. Similarly, $F$ is half-range periodic in $u$ and so $%
F(0)=F(\pi )$. By contrast, $A$ is continuous and half-range \emph{anti}%
-periodic in $u$, and therefore $A(0)=-A(\pi )$. (Note that the combination $%
F+sA$, which appears in the leading terms (\ref{em1}) and (\ref{em2}) in the
first-order equation of motion, is half-range periodic for $u\in [-\pi ,\pi
] $ as required. However, it is not necessarily continuous at $u=0$.)

Now, the $u\rightarrow 0^{+}$ lower limit of equation (\ref{FA-2}) reads:
\begin{equation}
F(0)+A(0)=-16\pi +8\pi \ln 2\pi +C_{2}  \label{FA(0)}
\end{equation}
while the $u\rightarrow \pi ^{-}$ upper limit reads:
\begin{equation}
F(\pi )+A(\pi )=8\pi -8\pi \ln (2\pi )-4\func{Si}(2\pi )+C_{2}\text{.}
\label{FA(pi)}
\end{equation}
(Note in particular that $\int_{2\pi }^{0}\ln (1-\cos \omega )\,d\omega
=2\pi \ln 2$ at the upper limit.)

Subtracting (\ref{FA(pi)}) from (\ref{FA(0)}) therefore gives:
\begin{equation}
2A(0)\equiv A(0)-A(\pi )=-24\pi +16\pi \ln 2\pi +4\func{Si}(2\pi )
\end{equation}
and so
\begin{equation}
A(0)=-12\pi +8\pi \ln 2\pi +2\func{Si}(2\pi )\approx 11.\,328\text{.}
\label{A(0)}
\end{equation}
Similarly, addition of (\ref{FA(0)}) and (\ref{FA(pi)}) gives:
\begin{equation}
2F(0)\equiv F(0)+F(\pi )=-8\pi -4\func{Si}(2\pi )+2C_{2}
\end{equation}
and so
\begin{equation}
F(0)=-4\pi -2\func{Si}(2\pi )+C_{2}\approx -15.\,403+C_{2}\text{.}
\label{F(0)}
\end{equation}

The constant of integration $C_{2}$ in (\ref{F(0)}) represents a vestige of
coordinate freedom in the specification of the correction function $F$ that,
in view of (\ref{A(0)}) is not present in the function $A$. In fact,
different values of $C_{2}$ correspond to the same physical configuration of
the evaporating loop, as the value of $C_{2}$ can be changed at will by
making a suitable Poincar\'{e} transformation (\ref{Poin-transf}). This
follows from the fact that the choice $\delta G=\delta A=0$ and $\delta F$
any arbitrary constant (representing a translation in the coordinate $t$ at
order $\mu $) automatically satisfies the Poincar\'{e} constraints (\ref
{Poin-const}). One natural gauge choice for $F$ is to set $F(0)=0$, in which
case $C_{2}=4\pi +2\func{Si}(2\pi )$.

Consider now equation (\ref{G-2}) for $G$. Taking the lower limit $%
u\rightarrow 0^{+}$ (or equivalently $W\rightarrow 2\pi ^{-}$) in this
equation gives:
\begin{equation}
G(0)=C_{1}\text{,}
\end{equation}
while taking the upper limit $u\rightarrow \pi ^{-}$ (equivalently $%
W\rightarrow 0^{+}$) gives:
\begin{equation}
G(\pi )=\pi \kappa +4\func{Ci}(2\pi )-4\ln 2\pi -4\gamma _{\text{E}}+C_{1}%
\text{.}
\end{equation}
So it follows that $G(0)=G(\pi )$ if and only if
\begin{equation}
\kappa =[4\ln 2\pi +4\gamma _{\text{E}}-4\func{Ci}(2\pi )]/\pi \approx
3.\,103\,7\text{.}  \label{kappa}
\end{equation}
That is, iff $\kappa =\gamma ^{0}/(4\pi )$ with $\gamma ^{0}\approx 39.0025$
the radiative efficiency of the ACO loop, as expected.

Unlike $C_{2}$, the integration constant $C_{1}$ in (\ref{G-2}) does have an
invariant physical significance, and it cannot be arbitrarily changed by
making a Poincar\'{e} or gauge transformation without affecting the form of $%
F$. In fact, different values of $C_{1}$ correspond to different values for
the length of the evaporating loop as measured along curves of constant
scale factor $e^{-\kappa \mu (u+v)}$, as is explained in more detail in
Appendix D.2. If the length $L_{\text{csf}}$ of the loop along such a curve
is assumed to be independent of $\mu $ (with $L_{\text{csf}}=\frac{1}{\sqrt{2%
}}L_{0}$ along the curve $u+v=0$) then
\begin{equation}
C_{1}=-28-6\gamma _{\text{E}}+10\ln 2\pi +6\func{Ci}(2\pi )+4\pi ^{-1}\func{%
Si}(2\pi )+4\pi ^{-1}\func{Si}(4\pi )\approx -9.\,\allowbreak 514\,3\text{.}
\label{C1}
\end{equation}

With $C_{1}$ assigned this value, the correction function $G$, which is the
half-range periodic extension of the function on the right-hand side of (\ref
{G-2}), is plotted for $u\in [-\pi ,\pi ]$ in Figure 5a. The half-range
periodic extension to $[-\pi ,\pi ]$ of the function $F+A$ defined by
equation (\ref{FA-2}) with $C_{2}=4\pi +2\func{Si}(2\pi )$ is plotted
against $u$ in Figure 5b. As mentioned at the end of Section 6.2, the
functions $F$ and $A$ are not separately uniquely determined, as different
choices of $F$ correspond to different choices of the gauge coordinate $u$.
However, the requirement that $A(0)=-A(\pi )=\lambda $ (with $\lambda \equiv
-12\pi +8\pi \ln 2\pi +2\func{Si}(2\pi )$) does place a restriction on the
choice of $F$. One natural gauge choice is to set $A$ proportional to the $z$%
-component $|u|-\frac{1}{2}\pi $ of the unperturbed position function. Then
\begin{equation}
A(u)=-\tfrac{2\lambda }{\pi }(|u|-\tfrac{1}{2}\pi )  \label{A-3}
\end{equation}
is continuous and half-range anti-periodic on $[-\pi ,\pi ]$ with $%
A(0)=\lambda $, while from (\ref{FA-2}) $F$ is the half-range periodic
extension to $[-\pi ,\pi ]$ of
\begin{eqnarray}
F(u) &=&\tfrac{1}{2}\kappa u(u-\pi )+\tfrac{2\lambda }{\pi }(u-\tfrac{1}{2}%
\pi )-8W^{-1}[1+\ln W-\ln (1-\cos W)](1-\cos W)  \nonumber \\
&&-4W(2-\ln W)-8(u\ln u-u)-4[K(u)\sin 2u-\Sigma (u)\cos 2u]  \nonumber \\
&&-4\int_{2\pi }^{W}\ln (1-\cos \omega )\,d\omega +4\pi +2\func{Si}(2\pi )%
\text{.}  \label{F}
\end{eqnarray}
The correction functions $A$ and $F$ in this gauge are plotted against $u$
in Figures 6a and 6b respectively.

To summarize, the position function (\ref{evap2}) with $F$, $G$ and $A$
suitable extensions of (\ref{F}), (\ref{G-2}) and (\ref{A-3}) is a solution
of the first-order string equation of motion (\ref{eqmo-first}) that reduces
to the ACO solution in the limit as $\mu \rightarrow 0$. Furthermore, if the
integration constant $C_{1}$ is chosen to take the value (\ref{C1}) then the
solution is the unique self-similar loop trajectory (up to gauge and
Poincar\'{e} transformations) with the property that the length of the loop
retains its ACO value $\frac{1}{\sqrt{2}}L_{0}$ along the curve $u+v=0$
(which to leading order in $\mu $ corresponds to the spacelike slice $t=0$).

\section{Matching the Solution to the Minkowski Vacuum}

By prescription, the evaporating ACO\ loop radiates away the last of its
energy at $t=t_{L}$, the moment when it shrinks to a point at the origin. It
is therefore to be expected that the metric induced by the loop reduces to
the Minkowski vacuum at all points in the chronological future of the
evaporation point $x^{a}=[t_{L},\mathbf{0}]^{a}$. If the evaporation point
is denoted by $p_{L}$, the boundary of the chronological future $%
I^{+}(p_{L}) $ is the future light cone $F_{L}$ of $p_{L}$, plus $p_{L}$
itself. The weak-field metric will match smoothly onto Minkowski spacetime
in $I^{+}(p_{L})$ if an appropriate set of junction conditions is satisfied
on $F_{L}$. I will show explicitly that these junction conditions are indeed
satisfied later in this section. (Note, however, that the weak-field metric
remains singular in every neighborhood of $p_{L}$, and cannot be matched
onto Minkowski spacetime at this one point.)

\subsection{The surface $t-t_{L}=|\mathbf{x}|$ is null to order $\mu $}

Before proceeding, it is important to first establish that the surface $%
F_{L} $ is, to order $\mu $, just the surface $t-t_{L}=|\mathbf{x}|$, which
is of course null with respect to the unperturbed metric tensor $\eta _{ab}$%
. To show that the surface $t-t_{L}=|\mathbf{x}|$ is also null with respect
to the first-order metric tensor $\eta _{ab}+h_{ab}$, it is sufficient to
show that $h_{ab}(t,\mathbf{x})\rightarrow 0$ as the field point $x^{a}=[t,%
\mathbf{x}]^{a}$ approaches the surface.

Suppose then that $\varepsilon $ is a number in $(0,1)$, and consider the
set of field points $[t,\mathbf{x}]$ lying to the past of the surface $%
t-t_{L}=|\mathbf{x}|$ and satisfying the conditions
\begin{equation}
|\mathbf{x}|-t+t_{L}=\varepsilon t_{L}\text{\qquad and\qquad }[%
(t-t_{L})^{2}+|\textbf{x}|^{2}]^{1/2}>\varepsilon ^{1/2}t_{L}\text{.}
\label{distance}
\end{equation}
The first condition implies that $\varepsilon $ is a dimensionless measure
of the distance from the point $[t,\mathbf{x}]$ to the surface $t-t_{L}=|%
\mathbf{x}|$ in the background Minkowski manifold. In fact, it is easily
seen that the \emph{Euclidean} distance from the point to the surface is
just $\tfrac{1}{\sqrt{2}}\varepsilon t_{L}$. The second condition ensures
that the field point avoids a certain neighborhood of the evaporation point $%
[t_{L},\mathbf{0}]$. (The scaling factor $\varepsilon ^{1/2}$ is chosen here
because the exclusion of a neighborhood about $[t_{L},\mathbf{0}]$ with the
smaller radius $\varepsilon t_{L}$ does not lead to a useful bound on the
components of $h_{ab}$, as will be seen shortly.) It is evident that as $%
\varepsilon \rightarrow 0$ the points satisfying (\ref{distance}) include
all points in some timelike past neighborhood of every point on the surface $%
t-t_{L}=|\mathbf{x}|$ except the evaporation point.

Eliminating $t$ from the conditions (\ref{distance}) yields one non-trivial
bound on $|\mathbf{x}|$ if $\varepsilon <1$, namely
\begin{equation}
|\mathbf{x}|>\tfrac{1}{2}[\varepsilon +(2\varepsilon -\varepsilon ^{2})^{1/2}%
]\,t_{L}\text{,}  \label{distance2}
\end{equation}
which in turn implies that $|\mathbf{x}|>\tfrac{1}{\sqrt{2}}\varepsilon
^{1/2}t_{L}$ (as can readily be verified by dividing the expression on the
right of (\ref{distance2}) by $\varepsilon ^{1/2}t_{L}$ and showing that the
minimum value of the resulting function is $1/\sqrt{2}$).

Now, the corresponding image point $[\bar{t},\mathbf{\bar{x}}]$ on the
future light cone $F_{0}$ of the origin has coordinates
\begin{equation}
\bar{t}=e^{\kappa \mu \psi }|\mathbf{x}|\text{\qquad and\qquad }\mathbf{\bar{%
x}}=e^{\kappa \mu \psi }R(-\psi )\mathbf{x}
\end{equation}
where
\begin{equation}
e^{\kappa \mu \psi }\equiv [(|\mathbf{x}|-t+t_{L})/t_{L}]^{-1}=\varepsilon
^{-1}
\end{equation}
and $R$ is the rotation matrix defined in (\ref{rot-mat}). So in particular
\begin{equation}
\bar{t}=\varepsilon ^{-1}|\mathbf{x}|>\tfrac{1}{\sqrt{2}}\varepsilon
^{-1/2}t_{L}\text{,}
\end{equation}
while of course $|\mathbf{\bar{x}}|=\bar{t}$.

Note here that $\bar{t}\rightarrow \infty $ as $\varepsilon \rightarrow 0$.
Also, according to (\ref{rotation}),
\begin{equation}
h_{ab}(t,\mathbf{x})=R_{a}^{\;c}(\psi )\,h_{cd}(\bar{t},\mathbf{\bar{x}}%
)\,R_{b}^{\;d}(\psi )
\end{equation}
where the components of the rotation operators $R_{b}^{a}$ are all bounded
functions of $\psi $. Hence, to prove that $h_{ab}(t,\mathbf{x})\rightarrow
0 $ as $\varepsilon \rightarrow 0$ in the conditions (\ref{distance}), it is
sufficient to show that $|h_{ab}(\bar{t},\mathbf{\bar{x}})|$ is bounded
above on $F_{0}$ by a function of $\bar{t}$ which goes to zero (uniformly
with respect to $\mathbf{\bar{x}}$) as $\bar{t}\rightarrow \infty $. It
should also be clear that a proof of the last statement would not be
sufficient if the factor $\varepsilon ^{1/2}$ in the second condition in (%
\ref{distance}) were replaced by $\varepsilon $ or a multiple of $%
\varepsilon $. For then (\ref{distance2}) would be replaced by an inequality
of the form $|\mathbf{x}|>c\varepsilon $ for some constant $c$, and the
image point $[\bar{t},\mathbf{\bar{x}}]$ would be placed under the much
weaker constraint $\bar{t}>c$, which does not guarantee that $h_{ab}(\bar{t},%
\mathbf{\bar{x}})\rightarrow 0$ as $\varepsilon \rightarrow 0$.

Now, given that the components of $\bar{\Psi}_{ab}$ in the equation (\ref
{h-ints}) for $h_{ab}(\bar{t},\mathbf{\bar{x}})$ all satisfy $|\bar{\Psi}%
_{ab}(s,v)|\leq 1$, it follows that
\begin{eqnarray}
|h_{ab}(\bar{t},\mathbf{\bar{x}})| &\leq &4\mu \int_{V_{0}}^{V_{1}+2\pi
}[T+(Z+\tfrac{1}{2}\pi )-v]^{-1}\,dv+4\mu \int_{V_{1}}^{V_{0}}[T-(Z+\tfrac{1%
}{2}\pi )-v]^{-1}\,dv  \nonumber \\
&\equiv &4\mu |I_{+}|+4\mu |I_{-}|\text{,}
\end{eqnarray}
where $[T,X,Y,Z]=4\pi L_{0}^{-1}[\bar{t},\mathbf{\bar{x}}]$ as before. So $%
|h_{ab}(\bar{t},\mathbf{\bar{x}})|$ will be bounded above by a function
which goes uniformly to zero as $\bar{t}\rightarrow \infty $ if the same is
true, separately, of $|I_{+}|$ and $|I_{-}|$.

Consider then
\begin{equation}
|I_{-}|=\int_{V_{1}}^{V_{0}}[E(1,v)]^{-1}\,dv\qquad \text{where\qquad }%
E(s,v)=T-s(Z+\tfrac{1}{2}\pi )-v\text{.}
\end{equation}
Since $E$ is positive semi-definite (see Appendix A.2), there are two
possible cases that need to be considered: either (i) $E(1,v)=0$ for some $%
v\in [V_{1},V_{0}]$, or (ii) $E(1,v)>0$ for all $v\in [V_{1},V_{0}]$.

Case (i) has been treated in some detail in Appendix A.2, where it is shown
that $E(1,v)=0$ for some $v$ only if the point $(X,Y)$ lies on the unit
circle (so that $T=(1+Z^{2})^{1/2}$) and $Z>-\tfrac{1}{2}\pi $. Furthermore,
if $Z>\tfrac{1}{2}\pi $ (and so $T>(1+\frac{1}{4}\pi ^{2})^{1/2}\approx
1.8621$, which will henceforth be assumed) then the values of $V_{0}$ and $%
V_{1}$ coincide, and according to (\ref{ints-minus})
\begin{equation}
|I_{-}|=\ln [1+\pi /(Z-\tfrac{1}{2}\pi )]\text{.}
\end{equation}
In view of the inequality $\ln (1+x)\leq x$ it follows that
\begin{equation}
|I_{-}|\leq \pi /(Z-\tfrac{1}{2}\pi )\equiv \pi /[(T^{2}-1)^{1/2}-\tfrac{1}{2%
}\pi ]
\end{equation}
where the function of $T$ appearing on the right is bounded above by $8\pi
/T $ if $T\geq \frac{1}{63}(4\pi +8\sqrt{16\pi ^{2}+63})\approx 2.0869$.
(The significance of the factor $8\pi $ will become apparent shortly.) So,
in this case,
\begin{equation}
|I_{-}|\leq 8\pi /T\text{\qquad if\qquad }T\gtrsim 2.0869\text{.}
\label{case1}
\end{equation}

In case (ii), $E(1,v)>0$ for all $v\in [V_{1},V_{0}]$. Since $E$ is a linear
decreasing function of $v$, there exists a number $V_{*}$ in $[V_{1},V_{0}]$
with the property that $E(1,v)>\frac{1}{2}T$ for all $v\in (V_{1},V_{*})$
and $0<E(1,v)<\frac{1}{2}T$ for all $v\in (V_{*},V_{0})$.\footnote{%
The value of $V_{*}$ will of course depend on $[T,X,Y,Z]$. Also, $V_{*}$ may
coincide with either $V_{0}$ or $V_{1}$, in which case one of the two
domains specified for $v$ will be empty.} An immediate consequence is that
\begin{equation}
|I_{-}|=\int_{V_{1}}^{V_{*}}[E(1,v)]^{-1}\,dv+%
\int_{V_{*}}^{V_{0}}[E(1,v)]^{-1}\,dv
\end{equation}
where
\begin{equation}
\int_{V_{1}}^{V_{*}}[E(1,v)]^{-1}\,dv<2(V_{*}-V_{1})/T\leq 4\pi /T\text{,}
\label{A1-bound}
\end{equation}
as $V_{*}-V_{1}\leq V_{0}-V_{1}\leq 2\pi $.

The integral over the remaining segment $(V_{*},V_{0})$ can be bounded above
by using equation (\ref{dv/du}) to replace $v$ with $u$ as the integration
variable, so that
\begin{equation}
\int_{V_{*}}^{V_{0}}[E(1,v)]^{-1}\,dv=\int_{0}^{u(V_{*})}[H(v)]^{-1}\,du
\label{remainder}
\end{equation}
where $H(v)=T-(u+v)+X\sin v-Y\cos v$ is a second positive semi-definite
function, examined in some detail in Appendix A.3. In particular, it is
evident from the first line of (\ref{A2-bound}) that
\begin{equation}
H(v)\geq |Z-(|u|-\tfrac{1}{2}\pi )|\geq Z-\tfrac{1}{2}\pi
\end{equation}
if $u$ is in $[0,\pi ]$ and $Z>\tfrac{1}{2}\pi $. Also, the condition $%
E(1,v)<\frac{1}{2}T$ corresponds to the inequality
\begin{equation}
Z>\tfrac{1}{2}T-\tfrac{1}{2}\pi -v\geq \tfrac{1}{2}T-\tfrac{3}{2}\pi -(1+%
\tfrac{1}{4}\pi ^{2})^{1/2}
\end{equation}
where $v$ here has been replaced by its maximum possible value, $\pi +(1+%
\tfrac{1}{4}\pi ^{2})^{1/2}$ [see Appendix A.1]. In particular, the
condition $Z>\tfrac{1}{2}\pi $ will be satisfied if $T\gtrsim $ $16.2906$.
Hence, if the latter condition is assumed,
\begin{equation}
\int_{0}^{u(V_{*})}[H(v)]^{-1}\,du<u(V_{*})\,[\tfrac{1}{2}T-2\pi -(1+\tfrac{1%
}{4}\pi ^{2})^{1/2}]^{-1}  \label{A2-int}
\end{equation}
where the maximum possible value of $u(V_{*})$ is $\pi $. The function of $T$
appearing on the right of (\ref{A2-int}) is therefore bounded above by $4\pi
/T$ if $T\geq 8\pi +2\sqrt{4+\pi ^{2}}\approx 32.5811$.

In summary, the inequalities (\ref{A1-bound}) and (\ref{A2-bound}) together
imply that
\begin{equation}
|I_{-}|\leq 8\pi /T\text{\qquad if\qquad }T\gtrsim 32.5811
\end{equation}
in case (ii), and so in view of (\ref{case1}) $|I_{-}|$ is bounded above by
a function that goes to zero uniformly as $T\rightarrow \infty $, as
required. Note that the factor of $8\pi $ in the bounding function arises as
a consequence of choosing $V_{*}$ to separate the domains $E\lessgtr \frac{1%
}{2}T$. The factor $\frac{1}{2}$ is to some extent arbitrary here, as any
factor less than $1$ will induce similar bounds on $|I_{-}|$. However, a
factor $1$ or greater is not viable, as it generates no useful bound on $Z$,
and hence no useful bound on the integral (\ref{remainder}).

The proof that the second contribution $|I_{+}|$ to $|h_{ab}(\bar{t},\mathbf{%
\bar{x}})|$ is also uniformly bounded above by $8\pi /T$ for sufficiently
large values of $T$ is formally identical to the proof just given, provided
that $|I_{-}|$ is everywhere replaced with $|I_{+}|$, $E(1,v)$ with $E(-1,v)$%
, $Z$ with $-Z$, $V_{0}$ with $V_{1}+2\pi $ and $V_{1}$ with $V_{0}$. Also,
the limit $u=0$ in the integral on the right of (\ref{remainder}) should
there and below be replaced with $u=-\pi $, and the bound $u(V_{*})\leq \pi $
with $u(V_{*})\leq 0$.

In conclusion, therefore, $|h_{ab}(\bar{t},\mathbf{\bar{x}})|\leq 64\pi \mu
/T\equiv 16L_{0}/\bar{t}$ for sufficiently large values of $\bar{t}$, and so
$h_{ab}(t,\mathbf{x})\rightarrow 0$ as $\varepsilon \rightarrow 0$ in the
conditions (\ref{distance}). It follows that the future light cone $F_{L}$
of the final evaporation point is, to order $\mu $, just the surface $%
t-t_{L}=|\mathbf{x}|$.

\subsection{Checking the junction conditions}

The junction conditions that apply when matching two spacetime regions
across a null hypersurface $\mathcal{N}$ were first usefully formulated by
Barrab\`{e}s and Israel \cite{Bar-Isr} and have been handily summarized by
Poisson in \cite{Poiss}. The first step in the Barrab\`{e}s-Israel method is
to parametrize the null hypersurface in the form $x^{a}=y^{a}(r,\theta
^{1},\theta ^{2})$, where $r$ is an arbitrary coordinate along the
generators of the hypersurface, and each generator is labeled by the two
remaining coordinates $\theta ^{A}$ (for $A=1,2$). This parametrization can
in turn be used to construct an orthonull basis $\{k^{a},\,e_{1}^{a},%
\,e_{2}^{a},\,N^{a}\}$ at each point on $\mathcal{N}$, with
\begin{equation}
k^{a}=\partial y^{a}/\partial r\text{,\qquad }e_{A}^{a}=\partial
y^{a}/\partial \theta ^{A}
\end{equation}
and the remaining (null) basis field $N^{a}$ uniquely determined by the
constraints
\begin{equation}
g_{ab}N^{a}k^{b}=1\text{\qquad and\qquad }g_{ab}N^{a}e_{A}^{b}=0\text{.}
\end{equation}

The hypersurface then has an associated mass density $\rho $, current
density $j^{A}$ and isotropic pressure $p$ given by
\begin{equation}
\rho =-\tfrac{1}{8\pi }\sigma ^{AB}[C_{AB}]\text{,\qquad }j^{A}=\tfrac{1}{%
8\pi }\sigma ^{AB}[C_{3B}]\text{\qquad and\qquad }p=-\tfrac{1}{8\pi }[C_{33}]%
\text{,}
\end{equation}
where $\sigma _{AB}=g_{ab}e_{A}^{a}e_{B}^{b}$ and $C_{\alpha \beta
}=e_{\alpha }^{a}e_{\beta }^{b}\nabla _{b}N_{a}$, with $\alpha ,\beta $
running from $1$ to $3$ and $e_{3}^{a}\equiv k^{a}$. Also, $[C_{\alpha \beta
}]$ denotes $C_{\alpha \beta }(\mathcal{N}^{+})-C_{\alpha \beta }(\mathcal{N}%
^{-})$ where $\mathcal{N}^{-}$ and $\mathcal{N}^{+}$ are the (timelike)
past- and future-facing sides of $\mathcal{N}$ respectively. (Note that $%
C_{\alpha \beta }$ typically has different values on the two sides of $%
\mathcal{N}$ because the limiting values of the covariant derivatives $%
\nabla _{b}N_{a}$ are different, even though the metric tensor $g_{ab}$ is
assumed here to be continuous across $\mathcal{N}$.) The two spacetime
regions can be matched across $\mathcal{N}$ without invoking an extraneous
source of stress-energy on $\mathcal{N}$ if $\rho $, $j^{A}$ and $p$ are all
zero.

In the case of the evaporating ACO loop, $\mathcal{N}$ is the null
hypersurface $F_{L}$ defined by $t-t_{L}=|\mathbf{x}|$, and $g_{ab}=\eta
_{ab}$ there. An obvious choice of parametrization is
\begin{equation}
y^{a}(r,\theta ^{1},\theta ^{2})=[t_{L},\mathbf{0}]^{a}+[r,r\sin \theta
^{1}\cos \theta ^{2},r\sin \theta ^{1}\sin \theta ^{2},r\cos \theta ^{1}]^{a}%
\text{.}  \label{parameter}
\end{equation}
The corresponding orthonull basis fields are:
\begin{equation}
k^{a}=[1,\sin \theta ^{1}\cos \theta ^{2},\sin \theta ^{1}\sin \theta
^{2},\cos \theta ^{1}]^{a}\text{,}
\end{equation}
\begin{equation}
e_{1}^{a}=[0,r\cos \theta ^{1}\cos \theta ^{2},r\cos \theta ^{1}\sin \theta
^{2},-r\sin \theta ^{1}]^{a}\text{,}
\end{equation}
\begin{equation}
e_{2}^{a}=[0,-r\sin \theta ^{1}\sin \theta ^{2},r\sin \theta ^{1}\cos \theta
^{2},0]^{a}
\end{equation}
and
\begin{equation}
N^{a}=\tfrac{1}{2}[1,-\sin \theta ^{1}\cos \theta ^{2},-\sin \theta ^{1}\sin
\theta ^{2},-\cos \theta ^{1}]^{a}\text{.}
\end{equation}
In particular, the line element $\sigma _{AB}\,d\theta ^{A}d\theta ^{B}$
intrinsic to the hypersurface is just that of a spatial 2-sphere:
\begin{equation}
\sigma _{AB}=-r^{2}\text{diag\thinspace }(1,\sin ^{2}\theta ^{1})\text{%
\qquad and\qquad }\sigma ^{AB}=-r^{-2}\text{diag\thinspace }(1,\csc
^{2}\theta ^{1})\text{.}
\end{equation}

To linear order in the derivatives of $h_{ab}$, the transverse curvature of $%
F_{L}$ is
\begin{equation}
C_{\alpha \beta }=e_{\alpha }^{a}e_{\beta }^{b}\partial _{b}N_{a}-\tfrac{1}{2%
}e_{\alpha }^{a}e_{\beta }^{b}N^{c}(\partial _{b}h_{ac}+\partial
_{a}h_{bc}-\partial _{c}h_{ab})\text{,}
\end{equation}
where the first term $e_{\alpha }^{a}e_{\beta }^{b}\partial _{b}N_{a}$ is
identical on the two sides of $F_{L}$, and the remaining terms are of course
zero on the Minkowski side $F_{L}^{+}$. Because $h_{ab}=0$ on $F_{L}^{-}$
the derivatives $e_{\alpha }^{a}\partial _{a}h_{bc}$ tangent to $F_{L}^{-}$
are all zero, but the limiting values of the transverse derivatives $%
N^{c}\partial _{c}h_{ab}$ are not necessarily zero there. So
\begin{equation}
\lbrack C_{\alpha \beta }]=-\tfrac{1}{2}e_{\alpha }^{a}e_{\beta
}^{b}N^{c}\partial _{c}h_{ab}|_{F_{L}^{-}}\text{.}
\end{equation}

Now, the derivatives of $h_{ab}$ can be evaluated formally by using (\ref
{rotation}) to write:
\begin{equation}
\partial _{c}h_{ab}(t,\mathbf{x})=[(\tfrac{d}{d\psi }R_{a}^{\;p})\,h_{pq}(%
\bar{t},\mathbf{\bar{x}})\,R_{b}^{\;q}+R_{a}^{\;p}h_{pq}(\bar{t},\mathbf{%
\bar{x}})\,(\tfrac{d}{d\psi }R_{b}^{\;q})]\,\partial _{c}\psi +(\partial
\bar{x}^{r}/\partial x^{c})R_{a}^{\;p}\bar{\partial}_{r}h_{pq}(\bar{t},%
\mathbf{\bar{x}})\,R_{b}^{\;q}\text{,}  \label{del-bar}
\end{equation}
where $\partial _{c}\psi $ and $\partial \bar{x}^{r}/\partial x^{c}$ can in
turn be calculated from (\ref{psi}) and (\ref{image}). However, as is shown
in Appendix A.3, the derivatives $\bar{\partial}_{r}h_{pq}(\bar{t},\mathbf{%
\bar{x}})\,$are divergent at image points that lie in the beaming directions
of the two kinks, and the loci of such points extend to future null infinity
on $F_{0}$. Although points of this type have $\bar{z}=\pm \frac{1}{8}L_{0}$%
, and so (since $z=e^{-\kappa \mu \psi }\bar{z}$) the only points on $F_{L}$
which are limit points of the singularities of $\partial _{c}h_{ab}$ lie on
the plane $z=0$, this feature of the metric makes (\ref{del-bar}) less than
ideal for calculating $[C_{\alpha \beta }]$.

Instead, it is more convenient when calculating $[C_{\alpha \beta }]$ to use
the formal definition of the derivative
\begin{equation}
\lim_{||\delta y||\rightarrow 0}||\delta y||^{-1}[f(y^{a}+\delta
y^{a})-f(y^{a})-\delta y^{c}\partial _{c}f]=0\text{,}  \label{def-deriv}
\end{equation}
where $f$ denotes any of the relevant combinations of the components of $%
h_{ab}$, and $||\cdot ||$ is the Euclidean norm. This approach has the
advantage of demonstrating that the relevant combinations of the metric
derivatives do in fact exist at all points on the null boundary (except at $%
p_{L}$ itself).

Consider then a fixed point $y^{a}\equiv [t_{L}+r,r\mathbf{\hat{r}}]^{a}$ on
$F_{L}\backslash \{p_{L}\}$ (with $\mathbf{\hat{r}}\equiv (\hat{x},\hat{y},%
\hat{z})$ and $|\mathbf{\hat{r}}|=1$), and let $x^{a}=y^{a}+\delta y^{a}$ be
a field point at a Euclidean distance $\varpi \,t_{L}$ from (and in the
timelike past of) $y^{a}$, so that $\delta y^{a}\equiv \varpi \,t_{L}[\delta
\hat{t},\delta \mathbf{\hat{r}}]^{a}$ where $\delta \mathbf{\hat{r}}\equiv
(\delta \hat{x},\delta \hat{y},\delta \hat{z})$ and $||[\delta \hat{t}%
,\delta \mathbf{\hat{r}}]||=1$. If $\varepsilon $ denotes $e^{-\kappa \mu
\psi }$ (as in the previous section) then
\begin{equation}
\varepsilon =|t_{L}^{-1}r\mathbf{\hat{r}}+\varpi \,\delta \mathbf{\hat{r}}%
|-t_{L}^{-1}r-\varpi \,\delta \hat{t}=(\mathbf{\hat{r}}\cdot \,\delta
\mathbf{\hat{r}}-\,\delta \hat{t})\varpi +O(\varpi ^{2})\text{.}
\label{epsilon}
\end{equation}
It is easily seen that $\varepsilon \leq \sqrt{2}\varpi $, and that $%
\varepsilon >0$ when $x^{a}$ lies in the past of $y^{a}$.

The image point $\bar{x}^{a}$ corresponding to $x^{a}$ has components
\begin{eqnarray}
\bar{t} &=&\varepsilon ^{-1}|r\mathbf{\hat{r}}+\varpi \,t_{L}\delta \mathbf{%
\hat{r}}|\text{,\qquad }\bar{x}=\varepsilon ^{-1}[(r\hat{x}+\varpi
\,t_{L}\delta \hat{x})\cos \psi +(r\hat{y}+\varpi \,t_{L}\delta \hat{y})\sin
\psi ]\text{,}  \nonumber \\
\bar{y} &=&\varepsilon ^{-1}[(r\hat{y}+\varpi \,t_{L}\delta \hat{y})\cos
\psi -(r\hat{x}+\varpi \,t_{L}\delta \hat{x})\sin \psi ]\text{\qquad
and\qquad }\bar{z}=\varepsilon ^{-1}(r\hat{z}+\varpi \,t_{L}\delta \hat{z})%
\text{.}  \nonumber \\
&&
\end{eqnarray}
If the non-dimensionalized components $[T,X,Y,Z]=4\pi L_{0}^{-1}[\bar{t},%
\bar{x},\bar{y},\bar{z}]$ are substituted into the equation (\ref{roots})
for $V_{k}$ then, after squaring both sides and rearranging, the equation
reads (for $k=0$ or $1$)
\begin{equation}
|r\mathbf{\hat{r}}+\varpi \,t_{L}\delta \mathbf{\hat{r}}|(V_{k}+k\pi )-r%
\mathbf{\hat{r}}\cdot \mathbf{X}_{k}=\varpi \,t_{L}\{\delta \mathbf{\hat{r}}%
\cdot \mathbf{X}_{k}+\tfrac{1}{2}\kappa \mu (\varepsilon /\varpi
)\,[(V_{k}+k\pi )^{2}-(1+\pi ^{2}/4)]\}\text{,}  \label{V-first}
\end{equation}
where the factor of $\kappa \mu $ arises because $\tfrac{1}{4\pi }%
L_{0}=\kappa \mu t_{L}$, and
\begin{equation}
\mathbf{X}_{k}\equiv (\cos (V_{k}+\psi ),\sin (V_{k}+\psi ),(k-\tfrac{1}{2}%
)\pi )\text{.}
\end{equation}

Equation (\ref{V-first}) suggests that, for small values of $\varpi $, the
root $V_{k}$ can be approximated by the solution of the equation $V_{k}+k\pi
=\mathbf{\hat{r}}\cdot \mathbf{X}_{k}$, which I will denote by $%
V_{k}^{\infty }$. In explicit form, the equations for $V_{0}^{\infty },_{1}$
read
\begin{equation}
V_{0}^{\infty }=\hat{x}\cos (V_{0}^{\infty }+\psi )+\hat{y}\sin
(V_{0}^{\infty }+\psi )-\tfrac{\pi }{2}\hat{z}
\end{equation}
and
\begin{equation}
V_{1}^{\infty }+\pi =\hat{x}\cos (V_{1}^{\infty }+\psi )+\hat{y}\sin
(V_{1}^{\infty }+\psi )+\tfrac{\pi }{2}\hat{z}\text{.}
\end{equation}
Note however that $V_{0}^{\infty },_{1}$ are typically still functions of $%
\varpi $, as $\psi =-(\kappa \mu )^{-1}\ln \varepsilon $ increases without
bound, and in particular $\lim_{\varpi \rightarrow 0}V_{k}^{\infty }$ does
not exist unless $\hat{x}=\hat{y}=0$.

Although $\lim_{\varpi \rightarrow 0}V_{k}^{\infty }$ does not generally
exist, it can be shown that $\lim_{\varpi \rightarrow 0}|V_{k}-V_{k}^{\infty
}|=0$, a fact that will prove important in establishing that certain
combinations of the metric components are differentiable on $F_{L}$. To show
that $\lim_{\varpi \rightarrow 0}|V_{k}-V_{k}^{\infty }|=0$, first rearrange
(\ref{V-first}) to give
\begin{eqnarray}
r(V_{k}+k\pi -\mathbf{\hat{r}}\cdot \mathbf{X}_{k}) &=&\varpi
\,t_{L}\{\delta \mathbf{\hat{r}}\cdot \mathbf{X}_{k}+\tfrac{1}{2}\kappa \mu
(\varepsilon /\varpi )\,[(V_{k}+k\pi )^{2}-(1+\pi ^{2}/4)]\}  \nonumber \\
&&+(r-|r\mathbf{\hat{r}}+\varpi \,t_{L}\delta \mathbf{\hat{r}}|)(V_{k}+k\pi )
\nonumber \\
&\equiv &\Theta (\varpi )\text{.}  \label{V-second}
\end{eqnarray}
Since $|V_{k}+k\pi |$ is bounded above (see Appendix A.1), it is clear that
the function $\Theta (\varpi )$ on the right-hand side of this equation
tends to zero as $\varpi \rightarrow 0$. The root $V_{k}^{\infty }$
satisfies the same equation (\ref{V-second}) with $\Theta (\varpi )=0$.

Subtracting the equation for $V_{k}^{\infty }$ from (\ref{V-second}) gives:
\begin{equation}
V_{k}-V_{k}^{\infty }-\hat{x}[\cos (V_{k}+\psi )-\cos (V_{k}^{\infty }+\psi )%
]-\hat{y}[\sin (V_{k}+\psi )-\sin (V_{k}^{\infty }+\psi )]=r^{-1}\Theta
(\varpi )\text{.}
\end{equation}
In view of the identities $\cos 2A-\cos 2B=2\sin (B-A)\sin (A+B)$ and $\sin
2A-\sin 2B=2\sin (A-B)\cos (A+B)$ this equation becomes
\begin{equation}
V_{k}-V_{k}^{\infty }+2[\hat{x}\sin (\tfrac{1}{2}V_{k}+\tfrac{1}{2}%
V_{k}^{\infty }+\psi )-\hat{y}\cos (\tfrac{1}{2}V_{k}+\tfrac{1}{2}%
V_{k}^{\infty }+\psi )]\sin [\tfrac{1}{2}(V_{k}-V_{k}^{\infty
})]=r^{-1}\Theta (\varpi )\text{,}
\end{equation}
and so
\begin{equation}
|V_{k}-V_{k}^{\infty }|-2(\hat{x}^{2}+\hat{y}^{2})^{1/2}|\sin [\tfrac{1}{2}%
(V_{k}-V_{k}^{\infty })]|\leq r^{-1}|\Theta (\varpi )|\text{.}
\end{equation}
Since $(\hat{x}^{2}+\hat{y}^{2})^{1/2}\leq 1$ and $|w|-2|\sin (\frac{1}{2}%
w)|\geq \tfrac{1}{48}|w|^{3}$ for all real $w$, it follows that
\begin{equation}
\tfrac{1}{48}|V_{k}-V_{k}^{\infty }|^{3}\leq r^{-1}|\Theta (\varpi )|
\end{equation}
and therefore $\lim_{\varpi \rightarrow 0}|V_{k}-V_{k}^{\infty }|=0$ as
claimed. Furthermore, because $\Theta (\varpi )$ goes to zero linearly in $%
\varpi $, $|V_{k}-V_{k}^{\infty }|$ is at worst of order $\varpi ^{1/3}$ for
small values of $\varpi $.

The asymptotic forms of the components of $h_{ab}$ can now be developed by
first substituting the expressions
\begin{equation}
\chi _{\pm }=4\pi L_{0}^{-1}\varepsilon ^{-1}(t\pm z)\pm \tfrac{\pi }{2}
\label{chi}
\end{equation}
into (\ref{I-ints})-(\ref{S,C-ints2}) to give
\begin{eqnarray}
I_{+} &=&\tfrac{1}{4\pi }L_{0}\varepsilon \tfrac{V_{0}-V_{1}-2\pi }{t+z}%
+O(\varepsilon ^{2})\text{,}\qquad I_{-}=\tfrac{1}{4\pi }L_{0}\varepsilon
\tfrac{V_{1}-V_{0}}{t-z}+O(\varepsilon ^{2})\text{,}  \nonumber \\
S_{\pm } &=&\pm \tfrac{1}{4\pi }L_{0}\varepsilon \tfrac{\cos V_{1}-\cos V_{0}%
}{t\pm z}+O(\varepsilon ^{2})\qquad \text{and\qquad }C_{\pm }=\pm \tfrac{1}{%
4\pi }L_{0}\varepsilon \tfrac{\sin V_{0}-\sin V_{1}}{t\pm z}+O(\varepsilon
^{2})  \nonumber \\
&&
\end{eqnarray}
(assuming for the moment that $t\pm z\neq 0$).

With $t=|r\mathbf{\hat{r}}+\varpi \,t_{L}\delta \mathbf{\hat{r}}|$ and $z=r%
\hat{z}+\varpi \,t_{L}\delta \hat{z}$, equation (\ref{metperts}) generates
the following expansion for the components of $h_{ab}$:
\begin{eqnarray}
&&h_{ab}(t,\mathbf{x})=R_{a}^{\;c}(\psi )\,h_{cd}(|\mathbf{\bar{x}}|,\mathbf{%
\bar{x}})\,R_{b}^{\;d}(\psi )  \nonumber \\
&=&\frac{2L_{0}\mu \varepsilon }{\pi r(1-\hat{z}^{2})}\left[
\begin{array}{cccc}
\Delta V\,\hat{z}-(1-\hat{z})\pi & (c_{0}-c_{1})\hat{z} & (s_{0}-s_{1})\hat{z%
} & -\Delta V-(1-\hat{z})\pi \\
(c_{0}-c_{1})\hat{z} & \Delta V\,\hat{z}-(1-\hat{z})\pi & 0 & c_{1}-c_{0} \\
(s_{0}-s_{1})\hat{z} & 0 & \Delta V\,\hat{z}-(1-\hat{z})\pi & s_{1}-s_{0} \\
-\Delta V-(1-\hat{z})\pi & c_{1}-c_{0} & s_{1}-s_{0} & \Delta V\,\hat{z}-(1-%
\hat{z})\pi
\end{array}
\right] _{ab}  \nonumber \\
&&+O(\varpi ^{2})\text{,}
\end{eqnarray}
where
\begin{equation}
c_{k}\equiv \cos (V_{k}+\psi )\text{,\qquad }s_{k}\equiv \sin (V_{k}+\psi
)\qquad \text{and\qquad }\Delta V\equiv V_{1}-V_{0}\text{.}
\end{equation}

In terms of the components of $\mathbf{\hat{r}}$ the relevant orthonull
basis vectors are
\begin{equation}
k^{a}=[1,\hat{x},\hat{y},\hat{z}]^{a}\text{,\qquad }e_{1}^{a}=r(1-\hat{z}%
^{2})^{-1/2}[0,\hat{x}\hat{z},\hat{y}\hat{z},-(\hat{x}^{2}+\hat{y}^{2})]%
\text{\qquad and\qquad }e_{2}^{a}=r[0,-\hat{y},\hat{x},0]^{a}\text{,}
\end{equation}
and so
\begin{equation}
k^{a}k^{b}h_{ab}=-4(L_{0}/r)\mu \varepsilon +O(\varpi ^{2})\text{,\qquad }%
k^{a}e_{2}^{b}h_{ab}=O(\varpi ^{2})
\end{equation}
\begin{equation}
k^{a}e_{1}^{b}h_{ab}=\tfrac{2}{\pi (1-\hat{z}^{2})^{3/2}}L_{0}\mu
\varepsilon (V_{1}+\pi -\mathbf{\hat{r}}\cdot \mathbf{X}_{1}-V_{0}+\mathbf{%
\hat{r}}\cdot \mathbf{X}_{0})+O(\varpi ^{2})\text{,}
\end{equation}
and
\begin{eqnarray}
\sigma ^{AB}e_{A}^{a}e_{B}^{b}h_{ab} &=&-r^{-2}[e_{1}^{a}e_{1}^{b}h_{ab}+(1-%
\hat{z}^{2})^{-1}e_{2}^{a}e_{2}^{b}h_{ab}]  \nonumber \\
&=&-\tfrac{4}{\pi (1-\hat{z}^{2})}(L_{0}/r)\mu \varepsilon [(V_{1}+\pi -%
\mathbf{\hat{r}}\cdot \mathbf{X}_{1}-V_{0}+\mathbf{\hat{r}}\cdot \mathbf{X}%
_{0})\hat{z}-\pi (1-\hat{z}^{2})]  \nonumber \\
&&+O(\varpi ^{2})\text{.}
\end{eqnarray}

Now, given that $\lim_{\varpi \rightarrow 0}|V_{k}-V_{k}^{\infty }|=0$ and $%
c_{k}$ and $s_{k}$ are continuous functions of $V_{k}$,
\begin{equation}
\lim_{\varpi \rightarrow 0}(V_{1}+\pi -\mathbf{\hat{r}}\cdot \mathbf{X}%
_{1})=\lim_{\varpi \rightarrow 0}[V_{1}^{\infty }+\pi -\hat{x}\cos
(V_{1}^{\infty }+\psi )-\hat{y}\sin (V_{1}^{\infty }+\psi )-\tfrac{\pi }{2}%
\hat{z}]=0\text{,}
\end{equation}
and similarly $\lim_{\varpi \rightarrow 0}(V_{0}-\mathbf{\hat{r}}\cdot
\mathbf{X}_{0})=0$. Hence, in view of the expansion (\ref{epsilon}),
\begin{eqnarray}
\lim_{\varpi \rightarrow 0}\varpi ^{-1}k^{a}k^{b}h_{ab} &=&-4(L_{0}/r)\mu (%
\mathbf{\hat{r}}\cdot \,\delta \mathbf{\hat{r}}-\,\delta \hat{t})\text{%
,\qquad }\lim_{\varpi \rightarrow 0}\varpi ^{-1}k^{a}e_{1}^{b}h_{ab}=0
\nonumber \\
\lim_{\varpi \rightarrow 0}\varpi ^{-1}k^{a}e_{2}^{b}h_{ab} &=&0\text{\qquad
and\qquad }\lim_{\varpi \rightarrow 0}\varpi ^{-1}\sigma
^{AB}e_{A}^{a}e_{B}^{b}h_{ab}=4(L_{0}/r)\mu (\mathbf{\hat{r}}\cdot \,\delta
\mathbf{\hat{r}}-\,\delta \hat{t})\text{.}  \nonumber \\
&&
\end{eqnarray}

Since $\delta y^{a}\equiv \varpi \,t_{L}[\delta \hat{t},\delta \mathbf{\hat{r%
}}]^{a}$ it follows from (\ref{def-deriv}) that $k^{a}k^{b}h_{ab}$, $%
k^{a}e_{A}^{b}h_{ab}$ and $\sigma ^{AB}e_{A}^{a}e_{B}^{b}h_{ab}$ are all
differentiable at each point on the null boundary $F_{L}^{-}$ with $\hat{z}%
^{2}\neq 1$, and that
\begin{equation}
k^{a}k^{b}\partial _{c}h_{ab}=-\sigma ^{AB}e_{A}^{a}e_{B}^{b}\partial
_{c}h_{ab}=4(L_{0}/r)t_{L}^{-1}\mu [1,-\mathbf{\hat{r}}]_{c}\text{\qquad
and\qquad }k^{a}e_{A}^{b}\partial _{c}h_{ab}=0
\end{equation}
there. Given that $N^{c}[1,-\mathbf{\hat{r}}]_{c}=1$, it seems therefore
that
\begin{equation}
\lbrack C_{33}]=-\sigma ^{AB}[C_{AB}]=-2(L_{0}/r)t_{L}^{-1}\mu \text{\qquad
and\qquad }\sigma ^{AB}[C_{3B}]=0\text{,}
\end{equation}
and hence that matching the evaporating ACO spacetime to the Minkowski
spacetime across $F_{L}$ induces a boundary layer with a non-zero density $%
\rho $ and isotropic pressure $p$.

However, it should be recalled that $L_{0}t_{L}^{-1}=4\pi \kappa \mu $, and
so the apparent density and pressure of this boundary layer,
\begin{equation}
\rho =-p=-\kappa \mu ^{2}/r\text{,}
\end{equation}
are quadratic in $\mu $.\footnote{%
Incidentally, it is to be expected that neither $\rho $ nor $p$ depend
separately on $L_{0}$ or $t_{L}$, as the values of the latter will vary as
the zero point of $t$ changes, but the stress-energy content of $F_{L}$
should remain invariant under time translations.} Thus $\rho $ and $p$ are
zero at the level of the first-order approximation used throughout this
paper, and the evaporating ACO spacetime can be matched to the Minkowski
spacetime without inducing a boundary layer \emph{at this level of
approximation}.

All that remains now is to show that the same conclusion holds at all points
on $F_{L}^{-}$ with $\hat{z}^{2}=1$ (that is, on the $z$-axis). In this
case, $\hat{x}=\hat{y}=0$ and $V_{k}$ tends to the limiting value
\begin{equation}
V_{k}^{\infty }=-k\pi +(k-\tfrac{1}{2})\pi \hat{z}\text{.}
\end{equation}
Also, it is convenient to use a slightly different asymptotic form for the
spatial components of the field point $x^{a}$, namely
\begin{equation}
x=\tfrac{1}{4\pi }L_{0}x^{*}\varepsilon \text{,}\qquad y=\tfrac{1}{4\pi }%
L_{0}y^{*}\varepsilon \qquad \text{and\qquad }z=r\hat{z}+\tfrac{1}{4\pi }%
L_{0}z^{*}\varepsilon \text{,}
\end{equation}
where $x^{*}$, $y^{*}$ and $z^{*}$ are all of order $\varepsilon ^{0}$.

Then
\begin{equation}
X=x^{*}\cos \psi +y^{*}\sin \psi \text{,\qquad }Y=y^{*}\cos \psi -x^{*}\sin
\psi \qquad \text{and\qquad }Z=4\pi L_{0}^{-1}\varepsilon ^{-1}r\hat{z}+z^{*}%
\text{,}
\end{equation}
and the analogue of (\ref{chi}) becomes
\begin{equation}
\chi _{\pm }=4\pi rL_{0}^{-1}(1\pm \hat{z})\varepsilon ^{-1}+z^{*}\hat{z}%
(1\pm \hat{z})\pm \tfrac{1}{2}\pi \allowbreak +\tfrac{1}{8\pi }%
L_{0}r^{-1}(\allowbreak X^{2}+Y^{2})\varepsilon -\tfrac{1}{2}(\tfrac{1}{4\pi
}L_{0})^{2}r^{-2}z^{*}\hat{z}(\allowbreak X^{2}+Y^{2})\varepsilon ^{2}
\end{equation}
to order $\varepsilon ^{2}$, while the analogue of equation (\ref{V-second})
for $V_{k}$ can be used to calculate the asymptotic expansion
\begin{eqnarray}
\chi _{\pm }-V_{k} &=&4\pi rL_{0}^{-1}(1\pm \hat{z})\varepsilon ^{-1}+z^{*}%
\hat{z}(1\pm \hat{z})\pm \tfrac{1}{2}\pi +k\pi -(k-\tfrac{1}{2})\pi \hat{z}
\nonumber \\
&&+\tfrac{1}{8\pi }L_{0}r^{-1}(\allowbreak X^{2}+Y^{2}+2Y\,\hat{z}%
+1)\varepsilon +\tfrac{1}{2}(\tfrac{1}{4\pi }L_{0})^{2}r^{-2}X\,(2Y\,+\hat{z}%
)\varepsilon ^{2}  \nonumber \\
&&+\tfrac{1}{2}(\tfrac{1}{4\pi }L_{0})^{2}r^{-2}\hat{z}[(k-\tfrac{1}{2})\pi
-z^{*}](\allowbreak X^{2}+Y^{2}+2Y\,\hat{z}+1)\varepsilon ^{2}
\end{eqnarray}
(plus terms of order $\varepsilon ^{3}$).

The expansions need to be developed to such a high order because if $\hat{z}%
=1$ (\textit{resp.} $\hat{z}=-1$) then the terms appearing at order $%
\varepsilon ^{-1}$ and $\varepsilon ^{0}$ make no contribution to $I_{-}$, $%
S_{-}$ or $C_{-}$ (\textit{resp.} $I_{+}$, $S_{+}$ or $C_{+}$), and the
terms of order $\varepsilon $ plus the terms proportional to $(k-\tfrac{1}{2}%
)$ at order $\varepsilon ^{2}$ dominate the asymptotic behavior.. Otherwise
(if $\hat{z}=-1$, \textit{resp.} $\hat{z}=1$), it is the terms at order $%
\varepsilon ^{-1}$ and $\varepsilon ^{0}$ that dominate. A straightforward
but lengthy calculation shows that
\begin{eqnarray}
I_{+} &=&I_{-}=-\tfrac{1}{4}(L_{0}/r)\varepsilon \text{,\qquad }S_{+}=-%
\tfrac{1}{8}(1-\hat{z})(L_{0}/r)\varepsilon \text{,}  \nonumber \\
S_{-} &=&\tfrac{1}{8}(1+\hat{z})(L_{0}/r)\varepsilon \text{\qquad and\qquad }%
C_{+}=C_{-}=0
\end{eqnarray}
to order $\varepsilon $, and hence from (\ref{metperts}) that
\begin{equation}
h_{ab}(t,\mathbf{x})=(L_{0}/r)\,\mu \varepsilon \,\left[ \allowbreak
\begin{array}{cccc}
-2 & \hat{z}\cos \psi  & \hat{z}\sin \psi  & 0 \\
\hat{z}\cos \psi  & -2 & 0 & -\cos \psi  \\
\hat{z}\sin \psi  & 0 & -2 & -\sin \psi  \\
0 & -\cos \psi  & -\sin \psi  & -2
\end{array}
\right] _{ab}+O(\varepsilon ^{2})\text{.}
\end{equation}

Furthermore, when $\hat{z}^{2}=1$ the spherical coordinates $(r,\theta
^{1},\theta ^{2})$ parametrizing the null surface $F_{L}$ are singular, as $%
\sigma _{AB}$ is degenerate. A more suitable parametrization of $F_{L}$ in
this case would align the poles of the spherical coordinate system with
(say) the $x$-axis. Then
\begin{equation}
k^{a}=[1,0,0,\hat{z}]^{a}\text{,\qquad }e_{1}^{a}=r[0,0,-1,0]\text{\qquad
and\qquad }e_{2}^{a}=r[0,\hat{z},0,0]^{a}\text{,}
\end{equation}
and $\sigma ^{AB}=-r^{-2}\delta ^{AB}$. With this choice of orthonull basis,
it is easily checked that
\begin{equation}
k^{a}k^{b}h_{ab}=-4(L_{0}/r)\mu \varepsilon +O(\varepsilon ^{2})\text{%
,\qquad }k^{a}e_{1}^{b}h_{ab}=O(\varepsilon ^{2})\text{,\qquad }%
k^{a}e_{2}^{b}h_{ab}=O(\varepsilon ^{2})
\end{equation}
and
\begin{equation}
\sigma ^{AB}e_{A}^{a}e_{B}^{b}h_{ab}=4(L_{0}/r)\mu \varepsilon
+O(\varepsilon ^{2})\text{,.}
\end{equation}
just as in the case $\hat{z}^{2}\neq 1$ considered earlier. So it can be
seen that the null boundary $F_{L}\backslash \{p_{L}\}$ separating the
evaporating ACO spacetime from the relic Minkowski spacetime contains no
stress-energy to first order in $\mu $.

\section{Dynamical Stability of the ACO Loop}

Realistic cosmic string loops would not, of course, have the exact shape of
an ACO\ loop, and would not evaporate in a strictly self-similar manner. It
is therefore important to be able to estimate the impact that deviations
from the ACO loop trajectory would have on the evaporation process. A full
analysis of the stability of the ACO loop in the presence of back-reaction
is beyond the scope of this paper (not least because the dynamical
consequences of the self-gravity of a string loop in the strong-field limit
are not yet understood), and in this section I will limit myself to some
observations about the dynamical stability of the flat-space ACO solution.

The trajectory of the flat-space ACO loop is fully described by the position
vector $\mathbf{X}(\tau ,\sigma )$ defined in (\ref{worldsheet}), where the
mode functions $\mathbf{a}$ and $\mathbf{b}$ are specified by equations (\ref
{a-mode}) and (\ref{b-mode}). For the purposes of the perturbation analysis
of this section I will refer to the unperturbed mode functions as $\mathbf{a}%
_{0}$ and $\mathbf{b}_{0}$, so that
\begin{equation}
\mathbf{a}_{0}(\sigma _{+})=(|\sigma _{+}|-\tfrac{1}{4}L)\,\mathbf{\hat{z}}%
\quad \quad \text{for\quad }-\tfrac{1}{2}L\leq \sigma _{+}\leq \tfrac{1}{2}L
\label{a0}
\end{equation}
and
\begin{equation}
\mathbf{b}_{0}(\sigma _{-})=\tfrac{L}{2\pi }[\cos (2\pi \sigma _{-}/L)\,%
\mathbf{\hat{x}}+\sin (2\pi \sigma _{-}/L)\,\mathbf{\hat{y}}]\text{,}
\end{equation}
where $\sigma _{\pm }=\tau \pm \sigma $. (Strictly speaking, of course, $%
\mathbf{a}_{0}$ is the even periodic extension of the function on the right
of (\ref{a0}).)

A generic perturbation of the ACO solution has mode functions of the form
\begin{equation}
\mathbf{a}(\sigma _{+})=\mathbf{a}_{0}(\sigma _{+})+\delta \mathbf{a}(\sigma
_{+})\text{\qquad and\qquad }\mathbf{b}(\sigma _{-})=\mathbf{b}_{0}(\sigma
_{-})+\delta \mathbf{b}(\sigma _{-})\text{,}  \label{a,b-modes}
\end{equation}
where $\delta \mathbf{a}$ and $\delta \mathbf{b}$ are continuous functions
of their arguments, and are both small, in the sense that $|\delta \mathbf{a}%
|\ll L$ and $|\delta \mathbf{b}|\ll L$ everywhere. Without loss of
generality it can be assumed that the perturbed solution has the same
parametric period, $L$, as the unperturbed ACO loop, and so $\delta \mathbf{a%
}$ and $\delta \mathbf{b}$ are periodic functions with period $L$. The only
other constraints on $\delta \mathbf{a}$ and $\delta \mathbf{b}$ are the
gauge conditions $|\mathbf{a}^{\prime }|^{2}=|\mathbf{b}^{\prime }|^{2}=1$,
which to linear order in the perturbations require that $\mathbf{a}%
_{0}^{\prime }\cdot \delta \mathbf{a}^{\prime }=0$ and $\mathbf{b}%
_{0}^{\prime }\cdot \delta \mathbf{b}^{\prime }=0$ at all points where these
derivatives exist.

Note in particular that the derivatives $\delta \mathbf{a}^{\prime }$ and $%
\delta \mathbf{b}^{\prime }$ need not be continuous, nor is it necessarily
true that either $|\delta \mathbf{a}^{\prime }|$ or $|\delta \mathbf{b}%
^{\prime }|$ is small. A simple way to visualize the effect of a
perturbation is to plot the unit vectors $\mathbf{a}^{\prime }$ and $\mathbf{%
b}^{\prime }$ on the surface of a unit sphere, an approach first popularized
by Kibble and Turok \cite{Kibb-Tur}. Figure 6 contains three plots of this
type. The first represents the unperturbed ACO loop, and shows the $\mathbf{b%
}^{\prime }$ curve following the equator of the sphere while $\mathbf{a}%
^{\prime }=\pm \mathbf{\hat{z}}$ is concentrated at just two points, the
north and south poles. In the second plot, the perturbations $\delta \mathbf{%
a}^{\prime }$ and $\delta \mathbf{b}^{\prime }$ are assumed to be both
continuous and small, with the result that the $\mathbf{a}^{\prime }$ and $%
\mathbf{b}^{\prime }$ curves depart only minimally from the unperturbed
curves. The $\mathbf{a}^{\prime }$ curve, in particular, is understood to
jump discontinuously from the small circle around the south pole to the
small circle around the north pole when $\sigma _{+}\approx 0$ then back
again when $\sigma _{+}\approx \pi $, so the two kinks are preserved under
perturbations of this form.

The third plot in Figure 6 shows the effect of a perturbation in which $%
\delta \mathbf{a}^{\prime }$ is continuous but no longer small. In fact, $%
|\delta \mathbf{a}^{\prime }|$ is assumed to be of order unity for a narrow
range of values of the parameter $\sigma _{+}$ centered on $\sigma
_{+}\approx 0$ and $\sigma _{+}\approx \pi $, creating two smooth segments
on the $\mathbf{a}^{\prime }$ curve linking the north and south poles. This
has the effect of removing the discontinuity in the $\mathbf{a}^{\prime }$
curve, and thus smoothing the two kinks, but an unavoidable consequence is
that the $\mathbf{a}^{\prime }$ and $\mathbf{b}^{\prime }$ curves now cross
at two points, indicating that a pair of cusps will momentarily form during
each oscillation of the perturbed loop.

So although the unperturbed ACO loop does not support cusps, it is evident
that there exist solutions of the flat-space equation of motion arbitrarily
close to the ACO loop in $\mathbf{a}$-$\mathbf{b}$ trajectory space that do
support cusps. The presence of cusps complicates the back-reaction problem
considerably. Because the bulk velocity of a string loop momentarily reaches
the speed of light $c$ at a cusp, a non-negligible fraction of the loop's
total energy can be concentrated there. A simple kinematic argument (\cite
{Anderson2}, Section 6.2) suggests that the total energy $M_{r}$ inside a
sphere of radius $r$ centered on a cusp scales as $r^{1/2}$, and so the
gravitational potential $M_{r}/r$ diverges as $r^{-1/2}$. This in turn is an
indication that the effects of gravitational back-reaction at a cusp cannot
be modeled accurately at the level of the weak-field approximation.\footnote{%
By contrast, because the bulk Lorentz factor of a string is bounded at a
kink, the energy $M_{r}$ inside a sphere of radius $r$ centered on a kink
typically scales as $r$, and the gravitational potential $M_{r}/r$ has no
problematic divergences. The breakdown of the weak-field approximation at a
kink is due not to a local spike in the energy content of the string, but to
the fact that the tangent vector $X_{(0)}^{a},_{u}$ is discontinuous (see
Section 9 below).}

Any cusps forming on a perturbed solution that is close to the ACO loop will
necessarily be very narrow, as the energy in such a cusp will be
proportional to the range of the parameter $\sigma _{+}$ covered by the
segment linking the two poles on the Kibble-Turok sphere. What effect
back-reaction would have on the cusps is still an open question. It is
possible that back-reaction would act to narrow the cusps even further, thus
driving the perturbed solution closer to the unperturbed ACO loop. The ACO
loop would then be dynamically stable.

Alternatively, the cusps could broaden in the presence of back-reaction and
ultimately expand to the size of ordinary macrocusps, which typically
contain a fraction of order $\mu $ of the total energy of the string loop (%
\cite{Anderson2}, Section 6.2). In this case, it seems likely that the
perturbed loop will evaporate very quickly, as cusps are known to beam away
copious amounts of gravitational radiation \cite{Casp-All}. Indeed, the
presumption alluded to in Section 3 that the ACO solution has the lowest
radiative efficiency of any cosmic string loop encourages the view that
dynamical instabilities in the trajectory of the ACO\ loop, if they exist,
are likely to be of only secondary importance. A perturbed trajectory that
diverges rapidly from the ACO loop will have an increased radiative
efficiency and will evaporate more quickly, whereas perturbations that
remain close to the ACO\ loop will survive longer.

A third possibility is that cusps on perturbed solutions close to the ACO
loop are unstable to gravitational collapse, with consequences that have not
yet been explored and can at present only be imagined.

Another potential source of dynamical instability is loop self-intersection.
String loops that intersect themselves are believed to split into two
daughter loops with high probability (\cite{Anderson2}, Section 2.7). The
ACO loop itself has no self-intersections, but there do exist trajectories
arbitrarily close to the ACO loop that intersect themselves, as will be seen
shortly.

A self-intersection occurs on a general flat-space loop when
\begin{equation}
\mathbf{a}(\sigma _{+}+\Delta )-\mathbf{a}(\sigma _{+})=\mathbf{b}(\sigma
_{-})-\mathbf{b}(\sigma _{-}-\Delta )  \label{self}
\end{equation}
for some values of $\sigma _{+}$. $\sigma _{-}$ and $\Delta $, with $\Delta
\in (0,L/2)$ (\cite{Anderson2}, Section 3.8). If $\mathbf{a}$ and $\mathbf{b}
$ are decomposed in accordance with (\ref{a,b-modes}), the $x$- and $y$%
-components of equation (\ref{self}) read:
\begin{equation}
\cos (2\pi \sigma _{-}/L)-\cos [2\pi (\sigma _{-}-\Delta )/L]=2\pi
L^{-1}(\delta _{\Delta }\mathbf{a}-\delta _{\Delta }\mathbf{b})\cdot \mathbf{%
\hat{x}}  \label{self-x}
\end{equation}
and
\begin{equation}
\sin (2\pi \sigma _{-}/L)-\sin [2\pi (\sigma _{-}-\Delta )/L]=2\pi
L^{-1}(\delta _{\Delta }\mathbf{a}-\delta _{\Delta }\mathbf{b})\cdot \mathbf{%
\hat{y}}  \label{self-y}
\end{equation}
respectively, where
\begin{equation}
\delta _{\Delta }\mathbf{a}\equiv \delta \mathbf{a}(\sigma _{+}+\Delta
)-\delta \mathbf{a}(\sigma _{+})\text{\qquad and\qquad }\delta _{\Delta }%
\mathbf{b}\equiv \delta \mathbf{b}(\sigma _{-})-\delta \mathbf{b}(\sigma
_{-}-\Delta )\text{.}
\end{equation}

Equations (\ref{self-x}) and (\ref{self-y}) can be solved to give:
\begin{equation}
\cos (2\pi \sigma _{-}/L)=\pi L^{-1}(\delta _{\Delta }\mathbf{a}-\delta
_{\Delta }\mathbf{b})\cdot [\mathbf{\hat{x}+}\tfrac{\sin (2\pi \Delta /L)}{%
1-\cos (2\pi \Delta /L)}\mathbf{\hat{y}}]
\end{equation}
and
\begin{equation}
\sin (2\pi \sigma _{-}/L)=\pi L^{-1}(\delta _{\Delta }\mathbf{a}-\delta
_{\Delta }\mathbf{b})\cdot [\mathbf{\hat{y}}-\tfrac{\sin (2\pi \Delta /L)}{%
1-\cos (2\pi \Delta /L)}\mathbf{\hat{x}}]\text{.}
\end{equation}
In particular,
\begin{equation}
1-\cos (2\pi \Delta /L)=2\pi ^{2}L^{-2}[|(\delta _{\Delta }\mathbf{a}-\delta
_{\Delta }\mathbf{b})\cdot \mathbf{\hat{x}}|^{2}+|(\delta _{\Delta }\mathbf{a%
}-\delta _{\Delta }\mathbf{b})\cdot \mathbf{\hat{y}}|^{2}]\text{,}
\end{equation}
and so $\Delta $ must be of order $|\delta _{\Delta }\mathbf{a}-\delta
_{\Delta }\mathbf{b}|$ or smaller. For small perturbations of the ACO loop,
therefore, self-intersections can only occur with $\Delta \ll L$.

Moreover, for small values of $\Delta $ equations (\ref{self-x}) and (\ref
{self-y}) reduce to
\begin{equation}
\Delta ^{-1}(\delta _{\Delta }\mathbf{a}-\delta _{\Delta }\mathbf{b})\cdot
\mathbf{\hat{x}}\approx -\sin (2\pi \sigma _{-}/L)\text{\qquad and\qquad }%
\Delta ^{-1}(\delta _{\Delta }\mathbf{a}-\delta _{\Delta }\mathbf{b})\cdot
\mathbf{\hat{y}}\approx \cos (2\pi \sigma _{-}/L)
\end{equation}
respectively, and can be satisfied only if either $\delta \mathbf{a}^{\prime
}$ or $\delta \mathbf{b}^{\prime }$ is discontinuous or of order unity at
the self-intersection point $(\sigma _{+},\sigma _{-})\approx (\sigma
_{+}+\Delta ,\sigma _{-}-\Delta )$. Hence, at least one of the $\mathbf{a}%
^{\prime }$ and $\mathbf{b}^{\prime }$ curves will deviate significantly
from the corresponding unperturbed ACO curve on the Kibble-Turok sphere.

The picture that emerges, therefore, is that a flat-space solution close to
the ACO loop can intersect itself only by pinching off a small daughter loop
with parametric period $\Delta \ll L$. A solution that does just this is
easily constructed by choosing (at time $t=0$ say) a small segment of the
unperturbed ACO loop with length of order $\Delta $ and deforming it so that
it crosses itself. But the same perturbation procedure could be applied to
any flat-space loop solution, so self-intersections of this type are in a
sense rather trivial.

Furthermore, the larger of the two daughter loops created by the
self-intersection will be little changed from the original parent loop. It
will have a marginally smaller parametric length $L-\Delta $, and in place
of the self-intersection there will be two narrow kinks traveling in
opposite directions around the loop. Analytic studies of kinks on long
strings strongly suggest that gravitational back-reaction will suppress
small-scale structure on a string with a characteristic length $r$ on a
timescale $\Delta t\sim r/(8\pi ^{2}\mu )$ (\cite{Anderson2}, Section 6.7),
and this estimate has been confirmed in numerical simulations of planar
loops carrying up to 32 kinks by Quashnock and Spergel \cite{Quash}. In the
case of a kink formed by the pinching off of a daughter loop with a
parametric period $\Delta $ the characteristic length $r$ will be of order $%
\Delta $, and the decay time $\Delta t$ for the kink will be much smaller
than the radiative lifetime $t_{L}=L/(4\pi \kappa \mu )$ of the ACO loop. So
it is to be expected that the two narrow kinks created by the
self-intersection will quickly dissipate, leaving behind a daughter loop
that very closely resembles the ACO loop.

To summarize, there is as yet no reason to believe that the ACO solution is
dynamically unstable, although a definitive statement on the matter will
depend on the results of full back-reaction stability analyses, both at the
weak-field level (without cusps) and in the strong-field limit (with cusps).
Nonetheless, it seems safe to conclude that the effect of possible
self-intersections on the stability of the ACO loop is negligible.
Furthermore, even if it were to turn out that the ACO loop is unstable to
certain mechanisms (such as cusp formation), this would not in itself
invalidate the importance of the self-similar evaporating ACO\ solution
described above, as the longest-lived loops in any ensemble of loops with
the same initial energy would be those that most closely resemble the ACO
loop -- on the presumption, of course, that the ACO solution is unique in
having the lowest radiative efficiency of any string loop.

I should add that at present there is an ongoing debate about the nature of
the scaling peaks in the production spectrum of cosmic string loops, fueled
by the conflicting results of recent numerical \cite{VOV, RSB} and analytic
\cite{Vanch, DPR} studies of the primordial string network. However, this
debate has no direct bearing on the stability or evolution of the ACO loop.

\section{Conclusions}

In this paper I have constructed a solution of the linearized Einstein and
Nambu-Goto equations describing a cosmic string loop that evaporates
self-similarly under the action of its own self-gravity, eventually
radiating away all its energy and momentum to leave a remnant flat
spacetime, and whose spacelike cross-sections are identical (to leading
order in the string's mass per unit length $\mu $) to those of the
Allen-Casper-Ottewill loop.

From a purely physical viewpoint, the ACO loop is one of the most important
flat-space cosmic string solutions, as it is long-lived and approximates
well the lowest-efficiency daughter loops observed in low-resolution string
network simulations \cite{All-Shell, Casp-All}\footnote{%
More recent high-resolution simulations indicate that string loops, when
they first form in an evolved string network, have an approximately fractal
structure, with a fractal dimension approaching 2 on large scales \cite{VHS,
Mart-Shell, Pol-Roch}. The ACO loop therefore approximates realistic
incipient string loops less closely than was first believed. Nonetheless, it
remains true that the ACO\ loop has the lowest known radiative efficiency of
any string loop. Furthermore, as was explained in Section 8, there are good
theoretical reasons for believing that radiative back-reaction acts to
preferentially eliminate small-scale structure from a loop on timescales
much shorter than the lifetime of the loop (although the principal decay
channel for structure on the very smallest scales would probably be particle
production \cite{VHS}). Hence, it seems likely that any fractal loop similar
in shape to an ACO\ loop on large scales would radiate away much of its
small-scale structure and quickly come to resemble the ACO\ loop on all
scales. It has also been suggested \cite{Vanch, DPR} that loop fragmentation
will quickly act to suppress fractal structure, by triggering a cascade of
daughter loops with progressively fewer cusps and kinks.}. The fact that its
evolution and ultimate evaporation are analytically tractable is an added
bonus, and makes the ACO loop an object of potentially great interest to
mathematical relativity.

However, much work needs to be done before it could be claimed with any
confidence that the evaporating ACO\ loop can be described by a
self-consistent solution of the Einstein equations. A first step in this
direction would be to construct a solution to the \emph{strong-field}
back-reaction problem that reduces to the bare ACO loop in the limit $\mu
\rightarrow 0$ -- or, at the very least, to demonstrate that such a solution
exists. A second desideratum would be to find a solution of the Abelian
Higgs field equations that can act as a suitable source for the evaporating
ACO metric. Neither task is likely to be easy.

One question that arises naturally in this context is whether the
self-similarity of the linearized solution would be preserved in the
corresponding fully non-linear solution (if indeed such a solution exists).
This question remains an open one, but it can be given a more rigorous
formulation as follows. If the original Minkowski coordinates $x^{a}=[t,%
\mathbf{x}]^{a}$ describing the linearized solution are written in terms of
the similarity coordinates $[\psi ,\mathbf{\bar{x}]}$, where $\psi $ and $%
\mathbf{\bar{x}}$ are defined by equations (\ref{psi}) and (\ref{image})
respectively, then it can be shown that the vector field
\begin{equation}
k^{a}\equiv \partial x^{a}/\partial \psi =[0,-y,x,0]^{a}-\kappa \mu
[t-t_{L},x,y,z]^{a}
\end{equation}
satisfies the equation
\begin{equation}
(\eta _{c(a}+h_{c(a})\partial _{b)}k^{c}+\tfrac{1}{2}k^{c}\partial
_{c}h_{ab}=-\kappa \mu \eta _{ab}
\end{equation}
to linear order in $\mu $. That is, $k^{a}$ satisfies the linearized version
of the equation $\nabla _{(a}k_{b)}=-\kappa \mu g_{ab}$ for a conformal
Killing vector field.

The corresponding non-linear solution will therefore preserve
self-similarity if it admits a vector field $k^{a}$ satisfying the \emph{%
exact} conformal Killing equation $\nabla _{(a}k_{b)}=-\kappa \mu g_{ab}$,
in which case it is always possible to find coordinates $[\psi ,\mathbf{\bar{%
x}]}$ in terms of which the metric tensor has the conformal form
\begin{equation}
g_{ab}(\psi ,\mathbf{\bar{x}})=e^{-2\kappa \mu \psi }p_{ab}(\mathbf{\bar{x}})%
\text{.}  \label{conform}
\end{equation}
(The conformal Killing field in this case is $k^{a}=\delta _{\psi }^{a}$.)
So the question of whether self-similarity would be preserved by a fully
non-linear solution can be recast as a question about the existence of
solutions to the strong-field back-reaction problem with the conformal form (%
\ref{conform}).

Another of the more problematic aspects of the ACO loop is the behavior of
the gravitational field near the two kink points. The discontinuity in the
null tangent vector $X_{(0)}^{a},_{u}$ at the kinks generates two apparent
anomalies in the weak-field analysis described in Section 6. The first is
that the term $X_{(0)}^{d},_{CD}$ appearing in the first-order equation of
motion (\ref{em2}) is undefined at the kinks, and so it cannot meaningfully
be said that the equation of motion is satisfied at these points. The second
anomaly is that the metric perturbations $h_{ab}$ fail to be differentiable
on a manifold of points extending from the trajectories of the kinks out to
future null infinity, as is explained in more detail in Appendix A.3.

The breakdown in the differentiability of $h_{ab}$ poses no intractable
difficulty for the linearized analysis of Section 6, because the ill-defined
terms make no overall contribution to the \textit{linearized} Einstein
tensor. However, the fully non-linear Einstein tensor contains products of
terms of the form $\partial _{c}g_{ab}$, and in the generic case it is
expected that a distributional interpretation of the Einstein equation is
possible only if the metric derivatives $\partial _{c}g_{ab}$ are locally
square integrable \cite{Geroch}. So the breakdown of differentiability in
the beaming directions of the kinks may bode ill for the prospects of a
strong-field solution.

A further problem is that the non-differentiability of $h_{ab}$ is
superimposed on the conical singularity marking the string worldsheet at the
kink points themselves. The singularity in $h_{ab}$ at the kinks was
examined briefly in Paper I, and is far more pathological than the simple
logarithmic singularity that appears at all other points on the string
worldsheet. An appreciation for the nature of the singularity can be gained
by repeating the analysis performed earlier, in Section 6.2, to calculate
the near-field metric perturbations at a kink point.

If the field point $x^{a}=\frac{1}{4\pi }L_{0}[T,X,Y,Z]^{a}$ is chosen to be
close to the lower of the kink points (at $u=0$) near the spacelike surface $%
t=0$, so that
\begin{equation}
\lbrack T,X,Y,Z]^{a}=[v,\cos v,\sin v,-\tfrac{1}{2}\pi ]^{a}+4\pi
L_{0}^{-1}\delta x^{a}
\end{equation}
for some $v\in (-\pi ,\pi )$, then it can be shown that $\chi _{\pm }=v+4\pi
L_{0}^{-1}(\delta t\pm \delta z)$, and that
\[
\chi _{+}-V_{1}-2\pi =\tfrac{1}{2\pi }(4\pi L_{0}^{-1})^{2}Q_{-}\text{%
,\qquad }\chi _{-}-V_{1}=2\pi -8\pi L_{0}^{-1}\delta z
\]
\begin{equation}
\text{and\qquad }\chi _{\pm }-V_{0}=4\pi L_{0}^{-1}(\delta t\pm \delta z-%
\tfrac{1}{2}\tfrac{\eta _{ab}\delta x^{a}\delta x^{b}}{\delta t+\delta x\sin
v-\delta y\cos v})  \label{args}
\end{equation}
to first or leading order in $\delta x^{a}$, where
\begin{eqnarray}
Q_{-} &\equiv &(\delta x)^{2}+(\delta y)^{2}+(\delta t+\delta z)(\delta
t+\delta z+2\delta x\sin v-2\delta y\cos v)  \nonumber \\
&\equiv &\bar{\Psi}_{ab}(-1,v)\,\delta x^{a}\delta x^{b}\text{.}
\end{eqnarray}

All four of the functions in (\ref{args}) appear as arguments of the
logarithmic and sine and cosine integral functions contributing to the terms
$I_{\pm }$, $S_{\pm }$ and $C_{\pm }$ in (\ref{metperts}). For points on the
future lightcone of the kink, where $\eta _{ab}\delta x^{a}\delta x^{b}=0$,
the expansions (\ref{args}) simplify considerably, as $\chi _{\pm }-V_{0}$
then becomes $4\pi L_{0}^{-1}(\delta t\pm \delta z)$, while $Q_{-}=2(\delta
t+\delta z)(\delta t+\delta x\sin v-\delta y\cos v)$. (Furthermore, the
function $\delta t+\delta x\sin v-\delta y\cos v$ appearing in $Q_{-}$ is
zero if and only if the field point $x^{a}+\delta x^{a}$ lies in the beaming
direction of the kink: see Appendix A.3.)

However, for the purposes of calculating the near-field derivatives $%
\partial _{c}h_{ab}$ it is necessary to retain the term $\eta _{ab}\delta
x^{a}\delta x^{b}$ in full, with the result that many of the derivatives are
undefined in the limit as $\delta x^{a}\rightarrow 0$. The near-field
expansion of $h_{ab}$ appearing in the footnote in Section 4.2 of Paper I is
as complicated as it is because the ``off-shell'' contributions have been
included so as to give a hint of the complexity of the derivatives.

It should be clear from the foregoing remarks that the breakdown of the
field equations at the kinks is not strictly speaking a strong-field effect,
but stems rather from the zero-thickness assumption underpinning the
Nambu-Goto action, which imposes infinite worldsheet curvature at the kinks.
It seems unlikely that the singularity at the kinks would be resolved in a
fully non-linear solution of the Einstein equations, and I suspect that the
kinks can be smoothed only by the inclusion of a finite-thickness
field-theoretic source. However, the fact -- mentioned in the footnote in
Section 8 -- that the Newtonian gravitational potential $M_{r}/r$ remains
bounded near the kinks (and in fact is likely to be of order $\mu $ there)
offers some hope that the kinks on the ACO loop could be incorporated into a
fully non-linear solution.

There are other topics of interest in this area that would repay further
investigation. For example, it would be useful to be able to offer a
definitive proof that the ACO loop does indeed have the lowest radiative
efficiency of all flat-space cosmic string loop solutions, or to demonstrate
that the evaporating ACO loop is at least neutrally stable to (cusp-free)
small perturbations.

I would like to thank the referee for highlighting the need for some helpful
clarifications.

\appendix

\section{Some Properties of the Roots $V_{1}$ and $V_{0}$}

\subsection{Bounds on $V_{1}$ and $V_{0}$}

If the vector $(X,Y,Z)$ corresponding to the spatial components $\mathbf{%
\bar{x}}=\frac{1}{4\pi }L_{0}(X,Y,Z)$ of the image point is represented in
standard spherical polar coordinates:
\begin{equation}
(X,Y,Z)=(r\sin \theta \cos \phi ,r\sin \theta \sin \phi ,r\cos \theta )\text{%
,}
\end{equation}
then since $T=r$ the equation (\ref{roots}) for the roots $V_{k}$
specialized to $V_{1}$ reads
\begin{equation}
V_{1}+\pi =r-[(r\sin \theta \cos \phi -\cos V_{1})^{2}+(r\sin \theta \sin
\phi -\sin V_{1})^{2}+(r\cos \theta -\tfrac{1}{2}\pi )^{2}]^{1/2}\text{.}
\label{V1-bound1}
\end{equation}

Differentiating the expression inside the square root with respect to $V_{1}$
indicates that $V_{1}+\pi $ takes on an extremal value whenever $\sin
(V_{1}-\phi )=0$ -- unless $r\sin \theta =0$, in which case equation (\ref
{V1-bound1}) solves explicitly to give $V_{1}+\pi =r-[1+(r\mp \tfrac{1}{2}%
\pi )^{2}]^{1/2}$ (with the sign choice $\mp $ corresponding to $\cos \theta
=\pm 1$ if $r\neq 0$). The case $r\sin \theta =0$ is of course equivalent to
$X=Y=0$. and is discussed in more detail at the end of Section 5.2 It is
easily verified that, in this case, $V_{1}+\pi $ is a strictly monotonically
increasing function of $r$, and so lies in the range $[-(1+\frac{1}{4}\pi
^{2})^{1/2},-\frac{1}{2}\pi )$ if $\cos \theta =+1$ and in $[-(1+\frac{1}{4}%
\pi ^{2})^{1/2},\frac{1}{2}\pi )$ if $\cos \theta =-1$.

If $\sin (V_{1}-\phi )\neq 0$ then the right-hand side of (\ref{V1-bound1})
is extremized if either $(\cos V_{1},\sin V_{1})=(\cos \phi ,\sin \phi )$ or
$(\cos V_{1},\sin V_{1})=(-\cos \phi ,-\sin \phi )$. It follows from (\ref
{V1-bound1}) that
\begin{equation}
r-[(r\sin \theta +1)^{2}+(r\cos \theta -\tfrac{1}{2}\pi )^{2}]^{1/2}\leq
V_{1}+\pi \leq r-[(r\sin \theta -1)^{2}+(r\cos \theta -\tfrac{1}{2}\pi
)^{2}]^{1/2}\text{.}  \label{V1-bound2}
\end{equation}
Again, differentiation with respect to $r$ of the lower and upper bounding
functions in (\ref{V1-bound2}) is sufficient to show that both are
monotonically increasing functions of $r$ (and furthermore the left-hand
function is strictly monotonic unless $\tan \theta =-2/\pi $, while the
right-hand function is strictly monotonic unless $\tan \theta =2/\pi $).
From this fact it is readily seen that
\begin{equation}
-(1+\tfrac{1}{4}\pi ^{2})^{1/2}\leq V_{1}+\pi \leq (1+\tfrac{1}{4}\pi
^{2})^{1/2}
\end{equation}
(the lower bound being achieved when $r=0$ or $\tan \theta =-2/\pi $, and
the upper bound when $\tan \theta =2/\pi $ and $r\geq (1+\frac{1}{4}\pi
^{2})^{1/2}$). Since the root $V_{0}$ lies in the interval $%
[V_{1},V_{1}+2\pi ]$, it is evident that $V_{0}$ is also bounded:
\begin{equation}
-(1+\tfrac{1}{4}\pi ^{2})^{1/2}-\pi \leq V_{0}\leq (1+\tfrac{1}{4}\pi
^{2})^{1/2}+\pi \approx 5.00369\text{.}
\end{equation}

The fact that $V_{0}$ and $V_{1}$ are bounded should come as no surprise, as
$V_{0}$ and $V_{1}+\pi $ are the values of the scaled time coordinate $T$ at
the points of intersection of the backwards light cone of the image point $[|%
\mathbf{\bar{x}}|,\mathbf{\bar{x}}]$ with the trajectories of the lower ($%
u=0 $) and upper ($u=\pm \pi $) kink points respectively.

\subsection{The zeroes of $T-s(Z+\tfrac{1}{2}\pi )-v$}

According to (\ref{dv/du}), the function $v(u)$ describing the curve of
intersection $\Gamma $ of the world sheet with the backwards light cone of $%
[|\mathbf{\bar{x}}|,\mathbf{\bar{x}}]\equiv \frac{1}{4\pi }L_{0}[T,X,Y,Z]$
has a possible stationary point whenever $E\equiv \,T-s(Z+\tfrac{1}{2}\pi
)-v=0$. (Recall here that $s\equiv \,$sgn$(u)$, or equivalently $s$ is the
sign of $\mathbf{\hat{z}}\cdot \mathbf{X},_{u}$ on the segment of the string
loop at which the intersection point occurs.) It is important to be able to
locate those points $[|\mathbf{\bar{x}}|,\mathbf{\bar{x}}]$ on the surface $%
F_{0}$ at which $E$ goes to zero for some value of $v$ in $[V_{1},V_{1}+2\pi
]$, because $E$ appears as the denominator in the integrals contributing to $%
h_{ab}(|\mathbf{\bar{x}}|,\mathbf{\bar{x}})$ in (\ref{h-ints}) and so the
components of $h_{ab}$ are potentially divergent at such a point.

As mentioned in section 5.2, $E$ is positive semi-definite by virtue of
equation (\ref{int3}), which can be rewritten in the form
\begin{eqnarray}
T-s(Z+\tfrac{1}{2}\pi )-v &=&u-s(Z+\tfrac{1}{2}\pi )+[(X-\cos v)^{2}
\nonumber \\
&&+(Y-\sin v)^{2}+\{Z-(|u|-\tfrac{1}{2}\pi )\}^{2}]^{1/2}  \nonumber \\
&\geq &u-s(Z+\tfrac{1}{2}\pi )+|Z-(|u|-\tfrac{1}{2}\pi )|\text{,}  \label{A1}
\end{eqnarray}
Here, equality occurs if and only if $X=\cos v$ and $Y=\sin v$, while the
function on the third line is zero if and only if $u\leq s(Z+\tfrac{1}{2}\pi
)$.

Suppose therefore that the point $(X,Y)$ lies on the unit circle, so that $%
(X,Y)=(\cos v_{\text{F}},\sin v_{\text{F}})$ for some angle $v_{\text{F}}$.
Then $T-s(Z+\tfrac{1}{2}\pi )-v_{\text{F}}$ will be zero whenever
\begin{equation}
Z=\lambda -\tfrac{1}{2}\pi \text{\qquad and\qquad }T=v_{\text{F}}+s\lambda
\text{,}  \label{Z,T-values}
\end{equation}
where $\lambda $ is any number satisfying $\lambda \geq u$ if $u>0$, or $%
\lambda \leq -u$ if $u<0$. (Note that if the null constraint $%
T=(X^{2}+Y^{2}+Z^{2})^{1/2}$ is also imposed then $v_{\text{F}}=[1+(\lambda -%
\tfrac{1}{2}\pi )^{2}]^{1/2}-s\lambda $, and for each value of $s$ there is
a one-parameter family of spacetime points on $F_{0}$ for which $E=0$ at
some point on $\Gamma $.)

If $0\leq \lambda \leq \pi $ then the spacetime point $\frac{1}{4\pi }%
L_{0}[T,X,Y,Z]$ coincides with the point $X^{a}(u,v)\equiv X^{a}(s\lambda
,v_{\text{F}})$ on the world sheet. It follows from the results of the
previous paragraph that $E$ will be zero for all values of $u$ in the range $%
[0,\lambda ]$ if $s=1$, and all values of $u$ in the range $[-\pi ,-\lambda
] $ if $s=-1$. This in turn means that the curve $\Gamma $ includes a
horizontal segment with $v\equiv v_{\text{F}}$ over the ranges of $u$
indicated.

More important for the purposes of calculating $h_{ab}(|\mathbf{\bar{x}}|,%
\mathbf{\bar{x}})$ are image points that lie off the worldsheet. These have $%
|Z|>\tfrac{1}{2}\pi $, and so either $\lambda >\pi $ if $s=1$ or $\lambda <0$
if $s=-1$. In geometric terms, such points either lie on the upwards ($Z>%
\tfrac{1}{2}\pi $) extension of the helical segment of the ACO loop with $%
u>0 $, or on the downwards ($Z<-\tfrac{1}{2}\pi $) extension of the helical
segment with $u<0$. Furthermore, it is evident from the preceding analysis
that in this case $E$ will be zero on the curve $\Gamma $ for all values of $%
u\in [0,\pi ]$ if $s=1$, and all values of $u\in [-\pi ,0]$ if $s=-1$, and
so $\Gamma $ includes a straight-line segment with $v\equiv v_{\text{F}}$
over one of these ranges. An immediate consequence is that $%
V_{0}=V_{1}\equiv v_{\text{F}}$ if $s=1$, and $V_{0}=V_{1}+2\pi \equiv v_{%
\text{F}}$ if $s=-1$. (This straight-line segment corresponds to the
``trivial solution'' $V_{k}=v$ mentioned in Section 6.1.)

It is evident now that if the image point $[|\mathbf{\bar{x}}|,\mathbf{\bar{x%
}}]$ lies on one of the two helical extensions of the loop, but not on the
loop itself, then the corresponding integral in the equation (\ref{h-ints})
for $h_{ab}(|\mathbf{\bar{x}}|,\mathbf{\bar{x}})$ is undefined. However, the
problematic integrals can be reformulated by making use of the equation (\ref
{dv/du}) for $dv/du$ to replace $v$ with $u$ as the variable of integration.
Furthermore, since $(X,Y)=(\cos v_{\text{F}},\sin v_{\text{F}})$ the
denominator $T-(u+v)+X\sin v-Y\cos v$ becomes simply $T-u-v_{\text{F}}$,
where $v_{\text{F}}=\chi _{\pm }\equiv $ $T\pm (Z+\tfrac{1}{2}\pi )$ if $%
s=\mp 1$. Hence,
\begin{eqnarray*}
&&\int_{V_{0}}^{V_{1}+2\pi }[T+(Z+\tfrac{1}{2}\pi )-v]^{-1}\bar{\Psi}%
_{ab}(-1,v)\,dv\text{\qquad is replaced by} \\
&&\int_{-\pi }^{0}[-(Z+\tfrac{1}{2}\pi )-u]^{-1}\bar{\Psi}_{ab}(-1,\chi
_{+})\,du
\end{eqnarray*}
and
\begin{eqnarray*}
&&\int_{V_{1}}^{V_{0}}[T-(Z+\tfrac{1}{2}\pi )-v]^{-1}\bar{\Psi}%
_{ab}(+1,v)\,dv\text{\qquad is replaced by} \\
&&\int_{0}^{\pi }(Z+\tfrac{1}{2}\pi -u)^{-1}\bar{\Psi}_{ab}(+1,\chi _{-})\,du%
\text{.}
\end{eqnarray*}

The new integrals are easily evaluated, giving
\begin{equation}
I_{+}=-\ln [(\tfrac{1}{2}\pi -Z)/(-\tfrac{1}{2}\pi -Z)]\text{,\qquad }%
S_{+}=I_{+}\sin (\chi _{+})\text{,\qquad }C_{+}=I_{+}\cos (\chi _{+})
\label{ints-plus}
\end{equation}
(if $s=-1$) and
\begin{equation}
I_{-}=-\ln [(Z+\tfrac{1}{2}\pi )/(Z-\tfrac{1}{2}\pi )]\text{,\qquad }%
S_{-}=I_{-}\sin (\chi _{-})\text{,\qquad }C_{-}=I_{-}\cos (\chi _{-})
\label{ints-minus}
\end{equation}
(if $s=1$) in place of the corresponding terms in (\ref{I-ints})-(\ref
{S,C-ints2}).

\subsection{The zeroes of $T-(u+v)+X\sin v-Y\cos v$}

A second consequence of (\ref{dv/du}) is that the graph of $v$ against $u$
on the intersection curve $\Gamma $ typically has a point of vertical
inflection whenever $H\equiv T-(u+v)+X\sin v-Y\cos v=0$. As was indicated in
Section 5.2, the factor $H$ is positive semi-definite, because
\begin{eqnarray}
T-(u+v)+X\sin v-Y\cos v &=&[(X\sin v-Y\cos v)^{2}+(X\cos v+Y\sin v-1)^{2}
\nonumber \\
&&+\{Z-(|u|-\tfrac{1}{2}\pi )\}^{2}]^{1/2}+X\sin v-Y\cos v  \nonumber \\
&\geq &|X\sin v-Y\cos v|+X\sin v-Y\cos v\text{,}  \label{A2-bound}
\end{eqnarray}
with equality occurring if and only if $X\cos v+Y\sin v=1$ and $Z=|u|-\tfrac{%
1}{2}\pi $. In particular, it follows that spacetime points with $H=0$
somewhere on $\Gamma $ have $|Z|\leq \tfrac{1}{2}\pi $. It was seen earlier
that spacetime points outside the world sheet with $E=0$ somewhere on $%
\Gamma $ have $|Z|>\tfrac{1}{2}\pi $, so the two classes of points are
disjoint outside the world sheet.

Suppose now that $H=0$ at the point $(u,v)=(u_{\text{F}},v_{\text{F}})$ on $%
\Gamma $. The equations $H=0$ and $X\cos v+Y\sin v=1$ (with $X\sin v-Y\cos
v\leq 0$) can then be solved exactly to give
\begin{eqnarray}
T &=&u_{\text{F}}+v_{\text{F}}+(R^{2}-1)^{1/2}\text{,}\qquad X=\cos v_{\text{%
F}}-\left( R^{2}-1\right) ^{1/2}\sin v_{\text{F}}  \nonumber \\
\text{and\qquad }Y &=&\sin v_{\text{F}}+\left( R^{2}-1\right) ^{1/2}\cos v_{%
\text{F}}\text{,}
\end{eqnarray}
while of course $Z=|u_{\text{F}}|-\tfrac{1}{2}\pi $. The value of $R\geq 1$
is fixed by the null constraint $T=(X^{2}+Y^{2}+Z^{2})^{1/2}$, which reads
explicitly
\begin{equation}
u_{\text{F}}+v_{\text{F}}=[R^{2}+(|u_{\text{F}}|-\tfrac{1}{2}\pi
)^{2}]^{1/2}-(R^{2}-1)^{1/2}\text{.}
\end{equation}
So there exists a two-parameter family of spacetime points with the property
that $H=0$ somewhere on $\Gamma $.

If $[T,X,Y,Z]$ is held constant and $u$ varied from $u_{\text{F}}$ to $u_{%
\text{F}}+\delta u$, a perturbation expansion of the equation (\ref
{backlight4}) for $v$ indicates that
\begin{equation}
v(u_{\text{F}}+\delta u)-v_{\text{F}}=-(6\delta u)^{1/3}
\end{equation}
to leading order in $\delta u$. Hence, $v$ remains a monotonic decreasing
function of $u$ even in the singular case where $H=0$ somewhere on $\Gamma $.

The importance of the zeroes of $H$ lies in the fact that the spacetime
derivatives $\bar{\partial}_{c}h_{ab}$ of the metric perturbations $h_{ab}(%
\bar{t},\mathbf{\bar{x}})$ are potentially divergent at any image point $[%
\bar{t},\mathbf{\bar{x}}]\equiv \frac{1}{4\pi }L_{0}[T,X,Y,Z]$ with the
property that $H=0$ when $v=V_{0}$ or $V_{1}$. This can be seen by first
differentiating the equation (\ref{h-ints}) for $h_{ab}$ and then
integrating by parts to give
\begin{eqnarray}
\bar{\partial}_{c}h_{ab} &=&-4\mu (\bar{\partial}_{c}\chi
_{+})\int_{V_{0}}^{V_{1}+2\pi }(\chi _{+}-v)^{-1}\Omega _{ab}(-1,v)\,dv
\nonumber \\
&&-4\mu (\bar{\partial}_{c}\chi _{-})\int_{V_{1}}^{V_{0}}(\chi
_{-}-v)^{-1}\Omega _{ab}(+1,v)\,dv\,  \nonumber \\
&&-4\mu \,(\chi _{+}-V_{1}-2\pi )^{-1}\bar{\Psi}_{ab}(-1,V_{1}+2\pi )\,\bar{%
\partial}_{c}(V_{1}-\chi _{+})  \nonumber \\
&&+4\mu (\chi _{-}-V_{1})^{-1}\bar{\Psi}_{ab}(+1,V_{1})\bar{\partial}%
_{c}(V_{1}-\chi _{-})  \nonumber \\
&&-4\mu (\chi _{-}-V_{0})^{-1}\bar{\Psi}_{ab}(+1,V_{0})\bar{\partial}%
_{c}(V_{0}-\chi _{-})  \nonumber \\
&&+4\mu (\chi _{+}-V_{0})^{-1}\bar{\Psi}_{ab}(-1,V_{0})\bar{\partial}%
_{c}(V_{0}-\chi _{+})\text{,}  \label{h-deriv}
\end{eqnarray}
where again $\chi _{\pm }=T\pm (Z+\tfrac{\pi }{2})$ and $\Omega
_{ab}(s,v)=\partial _{v}\bar{\Psi}_{ab}(s,v)$ as defined in Section 6.1.

The two integrals on the right-hand side of (\ref{h-deriv}) converge because
all the components of $\Omega _{ab}$ are identical to components of $\bar{%
\Psi}_{ab}$, and it was demonstrated in the previous sub-section that
integrals of this type will converge unless the field point lies on the
worldsheet itself. Also, the derivatives $\bar{\partial}_{c}\chi _{\pm
}=4\pi L_{0}^{-1}[1,\pm \mathbf{\hat{z}}]_{c}$ are everywhere well defined.

However, the terms involving $\bar{\partial}_{c}V_{1}$ and $\bar{\partial}%
_{c}V_{0}$ in (\ref{h-deriv}) are potentially problematic. Implicit
differentiation of the equation (\ref{backlight4}) defining $v$ gives:
\begin{eqnarray}
\bar{\partial}_{c}v &=&[T-(u+v)+X\sin v-Y\cos v]^{-1}  \nonumber \\
&&\times 4\pi L_{0}^{-1}[T-(u+v),-(X-\cos v),-(Y-\sin v),-\{Z-(|u|-\tfrac{1}{%
2}\pi )\}]_{c}\text{,}  \nonumber \\
&&  \label{div-v}
\end{eqnarray}
where the denominator is of course just $H$. The derivative $\bar{\partial}%
_{c}V_{k}$ therefore typically diverges if $H=0$ at $v=V_{k}$ (and in fact
the components of the 4-vector multiplying $H^{-1}$ in (\ref{div-v}) will
all vanish only if the point $\frac{1}{4\pi }L_{0}[T,X,Y,Z]$ lies on the
world sheet).\footnote{%
It might be supposed from (\ref{h-deriv}) that $\bar{\partial}_{c}h_{ab}$ is
also undefined whenever $\chi _{\pm }-V=0$, with $V=V_{1}$, $V_{0}$ or $%
V_{-1}$. That this is not so (at least for image points outside the world
sheet) can be seen by replacing $v$ with $u$ as the integration variable in
the equation (\ref{h-ints}) for $h_{ab}$ to give:
\[
h_{ab}(|\mathbf{\bar{x}}|,\mathbf{\bar{x}})=-4\mu \int_{-\pi }^{0}H^{-1}\bar{%
\Psi}_{ab}(-1,v)\,du-4\mu \int_{0}^{\pi }H^{-1}\bar{\Psi}_{ab}(+1,v)\,dv
\]
It is clear that $\bar{\partial}_{c}h_{ab}$ involves only terms of the form $%
\Omega _{ab}\bar{\partial}_{c}v$ and $H^{-2}\bar{\partial}_{c}H$, where $%
\bar{\partial}_{c}H$ is a linear combination of $\bar{\partial}_{c}T$, $\bar{%
\partial}_{c}X$, $\bar{\partial}_{c}Y$ and $\bar{\partial}_{c}v$, and
according to (\ref{div-v}) $\bar{\partial}_{c}v$ is undefined only when $H=0$%
.}

In fact, if $k=0$ or $1$ and the image point $\bar{x}^{a}=\frac{1}{4\pi }%
L_{0}[T,X,Y,Z]^{a}$ has the property that
\begin{equation}
X\cos V_{k}+Y\sin V_{k}=1\text{,}\qquad T=k\pi +V_{k}-X\sin V_{k}+Y\cos
V_{k}\qquad \text{and}\qquad Z=(|k|-\tfrac{1}{2})\pi  \label{TXYZ}
\end{equation}
(with $X\sin V_{k}-Y\cos V_{k}\leq 0$) then the perturbative solution of the
equation (\ref{roots}) for $V_{k}$ at a nearby point $\bar{x}^{a}+\delta
\bar{x}^{a}=\frac{1}{4\pi }L_{0}[T+\delta T,X+\delta X,Y+\delta Y,Z+\delta
Z]^{a}$ reads:
\begin{eqnarray}
V_{k}(\bar{x}^{a}+\delta \bar{x}^{a})-V_{k}(\bar{x}^{a}) &=&6^{1/3}(\delta
T+\delta X\sin V_{k}-\delta Y\cos V_{k})^{1/3}  \nonumber \\
&&+\tfrac{1}{4}(X\sin V_{k}-Y\cos V_{k})^{-1}6^{2/3}(\delta T+\delta X\sin
V_{k}-\delta Y\cos V_{k})^{2/3}\text{,}  \nonumber \\
&&  \label{V-expans}
\end{eqnarray}
plus terms linear in $\delta \bar{x}^{a}$.

Substitution of the expansion (\ref{V-expans}) into the equation (\ref
{h-ints}) for $h_{ab}$ then gives:
\begin{equation}
h_{tz}(\bar{x}^{a}+\delta \bar{x}^{a})-h_{tz}(\bar{x}^{a})=-8\hat{z}\mu
(\delta _{k}^{1/3}+\tfrac{1}{4}\delta _{k}^{2/3})+O(\delta \bar{x})\text{,}
\end{equation}
\begin{equation}
h_{xz}(\bar{x}^{a}+\delta \bar{x}^{a})-h_{xz}(\bar{x}^{a})=-8\hat{z}\mu
[\delta _{k}^{1/3}\sin V_{k}+\tfrac{1}{4}\delta _{k}^{2/3}(\sin
V_{k}+2D_{k}\cos V_{k})]+O(\delta \bar{x})
\end{equation}
and
\begin{equation}
h_{yz}(\bar{x}^{a}+\delta \bar{x}^{a})-h_{yz}(\bar{x}^{a})=8\hat{z}\mu
[\delta _{k}^{1/3}\cos V_{k}+\tfrac{1}{4}\delta _{k}^{2/3}(\cos
V_{k}-2D_{k}\sin V_{k})]+O(\delta \bar{x})\text{,}
\end{equation}
where
\begin{equation}
\delta _{k}\equiv 6D_{k}^{-3}(\delta T+\delta X\sin V_{k}-\delta Y\cos V_{k})%
\text{,}
\end{equation}
\begin{equation}
D_{k}\equiv -(X\sin V_{k}-Y\cos V_{k})\geq 0
\end{equation}
and $\hat{z}\equiv \,$sgn$(Z)\equiv \,$sgn$(k-\frac{1}{2})$, while all other
components of $h_{ab}(\bar{x}^{a}+\delta \bar{x}^{a})-h_{ab}(\bar{x}^{a})$
are of linear order or higher in the components of $\delta \bar{x}$.%
\footnote{%
Incidentally, the factor $D_{k}^{-3}$ appearing in the equation for $\delta
_{k}$ is guaranteed to be well defined if the field point lies away from the
string worldsheet, because if $X\sin V_{k}-Y\cos V_{k}=0$ then the previous
constraints (\ref{TXYZ}) imply that
\[
X=\cos V_{k}\text{,\quad }Y=\sin V_{k}\text{,\quad }T=k\pi +V_{k}\quad \text{%
and\quad }Z=|k\pi |-\tfrac{1}{2}\pi
\]
and so the field point is identical to the point $(u,v)=(k\pi ,V_{k})$ on
the worldsheet.}

In summary, if the field point $\bar{x}^{a}$ satisfies the constraints (\ref
{TXYZ}) for either $k=0$ or $k=1$ then the derivatives of the metric
components $h_{tz}$, $h_{xz}$ and $h_{yz}$ with respect to $\bar{t}$, $\bar{x%
}$ and $\bar{y}$ are all undefined there. Field points of this type have $%
\bar{z}=\pm \frac{1}{8}L_{0}$ and so have the same altitude as one of the
kink points on the string. The failure of $h_{ab}$ to be differentiable at
such points therefore has no effect on the analysis of the first-order
string equation of motion in Section 6, as the equation of motion is there
developed and solved only for points on the string away from the kinks.
However, the singularities in the derivatives of $h_{ab}$ do potentially
affect the matching of the weak-field solution to the Minkowski vacuum
across the null surface $F_{L}$. Fortunately, the relevant combinations of
the components of $h_{ab}$ are all differentiable on $F_{L}$, as is shown in
detail in Section 7.2.

It might seem puzzling that the weak-field metric induced by the ACO loop is
singular on a manifold of points that extends away from the worldsheet.
However, this pathological behavior has a simple physical explanation. It is
well known (\cite{Anderson2}, Section 6.3) that a generic kink on a
zero-thickness string emits a narrow beam of gravitational radiation in the
direction of its motion. Although the metric derivative is singular on the
locus of the beam, the singularity is nonetheless relatively benign, as the
geodesic equations for a free particle remain integrable across the beam.

In the case of the ACO loop, the kink points correspond to $u=k\pi $ for $%
k=0 $ or $1$ and so trace out the two families of spacetime points
\begin{equation}
X_{(0)}^{a}=\tfrac{1}{4\pi }L_{0}[k\pi +v,\cos v,\sin v,(|k|-\tfrac{1}{2}%
)\pi ]^{a}\text{.}
\end{equation}
The instantaneous 4-velocities of the kink points are therefore
\begin{equation}
\partial _{v}X_{(0)}^{a}=\tfrac{1}{4\pi }L_{0}[1,-\sin v,\cos v,0]^{a}\text{,%
}
\end{equation}
and the beams, which trace out null curves with tangent vector $\partial
_{v}X_{(0)}^{a}$, can be represented parametrically in the form
\begin{equation}
\lbrack T,X,Y,Z]=[k\pi +v,\cos v,\sin v,(|k|-\tfrac{1}{2})\pi ]+\lambda
[1,-\sin v,\cos v,0]\text{,}
\end{equation}
with $\lambda $ an affine parameter. Hence $Z=(|k|-\tfrac{1}{2})\pi $ and,
after eliminating $\lambda $,
\begin{equation}
T-(k\pi +v)+X\sin v-Y\cos v=0\text{.}
\end{equation}
The last equation is just the constraint $H=0$ with $u=k\pi $. So a
spacetime point lying off the worldsheet satisfies the constraints (\ref
{TXYZ}) if and only if it lies in the beaming direction of a kink.

\section{Calculating $h_{ab}$ and its Derivatives}

With $\chi _{\pm }\equiv v+(u\pm u)+4\pi L_{0}^{-1}(\delta t\pm \delta z)$
and
\[
\chi _{+}-V_{0}=2u+8\pi L_{0}^{-1}\delta z\text{,\qquad }\chi
_{+}-V_{1}=W+2u+4\pi L_{0}^{-1}(\delta t+\delta z-\delta V)
\]
\begin{equation}
\chi _{-}-V_{0}=\tfrac{1}{2}u^{-1}(4\pi L_{0}^{-1})^{2}Q_{+}\text{\qquad
and\qquad }\chi _{-}-V_{1}=W+4\pi L_{0}^{-1}(\delta t-\delta z-\delta V)%
\text{,}
\end{equation}
where $\delta V$ is of linear order in $\delta x^{a}$ and is defined in (\ref
{del-V}), while $Q_{+}$ is of quadratic order and is defined in (\ref{Q-plus}%
), the expressions (\ref{I-ints})-(\ref{S,C-ints2}) become, for $u>0$ and $%
\delta x^{a}$ small,
\begin{equation}
I_{+}=\allowbreak \ln (W+2u-2\pi )-\ln (2u)+4\pi L_{0}^{-1}\left( \frac{%
\delta t+\delta z-\delta V}{W+2u-2\pi }-\frac{\delta z}{u}\right) \text{,}
\label{small-1}
\end{equation}
\begin{equation}
I_{-}=\ln [(4\pi L_{0}^{-1})^{2}Q_{+}]-\ln (2uW)-4\pi L_{0}^{-1}\left( \frac{%
\delta t-\delta z-\delta V}{W}\right) \text{,}  \label{small-2}
\end{equation}
\begin{eqnarray}
\left[
\begin{array}{c}
S_{+} \\
C_{+}
\end{array}
\right] &=&\left[
\begin{array}{cc}
-\cos (2u+v) & \sin (2u+v) \\
\sin (2u+v) & \cos (2u+v)
\end{array}
\right] \left[
\begin{array}{c}
\allowbreak \func{Si}\left( W+2u-2\pi \right) -\func{Si}\left( 2u\right) \\
\allowbreak \func{Ci}\left( W+2u-2\pi \right) -\func{Ci}\left( 2u\right)
\end{array}
\right]  \nonumber \\
&&+4\pi L_{0}^{-1}\left[
\begin{array}{cc}
\sin (2u+v) & \cos (2u+v) \\
\cos (2u+v) & -\sin (2u+v)
\end{array}
\right] \left[
\begin{array}{c}
\allowbreak \func{Si}\left( W+2u-2\pi \right) -\func{Si}\left( 2u\right) \\
\allowbreak \func{Ci}\left( W+2u-2\pi \right) -\func{Ci}\left( 2u\right)
\end{array}
\right] (\delta t+\delta z)  \nonumber \\
&&-4\pi L_{0}^{-1}u^{-1}\left[ \allowbreak
\begin{array}{c}
\sin v \\
\cos v
\end{array}
\right] \delta z+4\pi L_{0}^{-1}(W+2u-2\pi )^{-1}\left[
\begin{array}{c}
\sin (v-W) \\
\cos (v-W)
\end{array}
\right] (\delta t+\delta z-\delta V)  \nonumber \\
&&  \label{small-3}
\end{eqnarray}
and
\begin{eqnarray}
\left[
\begin{array}{c}
S_{-} \\
C_{-}
\end{array}
\right] &=&\{\ln [(4\pi L_{0}^{-1})^{2}Q_{+}]+\gamma _{\text{E}}-\ln
(2u)\}\left[
\begin{array}{c}
\sin v \\
\cos v
\end{array}
\right] -\left[
\begin{array}{cc}
-\cos v & \sin v \\
\sin v & \cos v
\end{array}
\right] \left[
\begin{array}{c}
\func{Si}W \\
\func{Ci}W
\end{array}
\right]  \nonumber \\
&&+4\pi L_{0}^{-1}\{\ln [(4\pi L_{0}^{-1})^{2}Q_{+}]+\gamma _{\text{E}}-\ln
(2u)\}\left[
\begin{array}{c}
\cos v \\
-\sin v
\end{array}
\right] (\delta t-\delta z)  \nonumber \\
&&-4\pi L_{0}^{-1}\left[
\begin{array}{cc}
\sin v & \cos v \\
\cos v & -\sin v
\end{array}
\right] \left[
\begin{array}{c}
\func{Si}W \\
\func{Ci}W
\end{array}
\right] (\delta t-\delta z)  \nonumber \\
&&-4\pi L_{0}^{-1}W^{-1}\left[ \allowbreak
\begin{array}{c}
\sin (v-W) \\
\cos (v-W)
\end{array}
\right] (\delta t-\delta z-\delta V)\text{.}  \label{small-4}
\end{eqnarray}
Note here that $\func{Ci}x=\ln x+\gamma _{\text{E}}+O(x^{2})$ for small
values of $x$, where $\gamma _{\text{E}}\approx 0.5772$ is Euler's constant,
while $\func{Si}x=x+O(x^{3})$.

If the terms which tend to zero as $\delta x^{a}\rightarrow 0$ in (\ref
{small-1})-(\ref{small-4}) are neglected, and the resulting leading-order
expansions substituted into (\ref{metperts}), the metric perturbations $%
h_{ab}$ take on the form shown in equation (\ref{h-near}).

Also, given that
\begin{eqnarray}
\partial (\delta V)/\partial (\delta x^{a}) &=&(u-\pi +W-\sin W)^{-1}
\nonumber \\
&&\times [u-\pi +W,\{\cos (v-W)-\cos v\}\,\mathbf{\hat{x}}+\{\sin (v-W)-\sin
v\}\,\mathbf{\hat{y}}+(\pi -u)\,\mathbf{\hat{z}}]_{a}  \nonumber \\
&&
\end{eqnarray}
the terms in the small-distance expansions of the derivatives of $I_{\pm }$,
$S_{\pm }$ and $C_{\pm }$ which do not vanish in the limit as $\delta
x^{a}\rightarrow 0$ can be read off directly from equations (\ref{small-1})-(%
\ref{small-4}):
\begin{equation}
\partial _{a}I_{+}=4\pi L_{0}^{-1}\tfrac{1}{(W+2u-2\pi )(u-\pi +W-\sin W)}%
\,D_{a}-4\pi L_{0}^{-1}u^{-1}[0,\mathbf{\hat{z}}]_{a}  \label{deriv-1}
\end{equation}
\begin{equation}
\partial _{a}I_{-}=\partial _{a}\Lambda +4\pi L_{0}^{-1}\tfrac{1}{W(u-\pi
+W-\sin W)}\,E_{a}  \label{deriv-2}
\end{equation}
\begin{eqnarray}
\left[
\begin{array}{c}
\partial _{a}S_{+} \\
\partial _{a}C_{+}
\end{array}
\right] &=&4\pi L_{0}^{-1}\left[
\begin{array}{cc}
\sin (2u+v) & \cos (2u+v) \\
\cos (2u+v) & -\sin (2u+v)
\end{array}
\right] \left[
\begin{array}{c}
\func{Si}\left( W+2u-2\pi \right) -\func{Si}\left( 2u\right) \\
\func{Ci}\left( W+2u-2\pi \right) -\func{Ci}\left( 2u\right)
\end{array}
\right] [1,\mathbf{\hat{z}}]_{a}  \nonumber \\
&&-4\pi L_{0}^{-1}u^{-1}\left[
\begin{array}{c}
\sin v \\
\cos v
\end{array}
\right] [0,\mathbf{\hat{z}}]_{a}+4\pi L_{0}^{-1}\tfrac{1}{(W+2u-2\pi )(u-\pi
+W-\sin W)}\left[
\begin{array}{c}
\sin (v-W) \\
\cos (v-W)
\end{array}
\right] \,D_{a}  \nonumber \\
&&  \label{deriv-3}
\end{eqnarray}
and
\begin{eqnarray}
\left[
\begin{array}{c}
\partial _{a}S_{-} \\
\partial _{a}C_{-}
\end{array}
\right] &=&\partial _{a}\Lambda \left[
\begin{array}{c}
\sin v \\
\cos v
\end{array}
\right] +4\pi L_{0}^{-1}\{\Lambda +\gamma _{\text{E}}-\ln (2u)\}\left[
\begin{array}{c}
\cos v \\
-\sin v
\end{array}
\right] [1,-\mathbf{\hat{z}}]_{a}  \nonumber \\
&&+4\pi L_{0}^{-1}\partial _{a}\Lambda \left[
\begin{array}{c}
\cos v \\
-\sin v
\end{array}
\right] (\delta t-\delta z)-4\pi L_{0}^{-1}\left[
\begin{array}{cc}
\sin v & \cos v \\
\cos v & -\sin v
\end{array}
\right] \left[
\begin{array}{c}
\func{Si}W \\
\func{Ci}W
\end{array}
\right] [1,-\mathbf{\hat{z}}]_{a}  \nonumber \\
&&+4\pi L_{0}^{-1}\tfrac{1}{W(u-\pi +W-\sin W)}\left[
\begin{array}{c}
\sin (v-W) \\
\cos (v-W)
\end{array}
\right] \,E_{a}\text{.}  \label{deriv-4}
\end{eqnarray}
where $\Lambda =\ln [(4\pi L_{0}^{-1})^{2}Q_{+}]$,
\begin{equation}
D_{a}\equiv [-\sin W,-\{\cos (v-W)-\cos v\}\,\mathbf{\hat{x}}-\{\sin
(v-W)-\sin v\}\,\mathbf{\hat{y}}+(2u-2\pi +W-\sin W)\,\mathbf{\hat{z}}]_{a}
\end{equation}
and
\begin{equation}
E_{a}\equiv [\sin W,\{\cos (v-W)-\cos v\}\,\mathbf{\hat{x}}+\{\sin
(v-W)-\sin v\}\,\mathbf{\hat{y}}+(W-\sin W)\,\mathbf{\hat{z}}]_{a}\text{.}
\end{equation}

Again, the expressions (\ref{ht-near})-(\ref{hz-near}) for the spacetime
derivatives of $h_{ab}$ at points on the null surface $F_{0}$ and close to
the worldsheet of the loop follow directly by substituting the leading-order
expansions (\ref{deriv-1})-(\ref{deriv-4}) into (\ref{metperts}).

\section{Deriving the Equations of Motion}

If the position function $X^{a}$ of the evaporating loop is assumed to have
the form (\ref{evap2}) then
\begin{equation}
X_{(0)}^{a}(u,v)=\tfrac{1}{4\pi }L_{0}[u+v,\,\mathbf{r}(u,v)]^{a}
\label{X0-new}
\end{equation}
with
\begin{equation}
\,\mathbf{r}(u,v)=\cos (v)\,\mathbf{\hat{x}}+\sin (v)\,\mathbf{\hat{y}}+(|u|-%
\tfrac{1}{2}\pi )\,\mathbf{\hat{z}}\text{,}
\end{equation}
and the first-order perturbation $\delta X^{a}$ is as given in (\ref
{pos-first2}), namely
\begin{equation}
\delta X^{a}(u,v)=-\tfrac{1}{4\pi }L_{0}\kappa \mu [\tfrac{1}{2}%
(u+v)^{2},(u+v)\,\mathbf{r}(u,v)]^{a}+\tfrac{1}{4\pi }L_{0}\mu [-F,G\{\cos
(v)\,\mathbf{\hat{x}}+\sin (v)\,\mathbf{\hat{y}}\}\mathbf{+}A\,\mathbf{\hat{z%
}}]^{a}  \label{delX-new}
\end{equation}
The first derivatives of $X_{(0)}^{a}$ are therefore:
\begin{equation}
X_{(0)}^{a},_{u}=\tfrac{1}{4\pi }L_{0}[1,s\,\mathbf{\hat{z}}]^{a}\qquad
\text{and\qquad }X_{(0)}^{a},_{v}=\tfrac{1}{4\pi }L_{0}[1,-\sin (v)\,\mathbf{%
\hat{x}}+\cos (v)\,\mathbf{\hat{y}}]^{a}\text{,}
\end{equation}
with $s=\,$sgn$(u)$, while the only non-zero component of $X_{(0)}^{a},_{AB}$
is\footnote{%
The claim that $X_{(0)}^{a},_{uu}=0$ is true away from the kink points at $%
u=0$ and $u=\pi $, but it can be seen from the expression for $%
X_{(0)}^{a},_{u}$ that the $z$-component of $X_{(0)}^{a},_{uu}$ contains a
distributional term at the kinks. The possibility of pathological behavior
at the kinks will be ignored here, as the ramifications of the singularities
at the kinks have already been discussed in Paper I. It seems unlikely that
the dynamical role of the kinks can be clarified without developing a
solution of the Abelian Higgs field equations as a finite-thickness source
for the ACO loop.}:
\begin{equation}
X_{(0)}^{a},_{vv}=\tfrac{1}{4\pi }L_{0}[0,-\cos (v)\,\mathbf{\hat{x}}-\sin
(v)\,\mathbf{\hat{y}}]^{a}\text{.}
\end{equation}

Similarly, the first derivatives of $\delta X^{a}$ are:
\begin{equation}
\delta X^{a},_{u}=-\tfrac{1}{4\pi }L_{0}\kappa \mu [u+v,\,\mathbf{r}%
(u,v)+s(u+v)\hspace{0in}\,\mathbf{\hat{z}}]^{a}+\tfrac{1}{4\pi }L_{0}\mu
[-F_{u},G_{u}\{\cos (v)\,\mathbf{\hat{x}}+\sin (v)\,\mathbf{\hat{y}}\}%
\mathbf{+}A_{u}\,\mathbf{\hat{z}}]^{a}\text{,}
\end{equation}
\begin{eqnarray}
\delta X^{a},_{v} &=&-\tfrac{1}{4\pi }L_{0}\kappa \mu [u+v,\,\mathbf{r}%
(u,v)]^{a}+\tfrac{1}{4\pi }L_{0}\mu [-F_{v},G_{v}\{\cos (v)\,\mathbf{\hat{x}}%
+\sin (v)\,\mathbf{\hat{y}}\}\mathbf{+}A_{v}\,\mathbf{\hat{z}}]^{a}
\nonumber \\
&&+\tfrac{1}{4\pi }L_{0}\mu \{G-\kappa (u+v)\}[0,-\sin (v)\,\mathbf{\hat{x}}%
+\cos (v)\,\mathbf{\hat{y}}]^{a}
\end{eqnarray}
and the only second derivative of $\delta X^{a}$ contributing to the
first-order equation of motion is:
\begin{eqnarray}
\delta X^{a},_{uv} &=&-\tfrac{1}{4\pi }L_{0}\kappa \mu [1,s\hspace{0in}\,%
\mathbf{\hat{z}}]^{a}+\tfrac{1}{4\pi }L_{0}\mu (G_{u}-\kappa )[0,-\sin (v)\,%
\mathbf{\hat{x}}+\cos (v)\,\mathbf{\hat{y}}]^{a}  \nonumber \\
&&+\tfrac{1}{4\pi }L_{0}\mu [-F_{uv},G_{uv}\{\cos (v)\,\mathbf{\hat{x}}+\sin
(v)\,\mathbf{\hat{y}}\}\mathbf{+}A_{uv}\,\mathbf{\hat{z}}]^{a}\text{.}
\end{eqnarray}

The 4-dimensional tangent space at the point $x^{a}=X^{a}(u,v)$ is spanned
by the vectors $X_{(0)}^{a},_{u}$ and $X_{(0)}^{a},_{v}$, plus
\begin{equation}
M^{a}(s,v)=[1,-\sin (v)\,\mathbf{\hat{x}}+\cos (v)\,\mathbf{\hat{y}}+s%
\hspace{0in}\,\mathbf{\hat{z}}]^{a}\text{\qquad and\qquad }N^{a}(v)=[0,\cos
(v)\,\mathbf{\hat{x}}+\sin (v)\,\mathbf{\hat{y}}]^{a}\text{.}
\end{equation}
In calculating the various contributions to the first-order equation of
motion (\ref{eqmo-first}), it is useful to note that all the terms are
transvected with $q_{d}^{c}$, where $q_{ab}=-\bar{\Psi}_{ab}(s,v)$ is the
projection operator orthogonal to the worldsheet at $X^{a}(u,v)$, so that
\begin{equation}
q_{d}^{c}X_{(0)}^{d},_{u}=q_{d}^{c}X_{(0)}^{d},_{v}=0\text{,\qquad }%
q_{d}^{c}M^{d}=M^{c}\text{\qquad and\qquad }q_{d}^{c}N^{d}=N^{c}\text{.}
\end{equation}
Other helpful identities involving $q_{d}^{c}$ include:
\begin{equation}
q_{d}^{c}[0,-\sin (v)\,\mathbf{\hat{x}}+\cos (v)\,\mathbf{\hat{y}}]^{d}=M^{c}%
\text{,\qquad }q_{d}^{c}[1,\mathbf{0}]^{d}=-M^{c}\qquad \text{and\qquad }%
q_{d}^{c}[0,s\hspace{0in}\,\mathbf{\hat{z}}]^{d}=M^{c}\text{.}
\end{equation}

Thus, recalling that the only non-zero component of $\gamma ^{AB}$ (at
zeroth order in $\mu $) is $\gamma ^{uv}=(\tfrac{1}{4\pi }L_{0})^{-2}$, the
first term in (\ref{eqmo-first}) is:
\begin{eqnarray}
q_{d}^{c}\gamma ^{CD}\delta X^{d},_{CD} &=&2(\tfrac{1}{4\pi }%
L_{0})^{-2}q_{d}^{c}\delta X^{d},_{uv}  \nonumber \\
&=&2\mu (\tfrac{1}{4\pi }L_{0})^{-1}(F_{uv}+sA_{uv}+G_{u}-\kappa
)\,M^{c}+2\mu (\tfrac{1}{4\pi }L_{0})^{-1}G_{uv}\,N^{c}\text{.}  \nonumber \\
&&
\end{eqnarray}
The second term in (\ref{eqmo-first}) is identically zero by virtue of the
zeroth-order equation of motion, while the third term is:
\begin{eqnarray}
-2q_{d}^{c}\gamma ^{AC}\gamma ^{BD}\eta _{ab}X_{(0)}^{a},_{A}\delta
X^{b},_{B}X_{(0)}^{d},_{CD} &=&-2(\tfrac{1}{4\pi }%
L_{0})^{-4}q_{d}^{c}X_{(0)}^{d},_{vv}(\eta _{ab}X_{(0)}^{a},_{u}\delta
X^{b},_{u})  \nonumber \\
&=&2\mu (\tfrac{1}{4\pi }L_{0})^{-1}[\kappa s(|u|-\tfrac{1}{2}\pi
)-(F_{u}+sA_{u})]N^{c}\text{.}  \nonumber \\
&&
\end{eqnarray}

The remaining terms in (\ref{eqmo-first}) are more complicated, as they
involve $h_{ab}$ or its derivatives. However, evaluation of these terms is
simplified by making use of the following table, which represents the inner
products of the terms in the first column with those in the first row in the
form $(k_{1},k_{2},k_{3},k_{4})$, which in turn is shorthand for $%
k_{1}X_{(0)a},_{u}+k_{2}X_{(0)a},_{v}+k_{3}M_{a}+k_{4}N_{a}$. Also, $c_{2}$
and $s_{2}$ stand in for $\cos 2u$ and $\sin 2u$ respectively, while $c_{W}$
and $s_{W}$ are shorthand for $\cos W$ and $\sin W$.
\[
\begin{tabular}{||c|c|c||}
\hline\hline
& $X_{(0)}^{b},_{u}$ & $X_{(0)}^{b},_{v}$ \\ \hline
$\bar{\Psi}_{ab}(s,v)$ & $\mathbf{0}$ & $\mathbf{0}$ \\ \hline
$\bar{\Psi}_{ab}(-s,v)$ & $(0,4,-2,0)$ & $\mathbf{0}$ \\ \hline
$\Phi _{ab}(s,v)$ & $\mathbf{0}$ & $(-2,0,1,0)$ \\ \hline
$\Phi _{ab}(-s,2u+v)$ & $(-2,0,2,0)c_{2}+(0,0,0,-2)s_{2}$ & $%
(-2,-2,3,0)c_{2}+(0,0,0,-1)s_{2}$ \\ \hline
$\Omega _{ab}(s,v)$ & $\mathbf{0}$ & $(0,0,0,-1)$ \\ \hline
$\Omega _{ab}(-s,2u+v)$ & $(2,0,-2,0)s_{2}+(0,0,0,-2)c_{2}$ & $%
(2,2,-3,0)s_{2}+(0,0,0,-1)c_{2}$ \\ \hline
$\Pi _{ab}(s)$ & $\mathbf{0}$ & $(2,0,-1,0)$ \\ \hline
$\Pi _{ab}(-s)$ & $(2,4,-2,0)$ & $(2,2,-3,0)$ \\ \hline
$\bar{\Psi}_{ab}(s,v-W)$ & $\mathbf{0}$ & $%
(2,0,-1,0)(1-c_{W})+(0,0,0,1)s_{W} $ \\ \hline
$\bar{\Psi}_{ab}(-s,v-W)$ & $(2(1-c_{W}),4,-2(2-c_{W}),2s_{W})$ & $%
(2,2,-3,0)(1-c_{W})+(0,0,0,1)s_{W}$ \\ \hline\hline
\end{tabular}
\]
Furthermore, the vectors $X_{(0)}^{a},_{u}$ and $X_{(0)}^{a},_{v}$ are both
null and orthogonal to $M^{a}$ and $N^{a}$, with $X_{(0)},_{u}\cdot
X_{(0)},_{v}=(\tfrac{1}{4\pi }L_{0})^{2}$.

It should of course be remembered that the expansions (\ref{h-near}) and (%
\ref{ht-near})-(\ref{hz-near}) were developed for the case $u>0$ only, and
so $s=+1$ in the expressions in this table. Because the term involving $\ln
[(4\pi L_{0}^{-1})^{2}Q_{+}]\,$in the equation for $h_{ab}$ is proportional
to $\bar{\Psi}_{ab}(+1,v)$, the term containing $%
h_{ab}X_{(0)}^{a},_{A}X_{(0)}^{b},_{B}$ in (\ref{eqmo-first}) is free of
logarithmic singularities. Similarly, the singular terms proportional to $%
\bar{\Psi}_{ab}(+1,v)$ in (\ref{ht-near})-(\ref{hz-near}) do not contribute
to the term $X_{(0)}^{b},_{B}(h_{bd},_{a}-\tfrac{1}{2}h_{ab},_{d})$ on the
right-hand side of (\ref{eqmo-first}). However, there remain singular terms
proportional to $\Omega _{ab}(+1,v)$ in the derivatives of $h_{ab}$ whose
contributions to the first-order equation of motion need to be examined more
carefully.

Now, the fourth term in the equation of motion is:
\begin{equation}
-q_{d}^{c}\gamma ^{AC}\gamma
^{BD}h_{ab}X_{(0)}^{a},_{A}X_{(0)}^{b},_{B}X_{(0)}^{d},_{CD}=-(\tfrac{1}{%
4\pi }%
L_{0})^{-4}q_{d}^{c}X_{(0)}^{d},_{vv}(h_{ab}X_{(0)}^{a},_{u}X_{(0)}^{b},_{u})
\end{equation}
where the only non-zero term in $h_{ab}X_{(0)}^{a},_{u}X_{(0)}^{b},_{u}$ is
proportional to the coefficient of $\Pi _{ab}(-1)$:
\begin{equation}
h_{ab}X_{(0)}^{a},_{u}X_{(0)}^{b},_{u}=16\mu (\tfrac{1}{4\pi }L_{0})^{2}[\ln
\left( W+2u-2\pi \right) -\ln (2u)]\text{,}
\end{equation}
and so
\begin{equation}
-q_{d}^{c}\gamma ^{AC}\gamma
^{BD}h_{ab}X_{(0)}^{a},_{A}X_{(0)}^{b},_{B}X_{(0)}^{d},_{CD}=16\mu (\tfrac{1%
}{4\pi }L_{0})^{-1}[\ln \left( W+2u-2\pi \right) -\ln (2u)]N^{c}\text{.}
\end{equation}

Furthermore, the right-hand expression in (\ref{eqmo-first}) can be broken
into two parts:
\begin{equation}
-q^{cd}\gamma ^{AB}X_{(0)}^{a},_{A}X_{(0)}^{b},_{B}h_{bd},_{a}=-(\tfrac{1}{%
4\pi }%
L_{0})^{-2}q^{cd}(X_{(0)}^{a},_{u}X_{(0)}^{b},_{v}+X_{(0)}^{a},_{v}X_{(0)}^{b},_{u})h_{bd},_{a}
\end{equation}
and
\begin{equation}
\tfrac{1}{2}q^{cd}\gamma ^{AB}X_{(0)}^{a},_{A}X_{(0)}^{b},_{B}h_{ab},_{d}=(%
\tfrac{1}{4\pi }L_{0})^{-2}q^{cd}X_{(0)}^{a},_{u}X_{(0)}^{b},_{v}h_{ab},_{d}%
\text{.}
\end{equation}

Here, $q^{cd}X_{(0)}^{a},_{u}X_{(0)}^{b},_{v}h_{bd},_{a}$ involves terms of
the form
\begin{eqnarray}
-q^{cd}X_{(0)}^{b},_{v}\partial _{t}h_{bd} &=&4\mu (\func{Si}W\,)M^{c}+4\mu
(\delta t-\delta z)\,(\partial _{t}\Lambda )N^{c}+4\mu [\Lambda +\gamma _{%
\text{E}}-\func{Ci}W-\ln (2u)]N^{c}  \nonumber \\
&&+4\mu \Sigma (u)[-(3\cos 2u)M^{c}+(\sin 2u)N^{c}]+4\mu K(u)[(3\sin
2u)M^{c}+(\cos 2u)N^{c}]  \nonumber \\
&&+4\mu \tfrac{\sin W}{u-\pi +W-\sin W}\tfrac{1}{W}[(1-\cos W)M^{c}-(\sin
W)N^{c}]  \nonumber \\
&&-4\mu \tfrac{\sin W}{u-\pi +W-\sin W}\tfrac{1}{W+2u-2\pi }\,[3(1-\cos
W)M^{c}-(\sin W)N^{c}]
\end{eqnarray}
and
\begin{eqnarray}
-q^{cd}X_{(0)}^{b},_{v}\partial _{z}h_{bd} &=&-4\mu (\func{Si}W\,)M^{c}+4\mu
(\delta t-\delta z)\,(\partial _{z}\Lambda )N^{c}-4\mu [\Lambda +\gamma _{%
\text{E}}-\func{Ci}W-\ln (2u)]N^{c}  \nonumber \\
&&+4\mu \Sigma (u)[-(3\cos 2u)M^{c}+(\sin 2u)N^{c}]+4\mu K(u)[(3\sin
2u)M^{c}+(\cos 2u)N^{c}]  \nonumber \\
&&+4\mu \tfrac{W-\sin W}{u-\pi +W-\sin W}\tfrac{1}{W}[(1-\cos W)M^{c}-(\sin
W)N^{c}]  \nonumber \\
&&+4\mu \tfrac{2u-2\pi +W-\sin W}{u-\pi +W-\sin W}\tfrac{1}{W+2u-2\pi }%
\,[3(1-\cos W)M^{c}-(\sin W)N^{c}]\text{.}
\end{eqnarray}

So
\begin{eqnarray}
-(\tfrac{1}{4\pi }%
L_{0})^{-2}q^{cd}X_{(0)}^{a},_{u}X_{(0)}^{b},_{v}h_{bd},_{a} &=&-(\tfrac{1}{%
4\pi }L_{0})^{-1}q^{cd}X_{(0)}^{b},_{v}(\partial _{t}h_{bd}+\partial
_{z}h_{bd})  \nonumber \\
&=&4\mu (\tfrac{1}{4\pi }L_{0})^{-1}\{(\delta t-\delta z)[(\partial
_{t}\Lambda )+(\partial _{z}\Lambda )]N^{c}  \nonumber \\
&&+2\Sigma (u)[-(3\cos 2u)M^{c}+(\sin 2u)N^{c}]  \nonumber \\
&&+2K(u)[(3\sin 2u)M^{c}+(\cos 2u)N^{c}]  \nonumber \\
&&+\tfrac{1}{u-\pi +W-\sin W}[(1-\cos W)M^{c}-(\sin W)N^{c}]  \nonumber \\
&&+\tfrac{2u-2\pi +W-2\sin W}{u-\pi +W-\sin W}\tfrac{1}{W+2u-2\pi }%
\,[3(1-\cos W)M^{c}-(\sin W)N^{c}]\}\text{,}  \nonumber \\
&&  \label{first-part1}
\end{eqnarray}
where
\begin{eqnarray}
\,(\,\partial _{t}\Lambda )+(\partial _{z}\Lambda ) &=&Q_{+}^{-1}\partial
_{t}(\bar{\Psi}_{ab}\delta x^{a}\delta x^{b})+Q_{+}^{-1}\partial _{z}(\bar{%
\Psi}_{ab}\delta x^{a}\delta x^{b})  \nonumber \\
&=&2Q_{+}^{-1}(\bar{\Psi}_{at}+\bar{\Psi}_{az})\delta x^{a}=0\text{,}
\end{eqnarray}
with $\bar{\Psi}_{ab}$ here denoting $\bar{\Psi}_{ab}(+1,v)$.

Similarly, $q^{cd}X_{(0)}^{a},_{v}X_{(0)}^{b},_{u}h_{bd},_{a}$ is a sum of
three terms proportional to
\begin{eqnarray}
-q^{cd}X_{(0)}^{b},_{u}\partial _{t}h_{bd} &=&4\mu \Sigma (u)[-(2\cos
2u)M^{c}+(2\sin 2u)N^{c}]+4\mu K(u)[(2\sin 2u)M^{c}+(2\cos 2u)N^{c}]
\nonumber \\
&&-8\mu \tfrac{\sin W}{u-\pi +W-\sin W}\tfrac{1}{W+2u-2\pi }\,[(2-\cos
W)M^{c}-(\sin W)N^{c}]\text{,}
\end{eqnarray}
\begin{equation}
-q^{cd}X_{(0)}^{b},_{u}\partial _{x}h_{bd}=-8\mu \tfrac{\cos (v-W)-\cos v}{%
u-\pi +W-\sin W}\tfrac{1}{W+2u-2\pi }\,[(2-\cos W)M^{c}-(\sin W)N^{c}]
\end{equation}
and
\begin{equation}
-q^{cd}X_{(0)}^{b},_{u}\partial _{y}h_{bd}=-8\mu \tfrac{\sin (v-W)-\sin v}{%
u-\pi +W-\sin W}\tfrac{1}{W+2u-2\pi }\,[(2-\cos W)M^{c}-(\sin W)N^{c}]\text{.%
}
\end{equation}
So
\begin{eqnarray}
-(\tfrac{1}{4\pi }%
L_{0})^{-2}q^{cd}X_{(0)}^{a},_{v}X_{(0)}^{b},_{u}h_{bd},_{a} &=&-(\tfrac{1}{%
4\pi }L_{0})^{-1}q^{cd}X_{(0)}^{b},_{u}[\partial _{t}h_{bd}-(\sin v)\partial
_{x}h_{bd}+(\cos v)\partial _{y}h_{bd}]  \nonumber \\
&=&4\mu (\tfrac{1}{4\pi }L_{0})^{-1}\Sigma (u)[-(2\cos 2u)M^{c}+(2\sin
2u)N^{c}]  \nonumber \\
&&+4\mu (\tfrac{1}{4\pi }L_{0})^{-1}K(u)[(2\sin 2u)M^{c}+(2\cos 2u)N^{c}]%
\text{.}  \label{first-part2}
\end{eqnarray}
Adding together (\ref{first-part1}) and (\ref{first-part2}) gives equation (%
\ref{em4}).

The various terms contributing to the second part of the last term in (\ref
{eqmo-first}) are:
\begin{equation}
X_{(0)}^{a},_{u}X_{(0)}^{b},_{v}\partial _{t}h_{ab}=8\mu (\tfrac{1}{4\pi }%
L_{0})[K(u)\sin 2u-\Sigma (u)\cos 2u]-8\mu (\tfrac{1}{4\pi }L_{0})\tfrac{%
\sin W}{u-\pi +W-\sin W}\tfrac{1-\cos W}{W+2u-2\pi }\text{,}
\end{equation}
\begin{equation}
X_{(0)}^{a},_{u}X_{(0)}^{b},_{v}\partial _{x}h_{ab}=-8\mu (\tfrac{1}{4\pi }%
L_{0})\tfrac{\cos (v-W)-\cos v}{u-\pi +W-\sin W}\tfrac{1-\cos W}{W+2u-2\pi }%
\text{,}
\end{equation}
\begin{equation}
X_{(0)}^{a},_{u}X_{(0)}^{b},_{v}\partial _{y}h_{ab}=-8\mu (\tfrac{1}{4\pi }%
L_{0})\tfrac{\sin (v-W)-\sin v}{u-\pi +W-\sin W}\tfrac{1-\cos W}{W+2u-2\pi }
\end{equation}
and
\begin{equation}
X_{(0)}^{a},_{u}X_{(0)}^{b},_{v}\partial _{z}h_{ab}=8\mu (\tfrac{1}{4\pi }%
L_{0})[K(u)\sin 2u-\Sigma (u)\cos 2u]+8\mu (\tfrac{1}{4\pi }L_{0})\tfrac{%
2u-2\pi +W-\sin W}{u-\pi +W-\sin W}\tfrac{1-\cos W}{W+2u-2\pi }\text{.}
\end{equation}

Now, if $V_{d}$ is any contravariant vector and $s=+1$, then
\begin{equation}
q^{cd}V_{d}=-[V_{t}-(\sin v)V_{x}+(\cos v)V_{y}+V_{z}]\,M^{c}-[(\cos
v)V_{x}+(\sin v)V_{y}]\,N^{c}\text{.}
\end{equation}
Hence, the second part of the last term in (\ref{eqmo-first}) becomes
\begin{eqnarray}
(\tfrac{1}{4\pi }%
L_{0})^{-2}q^{cd}X_{(0)}^{a},_{u}X_{(0)}^{b},_{v}h_{ab},_{d} &=&-16\mu (%
\tfrac{1}{4\pi }L_{0})^{-1}[K(u)\sin 2u-\Sigma (u)\cos 2u]\,M^{c}  \nonumber
\\
&&-8\mu (\tfrac{1}{4\pi }L_{0})^{-1}\tfrac{W+2u-2\pi -\sin W}{u-\pi +W-\sin W%
}\,\tfrac{1-\cos W}{W+2u-2\pi }M^{c}  \nonumber \\
&&-8\mu (\tfrac{1}{4\pi }L_{0})^{-1}\tfrac{1-\cos W}{u-\pi +W-\sin W}\tfrac{%
1-\cos W}{W+2u-2\pi }N^{c}\text{.}
\end{eqnarray}

\section{The Intrinsic 2-Metric and the Length of the Evaporating ACO\ Loop}

\subsection{The intrinsic 2-metric $\gamma _{AB}$}

To linear order in $\mu $, the intrinsic 2-metric $\gamma
_{AB}=g_{ab}X^{a},_{A}X^{b},_{B}$ consists of three basic terms:
\begin{equation}
\gamma _{AB}=X_{(0)},_{A}\cdot X_{(0)},_{B}+2X_{(0)},_{(A}\cdot \delta
X,_{B)}+h_{ab}X_{(0)}^{a},_{A}X_{(0)}^{b},_{B}\text{,}
\end{equation}
with $X_{(0)}^{a}$ and $\delta X^{a}$ as given in (\ref{X0-new}) and (\ref
{delX-new}). In particular, $X_{(0)},_{u}\cdot
X_{(0)},_{u}=X_{(0)},_{u}\cdot X_{(0)},_{u}=0$ and $X_{(0)},_{u}\cdot
X_{(0)},_{v}=(\tfrac{1}{4\pi }L_{0})^{2}$, while
\begin{equation}
2X_{(0)},_{u}\cdot \delta X,_{u}=2\mu (\tfrac{1}{4\pi }L_{0})^{2}[\kappa (u-%
\tfrac{1}{2}\pi )-(F_{u}+A_{u})]\text{,}
\end{equation}
\begin{equation}
2X_{(0)},_{(u}\cdot \delta X,_{v)}=\mu (\tfrac{1}{4\pi }L_{0})^{2}[\kappa (u-%
\tfrac{1}{2}\pi )-2\kappa (u+v)-F_{u}-(F_{v}+A_{v})]  \label{del-dot_uv}
\end{equation}
and
\begin{equation}
2X_{(0)},_{v}\cdot \delta X,_{v}=-2\mu (\tfrac{1}{4\pi }L_{0})^{2}(G+F_{v})
\end{equation}
for $u>0$.

One of the components of the term proportional to $h_{ab}$, namely
\begin{equation}
h_{ab}X_{(0)}^{a},_{u}X_{(0)}^{b},_{u}=16\mu (\tfrac{1}{4\pi }L_{0})^{2}[\ln
\left( W+2u-2\pi \right) -\ln (2u)]\text{,}
\end{equation}
has already been evaluated in Appendix C. The other two components are:
\begin{equation}
h_{ab}X_{(0)}^{a},_{u}X_{(0)}^{b},_{v}=-8\mu (\tfrac{1}{4\pi }%
L_{0})^{2}\{K(u)\cos 2u+\Sigma (u)\sin 2u-[\ln \left( W+2u-2\pi \right) -\ln
(2u)]\}
\end{equation}
and
\begin{eqnarray}
h_{ab}X_{(0)}^{a},_{v}X_{(0)}^{b},_{v} &=&-8\mu (\tfrac{1}{4\pi }%
L_{0})^{2}\{\gamma _{\text{E}}-\func{Ci}W+K(u)\cos 2u+\Sigma (u)\sin 2u+\ln W
\nonumber \\
&&-[\ln \left( W+2u-2\pi \right) -\ln (2u)]\}\text{.}
\end{eqnarray}

So
\begin{equation}
\gamma _{uu}=(\tfrac{1}{4\pi }L_{0})^{2}e^{-2\kappa \mu (u+v)}[-2\mu
\{F_{u}+A_{u}-\kappa (u-\tfrac{1}{2}\pi )\}+16\mu \{\ln (W+2u-2\pi )-\ln
(2u)\}]\text{,}  \label{gamma-uu}
\end{equation}
\begin{eqnarray}
\gamma _{uv} &=&(\tfrac{1}{4\pi }L_{0})^{2}e^{-2\kappa \mu (u+v)}[1-\mu
\{F_{u}+F_{v}+A_{v}-\kappa (u-\tfrac{1}{2}\pi )\}  \nonumber \\
&&-8\mu \{-\ln (W+2u-2\pi )+\ln (2u)+K(u)\cos 2u+\Sigma (u)\sin 2u\}]
\nonumber \\
&&  \label{gamma-uv}
\end{eqnarray}
and
\begin{eqnarray}
\gamma _{vv} &=&(\tfrac{1}{4\pi }L_{0})^{2}e^{-2\kappa \mu (u+v)}[-2\mu
(G+F_{v})  \nonumber \\
&&-8\mu \{\gamma _{\text{E}}-\func{Ci}W+\ln W-\ln (W+2u-2\pi )+\ln
(2u)+K(u)\cos 2u+\Sigma (u)\sin 2u\}]  \nonumber \\
&&  \label{gamma-vv}
\end{eqnarray}
to linear order in $\mu $, provided that $\kappa \mu |u+v|\ll 1$. Here, the
factor of $e^{-2\kappa \mu (u+v)}$ inserted into $\gamma _{uv}$ absorbs the
term $-2\kappa (u+v)$ in (\ref{del-dot_uv}), while multiplying $\gamma _{uu}$
and $\gamma _{vv}$ by $e^{-2\kappa \mu (u+v)}$ commits an error of order $%
\mu ^{2}$.

Since $W$ is a function of $u$ alone by virtue of (\ref{u-W}), the
requirement that $\gamma _{AB}$ take on the self-similar form $(\tfrac{1}{%
4\pi }L_{0})^{2}e^{-2\kappa \mu (u+v)}\bar{\gamma}_{AB}(u)$ to linear order
in $\mu $, as explained in Section 4, places three constraints on the
correction functions $F$, $G$ and $A$, namely:
\begin{equation}
F_{uv}+A_{uv}=0\text{,\qquad }F_{uv}+F_{vv}+A_{vv}=0\qquad \text{and\qquad }%
G_{v}+F_{vv}=0\text{.}
\end{equation}

It should be noted that although equations (\ref{gamma-uu})-(\ref{gamma-vv})
were derived on the assumption that $\kappa \mu |u+v|\ll 1$, they do in fact
hold for all values of $u+v$. This can be seen by first taking the corrected
position vector (\ref{evap2}) and using this to calculate $X,_{A}\cdot
X,_{B} $, with the multiplicative scale factor $e^{-2\kappa \mu (u+v)}$
retained but all other terms truncated at order $\mu $. This recovers the
contribution of $X_{(0)},_{A}\cdot X_{(0)},_{B}+2X_{(0)},_{(A}\cdot \delta
X,_{B)}$ to (\ref{gamma-uu})-(\ref{gamma-vv}) as shown.

The contribution of the remaining term $%
h_{ab}X_{(0)}^{a},_{A}X_{(0)}^{b},_{B}$ to $\gamma _{AB}$ -- where $%
X_{(0)}^{a},_{A}$ is given by (\ref{tan-vectors}) -- can be evaluated by
referring back to the discussion of rotating self-similarity in Section 5.1.
As was demonstrated there, the value of $h_{ab}$ at an arbitrary point
\begin{equation}
x^{a}=X_{(0)}^{a}(u,v)\equiv \tfrac{1}{4\pi }L_{0}[t_{L}(1-e^{-\kappa \mu
(u+v)}),e^{-\kappa \mu (u+v)}\mathbf{r}(u,v)]^{a}
\end{equation}
on the worldsheet is related to its value $\bar{h}_{ab}$ at the image point
\begin{equation}
\bar{x}^{a}=\tfrac{1}{4\pi }L_{0}e^{\kappa \mu \psi -\kappa \mu (u+v)}[|%
\mathbf{r}(u,v)|,R(-\psi )\mathbf{r}(u,v)]^{a}
\end{equation}
on the future light cone $F_{0}$ of the origin through the formula
\begin{equation}
h_{ab}=R_{a}^{\;c}(\psi )\,\bar{h}_{cd}\,R_{b}^{\;d}(\psi )\text{,}
\end{equation}
where, from (\ref{image}),
\begin{equation}
\psi =u+v-(\kappa \mu )^{-1}\ln (1+\kappa \mu |\mathbf{r}(u,v)|)\text{.}
\end{equation}

Here, the image point
\begin{equation}
\bar{x}^{a}=\tfrac{1}{4\pi }L_{0}(1+\kappa \mu |\mathbf{r}(u,v)|)^{-1}[|%
\mathbf{r}(u,v)|,R(-\psi )\mathbf{r}(u,v)]^{a}
\end{equation}
is just the point $X_{(0)}^{a}(\bar{u},\bar{v})$ on the worldsheet with $%
\bar{u}=u$ and $\bar{v}=v-\psi \equiv (\kappa \mu )^{-1}\ln (1+\kappa \mu |%
\mathbf{r}(u,v)|)-u$, and it was shown in Section 5.2 that $\kappa \mu |\bar{%
u}+\bar{v}|\ll 1$ for reasonable values of $\mu $. So
\begin{equation}
X_{(0)}^{a},_{A}(u,v)=X_{(0)}^{a},_{A}(\bar{u},\bar{v}+\psi )=e^{-\kappa \mu
(u+v)}R_{b}^{\;a}(\psi )\,X_{(0)}^{b},_{A}(\bar{u},\bar{v})
\end{equation}
to leading order in $\mu $, and therefore
\begin{equation}
h_{ab}X_{(0)}^{a},_{A}X_{(0)}^{b},_{B}|_{(u,v)}=e^{-2\kappa \mu (u+v)}\bar{h}%
_{ab}X_{(0)}^{a},_{A}X_{(0)}^{b},_{B}|_{(\bar{u},\bar{v})}\text{.}
\end{equation}
That is, to linear order in $\mu $ the term $%
h_{ab}X_{(0)}^{a},_{A}X_{(0)}^{b},_{B}$ at a general point on the worldsheet
differs from its value at the image point $\bar{x}^{a}$ only by a
multiplicative factor of $e^{-2\kappa \mu (u+v)}$, and equations (\ref
{gamma-uu})-(\ref{gamma-vv}) are recovered in full.

\subsection{The length of the loop as measured along curves of constant
scale factor}

Equations (\ref{gamma-uu})-(\ref{gamma-vv}) can be used to calculate the
length of the evaporating loop as measured along any curve of constant scale
factor $e^{-\kappa \mu (u+v)}$. The length $L_{\text{csf}}$ of the loop
along the curve $u+v=u_{*}+v_{*}$ is given by
\begin{equation}
L_{\text{csf}}=2\int_{0}^{\pi }(2\gamma _{uv}-\gamma _{uu}-\gamma
_{vv})^{1/2}|_{v=u_{*}+v_{*}-u}\,du\text{,}
\end{equation}
where, to linear order in $\mu $,
\begin{eqnarray}
(2\gamma _{uv}-\gamma _{uu}-\gamma _{vv})^{1/2} &=&\sqrt{2}(\tfrac{1}{4\pi }%
L_{0})e^{-\kappa \mu (u+v)}[1+\tfrac{1}{2}\mu (G+A_{u})+2\mu \{\gamma _{%
\text{E}}-\func{Ci}W  \nonumber \\
&&+\ln (2uW)-\ln (W+2u-2\pi )-K(u)\cos 2u-\Sigma (u)\sin 2u\}]\text{.}
\nonumber \\
&&
\end{eqnarray}

Hence, in view of the expression (\ref{G-2}) for $G$:
\begin{eqnarray}
L_{\text{csf}} &=&2\sqrt{2}(\tfrac{1}{4\pi }L_{0})e^{-\kappa \mu
(u_{*}+v_{*})}\int_{0}^{\pi }[1+\tfrac{1}{2}\mu (\kappa u+C_{1}+A_{u})
\nonumber \\
&&+2\mu \{\gamma _{\text{E}}+\ln W-\func{Ci}W-2K(u)\cos 2u-2\Sigma (u)\sin
2u\}]\,du\text{,}  \nonumber \\
&&  \label{Lcsf}
\end{eqnarray}
where
\begin{eqnarray}
&&\int [\ln W-\func{Ci}W-2K(u)\cos 2u-2\Sigma (u)\sin 2u]\,du  \nonumber \\
&=&W^{-1}(2+\ln W)(1-\cos W)-\tfrac{1}{2}W(1+\ln W)-\tfrac{1}{2}%
W^{-1}(1-\cos 2W)-\tfrac{3}{2}\sin W  \nonumber \\
&&-(u-\pi )\func{Ci}W-\func{Si}W+\func{Si}(2W)-[K(u)\sin 2u-\Sigma (u)\cos
2u]
\end{eqnarray}
and so
\begin{equation}
\int_{0}^{\pi }[\ln W-\func{Ci}W-2K(u)\cos 2u-2\Sigma (u)\sin 2u]\,du=\pi
\ln 2\pi +\pi -\pi \func{Ci}(2\pi )-\func{Si}(4\pi )\text{.}  \label{int-1}
\end{equation}

Furthermore, it follows from (\ref{kappa}) and (\ref{A(0)}) that
\begin{equation}
\int_{0}^{\pi }\kappa u\,du=\tfrac{1}{2}\mu \kappa \pi ^{2}=2\pi \ln 2\pi
+2\pi \gamma _{\text{E}}-2\pi \func{Ci}(2\pi )]  \label{int-2}
\end{equation}
and
\begin{equation}
\int_{0}^{\pi }A_{u}du=A(\pi )-A(0)=24\pi -16\pi \ln 2\pi -4\func{Si}(2\pi )%
\text{.}  \label{int-3}
\end{equation}
Substituting (\ref{int-1}), (\ref{int-2}) and (\ref{int-3}) into (\ref{Lcsf}%
) therefore gives:
\begin{equation}
L_{\text{csf}}=(\tfrac{1}{\sqrt{2}}L_{0})e^{-\kappa \mu (u_{*}+v_{*})}[1+%
\tfrac{1}{2}\mu \{C_{1}+28+6\gamma _{\text{E}}-10\ln 2\pi -6\func{Ci}(2\pi
)-4\pi ^{-1}\func{Si}(2\pi )-4\pi ^{-1}\func{Si}(4\pi )\}]\text{.}
\end{equation}

Note here that $L_{\text{csf}}\rightarrow \tfrac{1}{\sqrt{2}}L_{0}$ in the
flat-space limit $\mu \rightarrow 0$. This is to be expected, as the curves
of constant scale factor are curves of constant $t$ in this limit, and the
speed $|\partial _{t}\mathbf{X}|$ is equal to $1/\sqrt{2}$ at all points on
the unperturbed ACO loop (except at the kink points at $u=0$ and $\pi $,
where it is undefined). The measured length $L_{\text{csf}}=\tfrac{1}{\sqrt{2%
}}L_{0}$ is therefore just the Lorentz-contracted fundamental length $L_{0}$.

If $\mu $ is non-zero then $L_{\text{csf}}$ will deviate from its ACO value $%
\tfrac{1}{\sqrt{2}}L_{0}$ on the curve $u+v=0$ by a term of order $\mu $
unless $C_{1}$ takes on the particular value
\begin{equation}
C_{1}=-28-6\gamma _{\text{E}}+10\ln 2\pi +6\func{Ci}(2\pi )+4\pi ^{-1}\func{%
Si}(2\pi )+4\pi ^{-1}\func{Si}(4\pi )\approx -9.\,\allowbreak 514\,3\text{.}
\end{equation}
There is of course no necessary reason to choose this value for $C_{1}$, but
it bears repeating that different values of $C_{1}$ will lead to different
``dressed'' values of $L_{\text{csf}}$ along any given curve of constant $%
u+v $.

\subsection{Invariant length of the evaporating loop}

Another measure of the length of the loop that is potentially of interest
here is the \emph{invariant length} $L_{\text{I}}$, which is a
gauge-invariant quantity defined at each point $(u_{*},v_{*})$ on the
worldsheet by
\begin{equation}
L_{\text{I}}(u_{*},v_{*})=2\left( \int_{\mathcal{D}(u_{*},v_{*})}\gamma
^{1/2}du\,dv\right) ^{1/2}\text{,}  \label{L-inv}
\end{equation}
where $\mathcal{D}(u_{*},v_{*})$ is the closure of the subset of the
worldsheet that is causally disconnected from the reference point $%
(u_{*},v_{*})$ \cite{Anderson4}. Calculating $L_{\text{I}}$ to order $\mu $
at a general point on the trajectory (\ref{evap2}) of the evaporating ACO
loop is a lengthy process whose details are of marginal importance, so I
will offer only an abbreviated summary here.

In the flat-space limit $\mu =0$ the domain $\mathcal{D}(u_{*},v_{*})$ is
bounded by the null curves $u=u_{*}$ and $v=v_{*}$ which, in view of the
identification of $(u,v)$ with $(u+2\pi k,v-2\pi k)$ for all integers $k$,
intersect at the points $(u_{*},v_{*}+2\pi )\equiv (u_{*}+2\pi ,v_{*})$ and $%
(u_{*},v_{*}-2\pi )\equiv (u_{*}-2\pi ,v_{*})$. The boundary of $\mathcal{D}%
(u_{*},v_{*})$ is therefore completed by the curves
\begin{equation}
v=u+2\pi -u_{*}+v_{*}\text{\qquad and\qquad }v=u-2\pi -u_{*}+v_{*}
\end{equation}
and so
\begin{eqnarray}
\lim_{\mu \rightarrow 0}L_{\text{I}}(u_{*},v_{*}) &=&2\left(
\int_{u_{*}-2\pi }^{u_{*}}\int_{v_{*}}^{u+2\pi -u_{*}+v_{*}}(\tfrac{1}{4\pi }%
L_{0})^{2}dv\,du+\int_{u_{*}}^{u_{*}+2\pi }\int_{u-2\pi
-u_{*}+v_{*}}^{v_{*}}(\tfrac{1}{4\pi }L_{0})^{2}dv\,du\right) ^{1/2}
\nonumber \\
&=&L_{0}
\end{eqnarray}

In order to calculate $L_{\text{I}}$ when $\mu $ is non-zero, it is useful
to first transform to a set of gauge coordinates $(u^{\prime },v^{\prime })$
with the property that the curves $u^{\prime }=u_{*}$ and $v^{\prime }=v_{*}$
are (to order $\mu $) null curves through the reference point $%
(u,v)=(u_{*},v_{*})$. A suitable choice is
\begin{equation}
u^{\prime }=u-\mu (v-v_{*})\mathcal{G}(u)\text{\qquad and\qquad }v^{\prime
}=v-\mu \mathcal{F}(u;u_{*})\text{,}
\end{equation}
where
\begin{eqnarray}
\mathcal{G}(u) &\equiv &G(u)+4\{\gamma _{\text{E}}-\func{Ci}W+\ln W-\ln
(W+2u-2\pi )+\ln (2u)  \nonumber \\
&&+K(u)\cos 2u+\Sigma (u)\sin 2u\}  \nonumber \\
&=&\allowbreak \kappa u+4(\gamma _{\text{E}}-\func{Ci}W+\ln W)+C_{1}
\label{G-int}
\end{eqnarray}
and
\begin{equation}
\mathcal{F}(u;u_{*})\equiv \int_{u_{*}}^{u}[\{F_{u}+A_{u}-\kappa (u-\tfrac{1%
}{2}\pi )\}-8\{\ln (W+2u-2\pi )-\ln (2u)\}]\,du\text{.}  \label{F-int}
\end{equation}
In the primed gauge the 2-metric components (\ref{gamma-uu})-(\ref{gamma-vv}%
) become $\gamma _{u^{\prime }u^{\prime }}=\gamma _{v^{\prime }v^{\prime
}}=0 $ and
\begin{eqnarray}
\gamma ^{1/2} &\equiv &\gamma _{u^{\prime }v^{\prime }}=(\tfrac{1}{4\pi }%
L_{0})^{2}e^{-2\kappa \mu (u+v)}[1-\mu \{F_{u}-\kappa (u-\tfrac{1}{2}\pi
)\}+\mu (v-v_{*})\partial _{u}\mathcal{G}(u)  \nonumber \\
&&-8\mu \{-\ln (W+2u-2\pi )+\ln (2u)+K(u)\cos 2u+\Sigma (u)\sin 2u\}]
\nonumber \\
&&  \label{met-det}
\end{eqnarray}
to linear order in $\mu $.\footnote{%
Strictly speaking, the functions of $u$ on the right-hand side of (\ref
{G-int}), (\ref{F-int}) and (\ref{met-det}) should be the half-range
periodic extensions of the expressions shown. However, all three equations
are accurate if $u$ and $u_{*}$ both lie in $[0,\pi ]$.}

Recalling again that the point $(u,v)$ is identified with $(u\pm 2\pi ,v\mp
2\pi )$, the curves $u^{\prime }=u_{*}$ and $v^{\prime }=v_{*}$ intersect at
the points
\begin{equation}
(u,v)=(u_{*}-\Delta u,v_{*}+2\pi -\Delta v)\equiv (u_{*}+2\pi -\Delta
u,v_{*}-\Delta v)
\end{equation}
and
\begin{equation}
(u,v)=(u_{*}+\Delta u,v_{*}-2\pi +\Delta v)\equiv (u_{*}-2\pi +\Delta
u,v_{*}+\Delta v)\text{,}
\end{equation}
where
\begin{equation}
\Delta u\equiv -2\pi \mu \mathcal{G}(u_{*})
\end{equation}
and
\begin{eqnarray}
\Delta v &\equiv &-\mu \mathcal{F}(u_{*}+2\pi ;u_{*})  \nonumber \\
&=&-2\mu \int_{0}^{\pi }[\{F_{u}+A_{u}-\kappa (u-\tfrac{1}{2}\pi )\}-8\{\ln
(W+2u-2\pi )-\ln (2u)\}]\,du  \nonumber \\
&=&-[16\pi -8\func{Si}(2\pi )]\mu \approx -38.92\mu \text{.}  \label{del-v}
\end{eqnarray}
The deviation of $v$ by the amount $\pm 38.92\mu $ from its flat-space
values $v_{*}$ and $v_{*}\pm 2\pi $ at the four intersection points accounts
for the rotational phase shift of the same magnitude, mentioned at the end
of Section 3, that is induced by gravitational back reaction on the ACO loop
over the course of a single oscillation period.

The boundary of $\mathcal{D}(u_{*},v_{*})$ is completed by the straight line
segments (in $(u,v)$ coordinates) joining $(u_{*}-\Delta u,v_{*}+2\pi
-\Delta v)$ to $(u_{*}-2\pi +\Delta u,v_{*}+\Delta v)$, and $(u_{*}+2\pi
-\Delta u,v_{*}-\Delta v)$ to $(u_{*}+\Delta u,v_{*}-2\pi +\Delta v)$. In
the primed gauge, these segments become
\begin{equation}
v^{\prime }=v_{*}+(u^{\prime }-u_{*}+2\pi )[1-\pi ^{-1}\Delta v+\pi
^{-1}\Delta u+\mu \mathcal{G}(u^{\prime })]+\Delta v-\Delta u-\mu \mathcal{F}%
(u^{\prime };u_{*})\allowbreak  \label{V1}
\end{equation}
and
\begin{equation}
v^{\prime }=v_{*}+(u^{\prime }-u_{*}-2\pi )[1-\pi ^{-1}\Delta v+\pi
^{-1}\Delta u+\mu \mathcal{G}(u^{\prime })]-\Delta v+\Delta u-\mu \mathcal{F}%
(u^{\prime };u_{*})\allowbreak  \label{V2}
\end{equation}
to first order in $\mu $.

The equation (\ref{L-inv}) therefore becomes:
\begin{equation}
L_{\text{I}}(u_{*},v_{*})=2\left( \int_{u_{*}-2\pi +\Delta
u}^{u_{*}}\int_{v_{*}}^{v_{1}(u^{\prime })}\gamma ^{1/2}dv^{\prime
}\,du^{\prime }+\int_{u_{*}}^{u_{*}+2\pi -\Delta u}\int_{v_{2}(u^{\prime
})}^{v_{*}}\gamma ^{1/2}dv^{\prime }\,du^{\prime }\right) ^{1/2}
\label{L-inv2}
\end{equation}
where $v_{1}$ and $v_{2}$ are the functions on the right-hand sides of (\ref
{V1}) and (\ref{V2}) respectively, and the form of $\gamma ^{1/2}$ is given
in (\ref{met-det}), with $u$ and $v$ everywhere replaced by $u^{\prime }$
and $v^{\prime }$. If the scale factor $e^{-2\mu (u+v)}$ in (\ref{met-det})
is expanded as $e^{-2\mu (u_{*}+v_{*})}[1-2\mu (u^{\prime }-u_{*}+v^{\prime
}-v_{*})]$ then the integrals inside the square root in (\ref{L-inv2}) can
be broken into two parts:
\begin{equation}
(\tfrac{1}{4\pi }L_{0})^{2}e^{-2\kappa \mu (u_{*}+v_{*})}\left(
\int_{u_{*}-2\pi +\Delta u}^{u_{*}}[v_{1}(u^{\prime })-v_{*}]\,du^{\prime
}+\int_{u_{*}}^{u_{*}+2\pi -\Delta u}[v_{*}-v_{2}(u^{\prime })]\,du^{\prime
}\right)  \label{gamma-int1}
\end{equation}
and
\begin{equation}
(\tfrac{1}{4\pi }L_{0})^{2}e^{-2\kappa \mu (u_{*}+v_{*})}\left(
\int_{u_{*}-2\pi }^{u_{*}}\int_{v_{*}}^{u^{\prime }+2\pi -u_{*}+v_{*}}\mu
\Delta dv^{\prime }\,du^{\prime }+\int_{u_{*}}^{u_{*}+2\pi }\int_{u^{\prime
}-2\pi -u_{*}+v_{*}}^{v_{*}}\mu \Delta dv^{\prime }\,du^{\prime }\right)
\text{,}  \label{gamma-int2}
\end{equation}
where
\begin{eqnarray}
\Delta &\equiv &-2(u^{\prime }-u_{*}+v^{\prime }-v_{*})-\{F_{u^{\prime
}}-\kappa (u^{\prime }-\tfrac{1}{2}\pi )\}+(v^{\prime }-v_{*})\partial
_{u^{\prime }}\mathcal{G}(u^{\prime })  \nonumber \\
&&-8\{-\ln (W+2u^{\prime }-2\pi )+\ln (2u^{\prime })+K(u^{\prime })\cos
2u^{\prime }+\Sigma (u^{\prime })\sin 2u^{\prime }\}\text{.}  \nonumber \\
&&
\end{eqnarray}

Now, it is easily seen that the contribution of the term $-2(u^{\prime
}-u_{*}+v^{\prime }-v_{*})$ in $\Delta $ to the integrals in (\ref
{gamma-int2}) is zero, and since the remaining terms in $\Delta $ are all
half-range periodic functions of $u^{\prime }$ it turns out that
\begin{eqnarray}
&&\int_{u_{*}-2\pi }^{u_{*}}\int_{v_{*}}^{u^{\prime }+2\pi -u_{*}+v_{*}}\mu
\Delta dv^{\prime }\,du^{\prime }+\int_{u_{*}}^{u_{*}+2\pi }\int_{u^{\prime
}-2\pi -u_{*}+v_{*}}^{v_{*}}\mu \Delta dv^{\prime }\,du^{\prime }  \nonumber
\\
&=&2\pi \mu \int_{u_{*}}^{u_{*}+2\pi }(u^{\prime }-u_{*}-\pi )\partial
_{u^{\prime }}\mathcal{G}(u^{\prime })\,du-4\pi \mu \int_{0}^{\pi
}\{F_{u}-\kappa (u-\tfrac{1}{2}\pi )\}\,du  \nonumber \\
&&-32\pi \mu \int_{0}^{\pi }\{-\ln (W+2u^{\prime }-2\pi )+\ln (2u^{\prime
})+K(u^{\prime })\cos 2u^{\prime }+\Sigma (u^{\prime })\sin 2u^{\prime
}\}\,du^{\prime }\text{.}  \nonumber \\
&&  \label{gamma-int3}
\end{eqnarray}
Furthermore, the contribution of the second integral here is zero, as $F(\pi
)=F(0)$ and $u-\tfrac{1}{2}\pi $ integrates to zero.

Turning to the integrals in (\ref{gamma-int1}), the terms in $%
v_{1}(u^{\prime })-v_{*}$ and $v_{*}-v_{2}(u^{\prime })$ of order $\mu ^{0}$
contribute $4\pi ^{2}$ to the integrals as expected, while the terms
proportional to $\Delta u$ and $\Delta v$ integrate to give zero to order $%
\mu $. Also, since $\mathcal{G}$ is half-range periodic in $u^{\prime }$,
and $\mu \mathcal{F}(u^{\prime };u_{*})-\mu \mathcal{F}(u^{\prime }-2\pi
;u_{*})=-\Delta v$ from (\ref{F-int}) and (\ref{del-v}), it turns out that
\begin{equation}
\int_{u_{*}-2\pi +\Delta u}^{u_{*}}[v_{1}(u^{\prime })-v_{*}]\,du^{\prime
}+\int_{u_{*}}^{u_{*}+2\pi -\Delta u}[v_{*}-v_{2}(u^{\prime })]\,du^{\prime
}=4\pi ^{2}+2\pi \mu \int_{u_{*}}^{u_{*}+2\pi }\mathcal{G}(u^{\prime
})\,du^{\prime }-2\pi \Delta v\text{.}  \label{gamma-int4}
\end{equation}
Combining (\ref{gamma-int3}) and (\ref{gamma-int4}) gives:
\begin{eqnarray}
&&\int_{u_{*}-2\pi +\Delta u}^{u_{*}}\int_{v_{*}}^{v_{1}(u^{\prime })}\gamma
^{1/2}dv^{\prime }\,du^{\prime }+\int_{u_{*}}^{u_{*}+2\pi -\Delta
u}\int_{v_{2}(u^{\prime })}^{v_{*}}\gamma ^{1/2}dv^{\prime }\,du^{\prime }
\nonumber \\
&=&4\pi ^{2}+4\pi ^{2}\mu \mathcal{G}(u_{*})\,+2\pi \mu [16\pi -8\func{Si}%
(2\pi )]  \nonumber \\
&&-32\pi \mu \int_{0}^{\pi }\{-\ln (W+2u^{\prime }-2\pi )+\ln (2u^{\prime
})+K(u^{\prime })\cos 2u^{\prime }+\Sigma (u^{\prime })\sin 2u^{\prime
}\}\,du^{\prime }  \nonumber \\
&&
\end{eqnarray}
where the value of the integral in the final term is $2\pi \ln 2\pi +\func{Si%
}(2\pi )-3\pi $.

So, to first order in $\mu $, the invariant length of the loop is:
\begin{eqnarray}
L_{\text{I}}(u_{*},v_{*}) &=&2(\tfrac{1}{4\pi }L_{0})e^{-\kappa \mu
(u_{*}+v_{*})}\{4\pi ^{2}+4\pi ^{2}\mu [\allowbreak \kappa u_{*}+4(\gamma _{%
\text{E}}-\func{Ci}W_{*}+\ln W_{*})+C_{1}]  \nonumber \\
&&-16\pi \mu [-8\pi +4\pi \ln 2\pi +3\func{Si}(2\pi )]\}^{1/2}  \nonumber \\
&=&L_{0}\,e^{-\kappa \mu (u_{*}+v_{*})}\{1+\mu [\allowbreak \tfrac{1}{2}%
\kappa u_{*}+2(\gamma _{\text{E}}-\func{Ci}W_{*}+\ln W_{*})+C_{1}]  \nonumber
\\
&&-2\mu [-8+4\ln 2\pi +3\pi ^{-1}\func{Si}(2\pi )]\}
\end{eqnarray}
for $u_{*}\in (0,\pi )$. For values of $u_{*}$ outside this range, the term
multiplying the scale factor $e^{-\kappa \mu (u_{*}+v_{*})}$ should be
replaced by its half-range periodic extension. Furthermore the $u_{*}$%
-dependent part of this term, $\tfrac{1}{2}\kappa u_{*}+2(\gamma _{\text{E}}-%
\func{Ci}W_{*}+\ln W_{*})$, tends to $2(\gamma _{\text{E}}-\func{Ci}2\pi
+\ln 2\pi )$ as $u_{*}\rightarrow 0^{+}$, and to $\tfrac{1}{2}\kappa \pi $
as $u_{*}\rightarrow \pi ^{-}$. In view of the value (\ref{kappa}) deduced
for $\kappa $ in Section 6.3, the two limits are the same, and so $L_{\text{I%
}}$ is continuous for all values of $u_{*}$ and $v_{*}$.

LIST\ OF\ FIGURE CAPTIONS:\bigskip

Figure 1: $y$-$z$ projection of the Allen-Casper-Ottewill loop.\bigskip

Figure 2: Schematic representation of the evaporation of the ACO
loop.\bigskip

Figure 3: Schematic picture of the relationship between the field point $[t,%
\mathbf{x}]$

and its image point $[\bar{t},\mathbf{\bar{x}}]$.\bigskip

Figure 4a: Variation of $h_{tt}/(4\mu )$ with $Z$ along the $\bar{z}$%
-axis.\bigskip

Figure 4b: Variation of $h_{tx}/(4\mu )$ with $Z$ along the $\bar{z}$%
-axis.\bigskip

Figure 4c: Variation of $h_{ty}/(4\mu )$ with $Z$ along the $\bar{z}$%
-axis.\bigskip

Figure 4d: Variation of $h_{zx}/(4\mu )$ with $Z$ along the $\bar{z}$%
-axis.\bigskip

Figure 4e: Variation of $h_{zy}/(4\mu )$ with $Z$ along the $\bar{z}$%
-axis.\bigskip

Figure 5a: The correction $G$ as a function of $u$ if $C_{1}\approx
-9.\,\allowbreak 514\,3$.\bigskip

Figure 5b: The combination $F+\!$sgn$(u)\,A$ as a function of $u$ if $%
C_{2}\approx 15.\,\allowbreak 403$.\bigskip

Figure 5c: The ``natural'' correction function $A(u)=-\tfrac{2\lambda }{\pi }%
(|u|-\tfrac{1}{2}\pi )$.\bigskip

Figure 5d: The correction $F$ as a function of $u$ if $A$ has its natural
form.\bigskip

Figure 6: The Kibble-Turok sphere of the ACO loop and two possible types

of perturbation.

\newpage

\begin{figure}
\epsfig{file=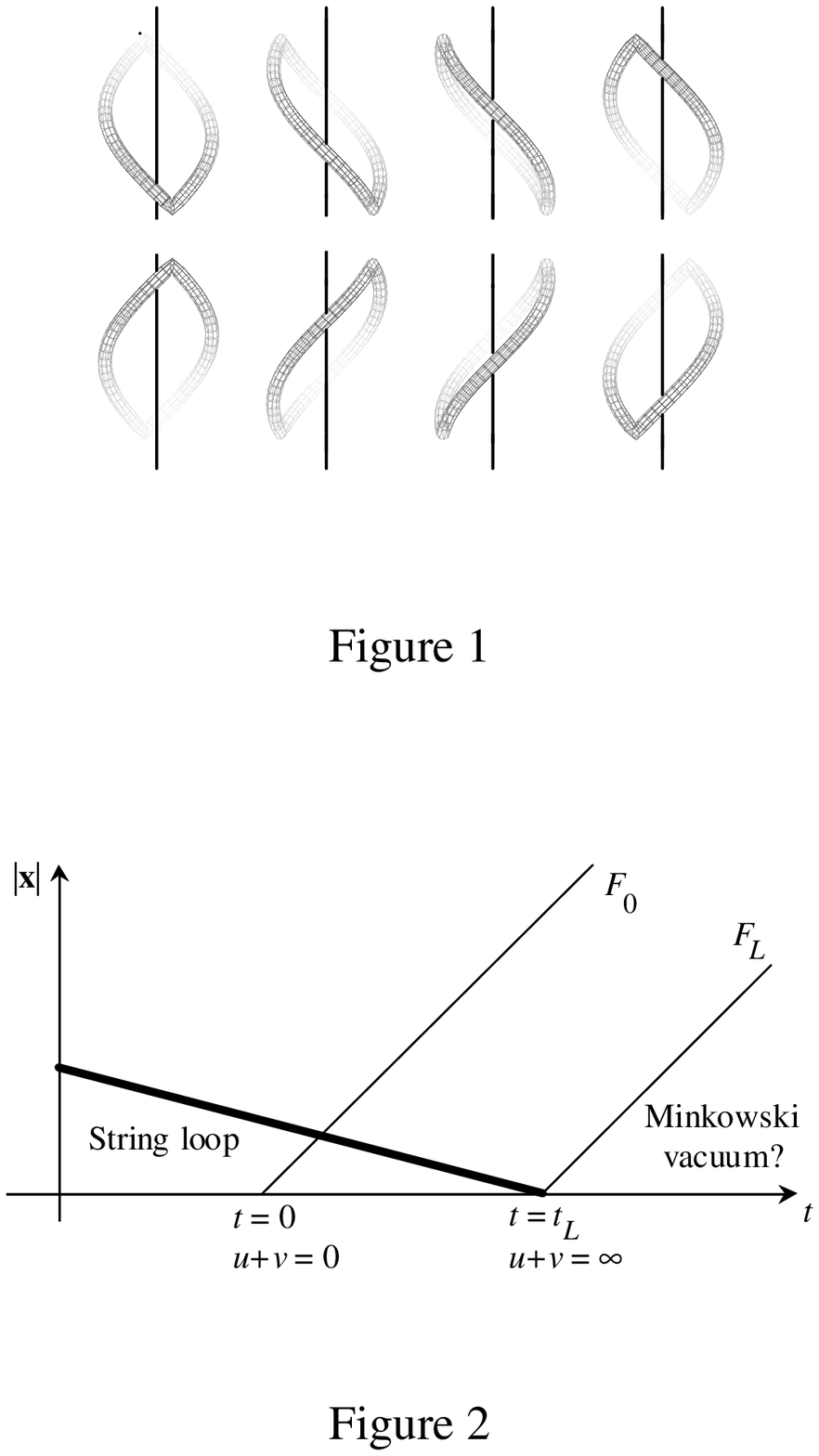, width=160mm}
\end{figure}

\newpage

\begin{figure}
\epsfig{file=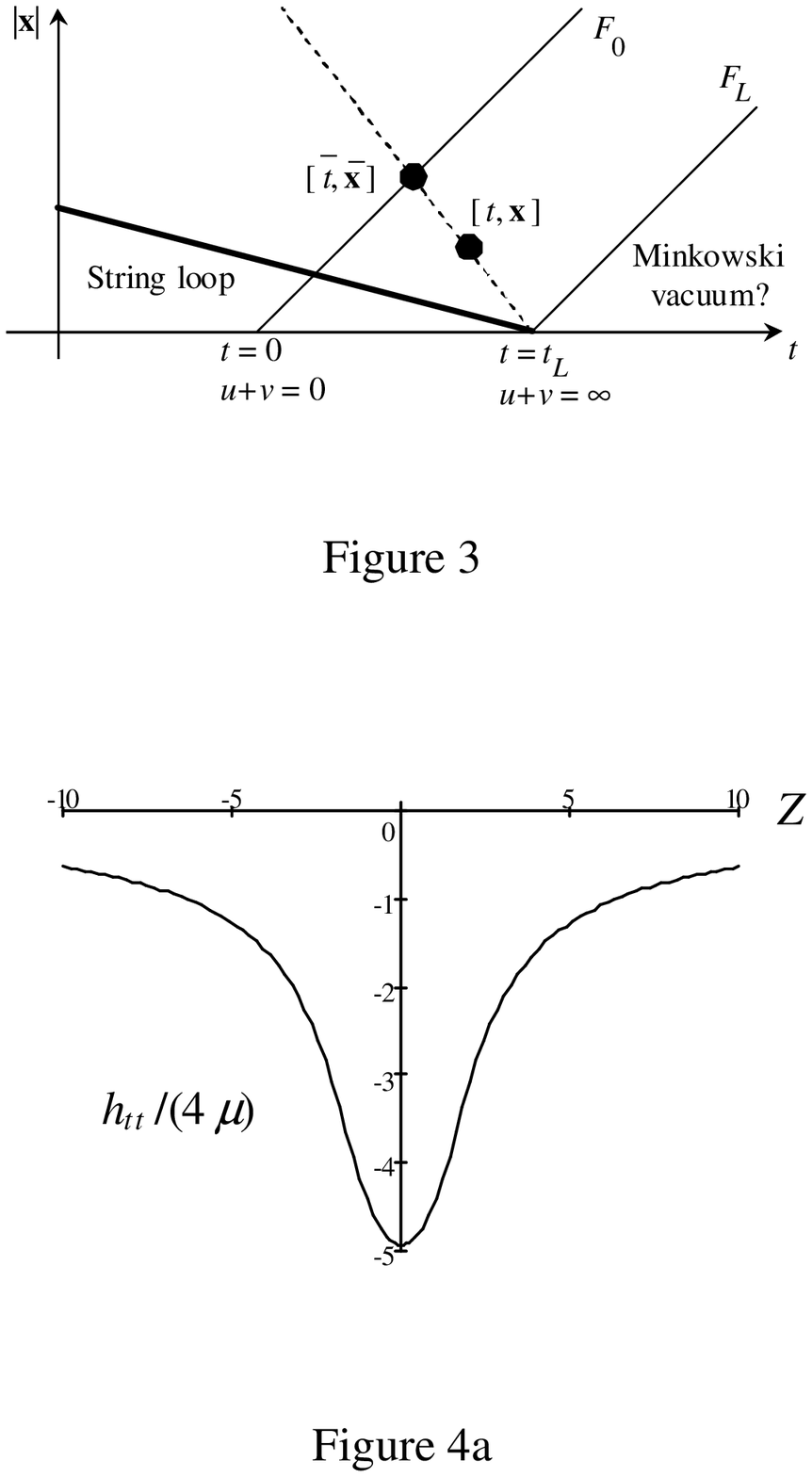, width=160mm}
\end{figure}

\newpage

\begin{figure}
\epsfig{file=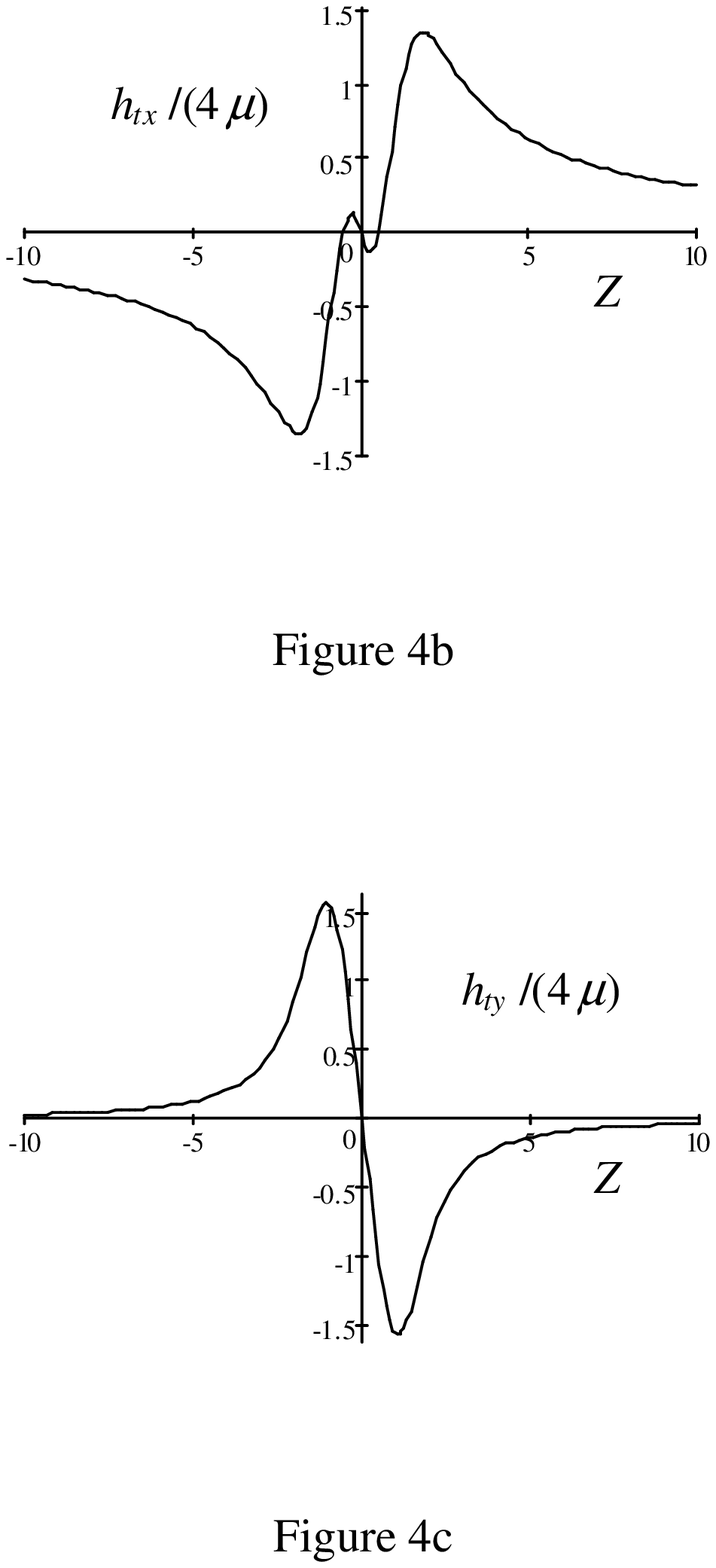, width=160mm}
\end{figure}

\newpage

\begin{figure}
\epsfig{file=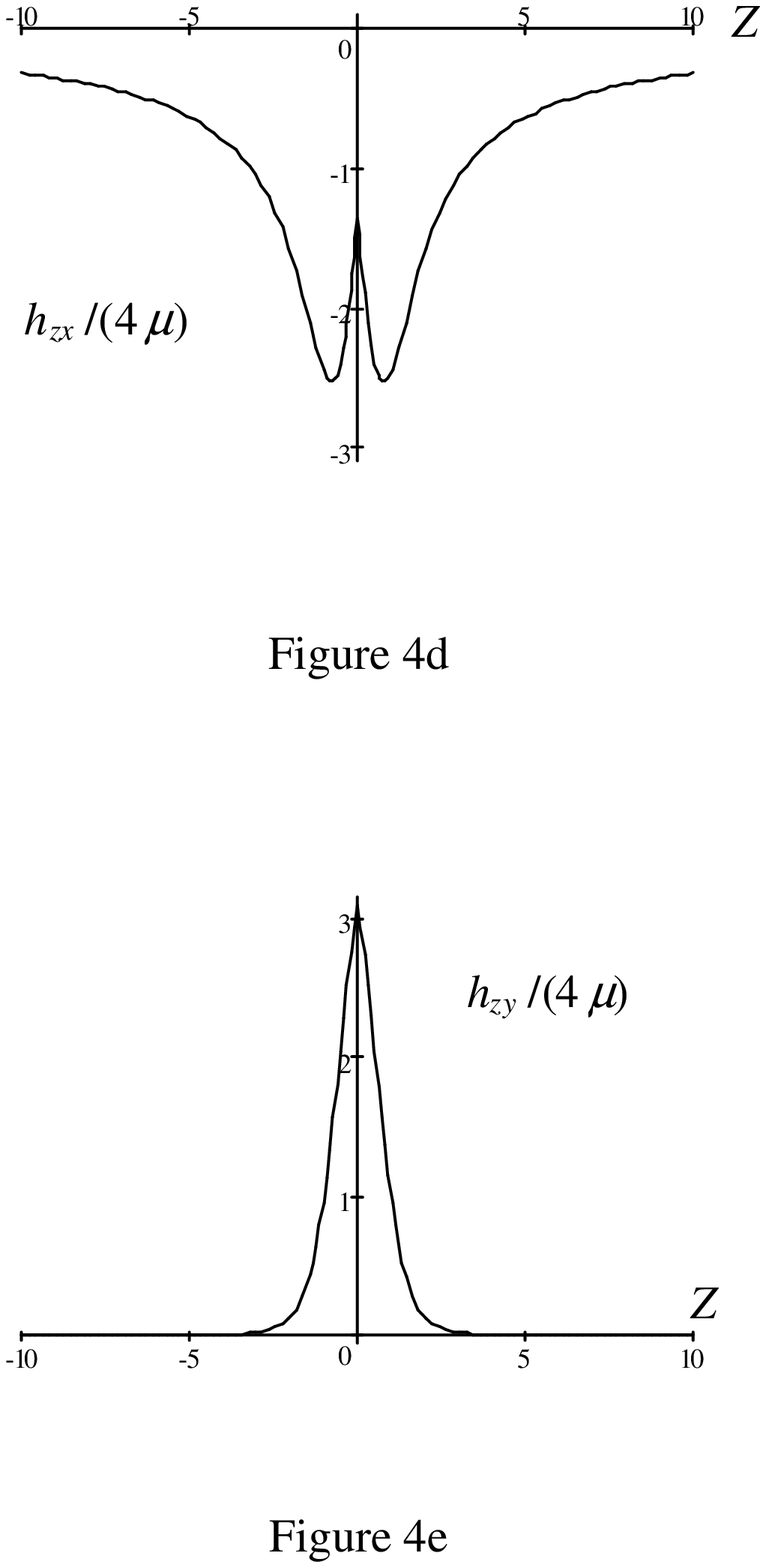, width=160mm}
\end{figure}

\newpage

\begin{figure}
\epsfig{file=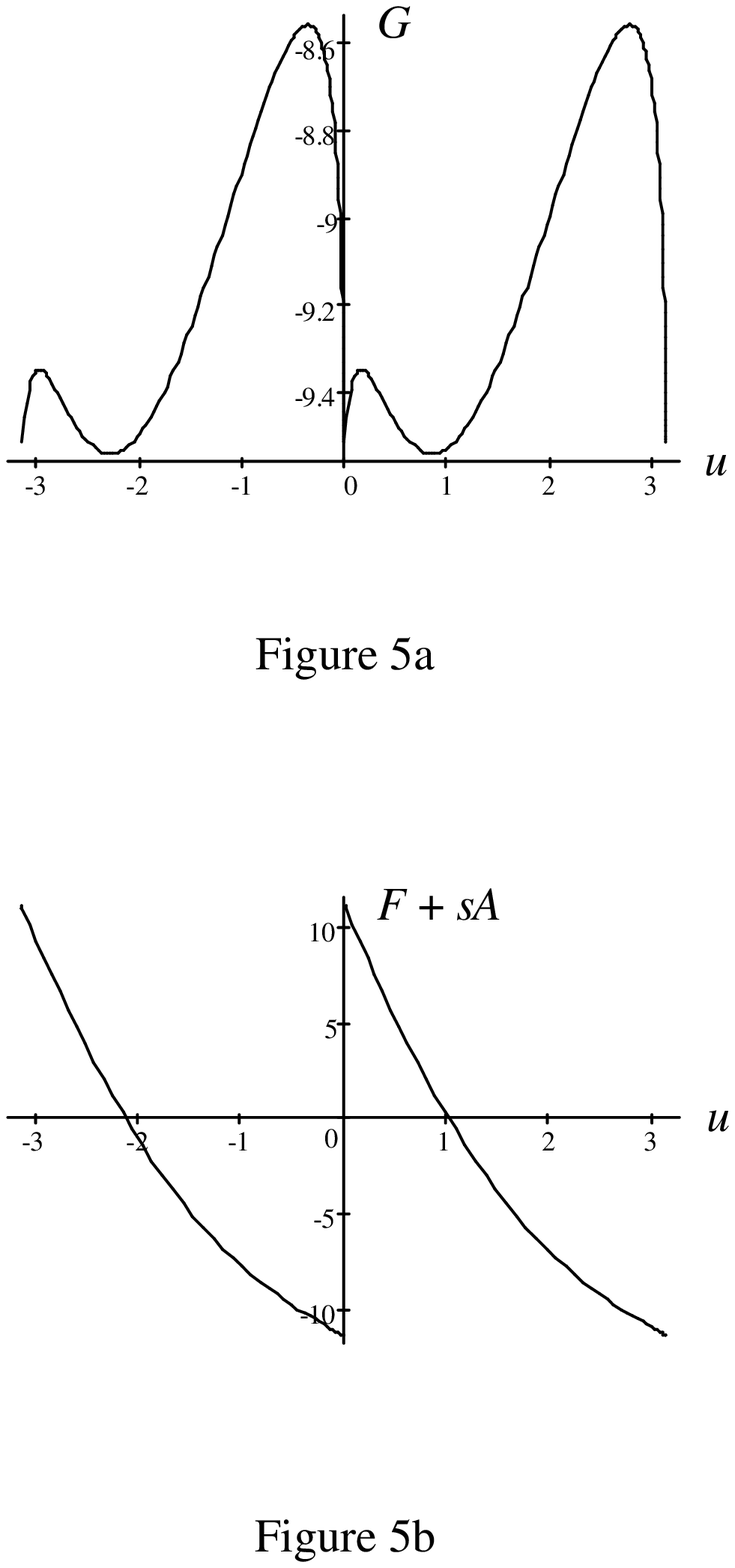, width=160mm}
\end{figure}

\newpage

\begin{figure}
\epsfig{file=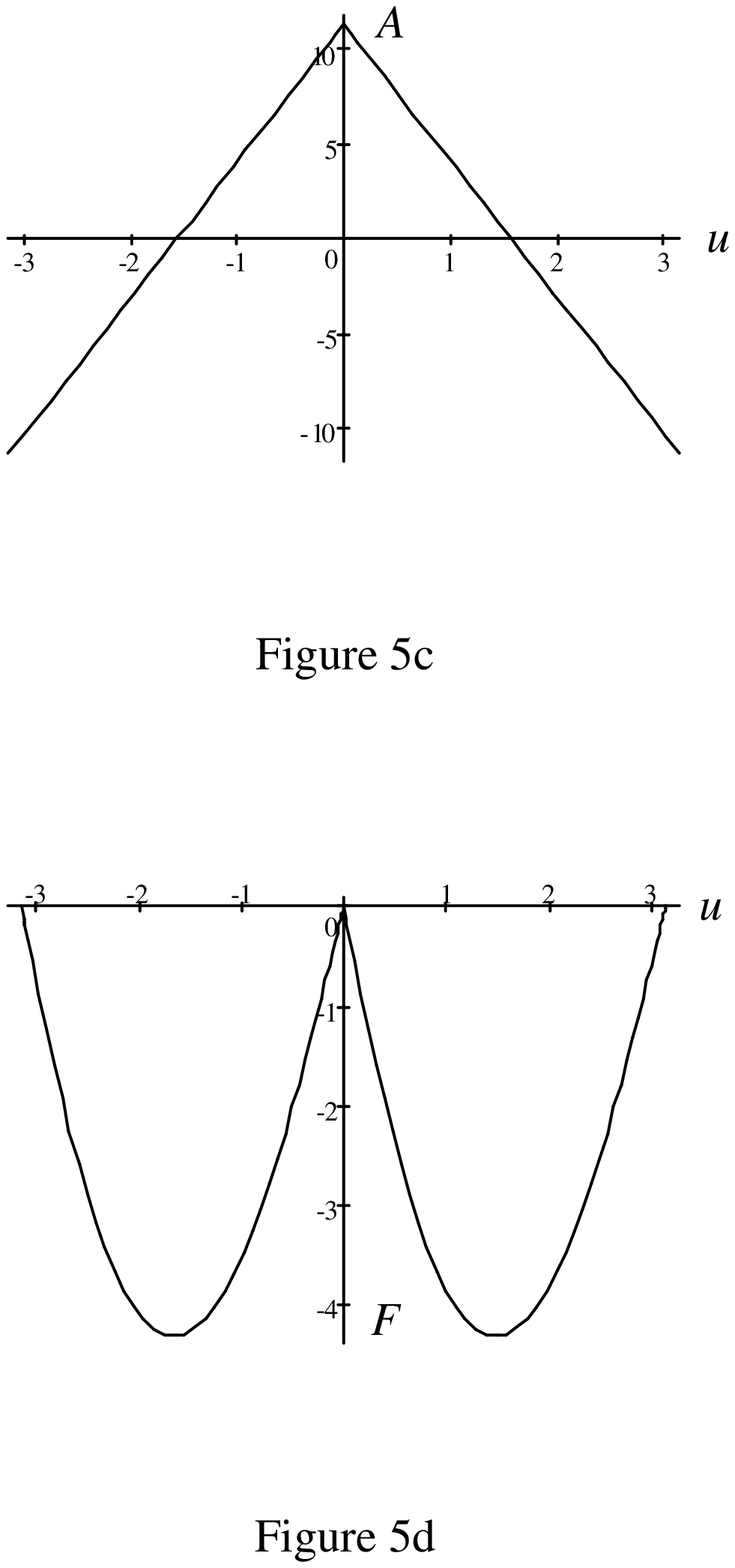, width=160mm}
\end{figure}

\newpage

\begin{figure}
\epsfig{file=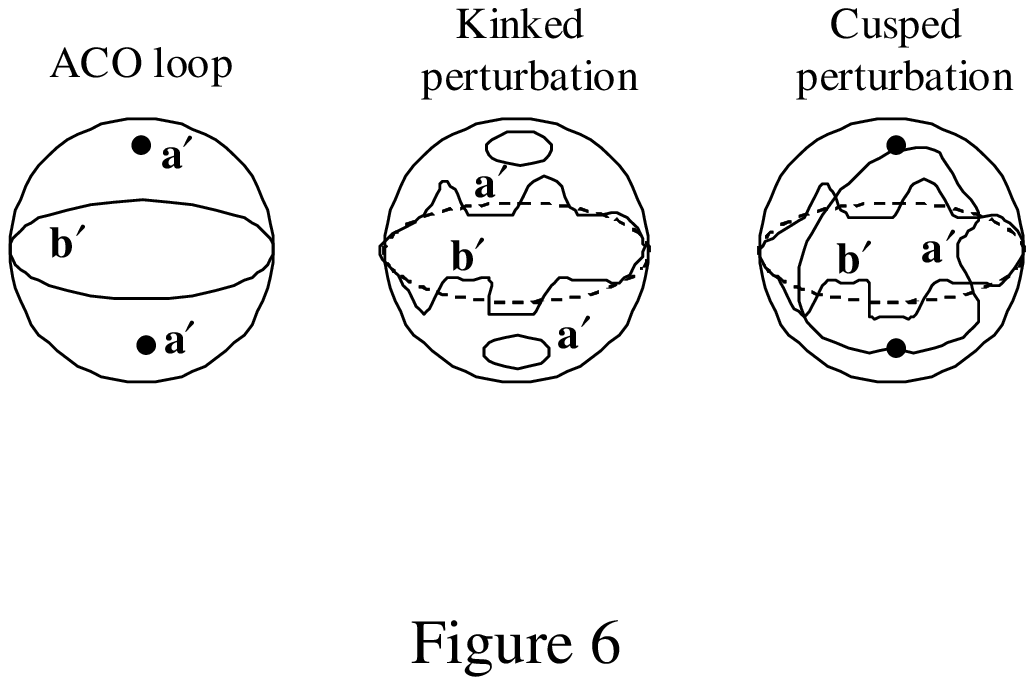, width=160mm}
\end{figure}

\end{document}